\documentclass[a4paper,11pt]{article}

\usepackage{amsmath,amssymb,amsfonts,amsthm,bm}    
\usepackage{graphicx}                           
\usepackage{hyperref}                          
\usepackage[authoryear]{natbib}                 
\usepackage{pdfpages}
\usepackage{bbm}
\usepackage{placeins}
\usepackage{amsmath}
\usepackage{amsfonts}
\usepackage{amssymb}
\usepackage{natbib}
\usepackage{graphicx}
\usepackage{pdfpages}
\usepackage[onehalfspacing]{setspace}
\usepackage[margin=1.5in]{geometry}
\usepackage{subcaption}
\usepackage{setspace}
\usepackage{placeins}
\usepackage{url}
\usepackage{xcolor}
\usepackage{mathtools}
\usepackage{titlesec}
\usepackage{algorithm2e}

\usepackage{graphicx}
\usepackage{pdfpages}
\usepackage[onehalfspacing]{setspace}
\usepackage[margin=1.5in]{geometry}
\usepackage{subcaption}
\usepackage{setspace}
\usepackage{placeins}
\usepackage{url}
\usepackage{xcolor}
\usepackage{mathtools}
\usepackage{titlesec}

\makeatletter
\def\BState{\State\hskip-\ALG@thistlm}
\makeatother

\titlelabel{\thetitle.\quad}

\begin{document}
	
	\doublespacing
	
	\def\spacingset#1{\renewcommand{\baselinestretch}%
		{#1}\small\normalsize} \spacingset{1}

	
	\title{\bf In Search of Lost Edges: A Case Study on Reconstructing Financial Networks\thanks{\textbf{Disclaimer} Data relating to SWIFT messaging flows is published with permission of S.W.I.F.T. SCRL. SWIFT\textsuperscript{\textcopyright}
			2019. All rights reserved. Because financial institutions have multiple means to exchange information about their financial transactions, SWIFT statistics on financial flows do not represent complete market or industry statistics. SWIFT disclaims all liability for any decisions based, in full or in part, on SWIFT statistics, and for their consequences.}   }
	\author{
		Michael Lebacher\thanks{Department of Statistics, Ludwig-Maximilians Universit\"at M\"unchen, michael.lebacher@stat.uni-muenchen.de}\hspace{.2cm}, Samantha Cook\thanks{Financial Network Analytics, Ltd. (FNA), sam@fna.fi}\hspace{.2cm}, Nadja Klein\thanks{ School of Business and Economics, Humboldt-Universit\"at zu Berlin, nadja.klein@hu-berlin.de }\hspace{.2cm}  and 
		G\"oran Kauermann\thanks{Department of Statistics, Ludwig-Maximilians Universit\"at M\"unchen, goeran.kauermann@stat.uni-muenchen.de}}
\date{}
	\maketitle
	
	\bigskip
	\normalsize
	\begin{abstract}
		\noindent 
		To capture the systemic complexity of international financial systems,  network data is an important prerequisite. However, dyadic data is often not available, raising the need for methods that allow for reconstructing networks based on limited information.
		In this paper, we are reviewing different methods that are designed for the estimation of matrices from their marginals and potentially exogenous information. This includes a  general discussion of the available methodology that provides edge probabilities as well as models that are focussed on the reconstruction of edge values. Besides summarizing the advantages, shortfalls and computational issues of the approaches, we put them into a competitive comparison using the SWIFT (Society for Worldwide Interbank Financial Telecommunication) MT 103  payment messages network (MT 103: Single Customer Credit Transfer).  This network is not only economically meaningful but also fully observed which allows for an extensive competitive horse race of methods.
		The comparison concerning the binary reconstruction is divided into an evaluation of the edge probabilities and the quality of the reconstructed degree structures. Furthermore, the accuracy of the predicted edge values is investigated.     
		To test the methods on different topologies, the application is split into two parts. The first part considers the full MT 103 network, being an illustration for the reconstruction of large, sparse financial networks. The second part is concerned with reconstructing a subset of the full network, representing a dense medium-sized network. 
		Regarding substantial outcomes, it can be found that no method is superior in every respect and that the preferred model choice highly depends on the goal of the analysis,  the presumed network structure and the availability of exogenous information.
	\end{abstract}
	\noindent%
	\vfill
	
	\newpage
	\section{Introduction} 	
	In recent years, interest in applying network-based methodology to financial data has strongly increased (see e.g.\ \citealp{soramaki2007topology}, \citealp{schweitzer2009economic}, \citealp{imakubo2010transaction},  \citealp{baek2014network},  \citealp{battiston2016complexity}).  A huge amount of this research effort is directed to the study and assessment of systemic risk (see e.g.\ \citealp{gai2010contagion}, \citealp{kaue2012structure}, \citealp{billio2012econometric}, \citealp{chinazzi2013post}, \citealp{thurner2013debtrank}, \citealp{soramaki2013sinkrank},  \citealp{bardoscia2017pathways} and \citealp{caccioli2018network}).  This focus stems from the fact that in the aftermath of the financial crisis it became clear that the banking system forms a complex network with inherent interdependencies and feedback loops. As a consequence, the centrality and connectedness of a financial institution can be just as important as size for its potential to wreak havoc on the system overall (\citealp{markose2012too}, \citealp{liu2015banking}). \citet{battiston2012debtrank} even suggest to add the term ``too-central-to-fail'' to the discussion of ``too-big-to-fail'' institutions.     Given that, investigation of the topologies of financial networks is very important for regulators, central banks and other institutions concerned with the stability of the financial system. Although considerable effort is put into modeling system risk,  those methods generally require information from the full network that is most often not observed. This raises the need for a methodology that allows providing an accurate reconstruction of the networks derived from the limited information available.
	
	The canonical examples of network reconstruction in finance are exposure networks created by interbank loans. In these networks, the total assets and liabilities of a given bank are mostly known, but the actual loans made to other banks, i.e.\ the binary edge structure (existence or non-existence of loans) and their corresponding edge weights (loan volume), are unobserved.  Knowledge of the edges and their values is nevertheless crucial to measure the systemic risk in the exposure network. If one bank fails to meet its' obligations that could lead its' creditor(s) unable to make their obligations which leads to further contagion, potentially affecting all banks or a large portion of the network. An example of how to process such information, if available, is  \textsl{DebtRank} (\citealp{battiston2012debtrank}), being a popular metric for assessing systemic importance in exposure networks based on the values of loans between bank pairs. 
	
	Although the reconstruction problem is introduced here as a task that belongs to the realms of Finance or Economics, it emerges in many different disciplines. In order to get an overview, from the perspective of Economics, see for example \citet{sheldon1998}, \citet{upper2011} and \citet{elsinger2013}. The article by \citet{SQUARTINI20181} provides a very broad overview from a methodological perspective, based on maximum-entropy methods and Statistical Physics (see also \citealp{cimini2015} and \citealp{mastrandrea2014}). In Computer Sciences and Statistics, a similar problem is often called traffic matrix estimation or network tomography and this research branch developed its' own methodological toolkit (e.g.\ \citealp{castro2004network}, \citealp{zhang2003}, \citealp{airoldi2013}, \citealp{ZHOU2016220} and \citealp{nie2017modeling}). In the given paper, not all models proposed in different research fields can be included but we have selected the ones that are feasible and potentially useful for the given data situation.
	
	A good reference point for this paper is certainly the extensive study by \citet{anand2018}. In their paper, they employ seven different reconstruction methods to 25 different networks. Although we are not that ambitious regarding the variety of use cases, our approach can be seen as a related paper that focusses on other aspects. First of all, we do not restrict our methodology to methods that rely only on aggregated row- and column sums but also include density-calibrated methods and models that are capable of incorporating exogenous covariates. Further, it is tried to propose regularized least-squares models inspired by the network tomography literature and new methodology not considered by \citet{anand2018}. Additionally, we provide a more detailed technical exposition of the models in a manner that is comprehensible for practitioners.  Regarding the evaluation techniques, we separate the evaluation of the binary and valued reconstruction more clearly and employ measures that are more standard in Statistics and Machine Learning.
	
	To compare the different models, we use data provided by the Society for Worldwide Interbank Financial Telecommunication (SWIFT, \url{www.swift.com}). SWIFT acts as an infrastructure for financial institutions and enables them to send and receive information about financial transactions encoded in the form of secure standardized messages. 
	One of the most important types of messages is the \textsl{MT 103 single customer credit transfer}, representing payments sent between clients of financial institutions. 	The MT 103 data under study consist of monthly bilateral message counts aggregated at the country level between January 2003 and February 2018.  Note that the concept of a country here is not limited to independent, passport-granting states, but also includes territories (e.g.\ Turks and Caicos Islands), dependencies (e.g.\ Guernsey and Jersey) and autonomous constituent states (e.g.\ Greenland).
	
	The SWIFT network is especially suitable for testing network reconstruction methods because it is an economically meaningful data set (see \citealp{cook2014global} for an extensive investigation). Further, the data provides a long time series available for testing with full link data available. Hence, in this dataset, it is known exactly how well different methods work allowing to compare different models.
	
	The article is structured as follows. In Section \ref{sec:notation} we formalize the problem and give general notation for the paper. This is  followed by a description of the SWIFT data in Section \ref{sec:description}. In Section \ref{sec:models} we introduce the models under study and their evaluation is provided in Section \ref{sec:eval}. Section \ref{sec:conclusion} discusses the results and  concludes the paper.\footnote{We provide the code online at Github \url{https://github.com/lebachem/lost_edges} . Because the used data set is confidential, the code is not accompanied with the actual dataset but with a ``fake dataset'' that does \textit{not} represent the original data but only the same dimension and a similar density.}

	\section{Notation} \label{sec:notation}
	The SWIFT MT 103 messages can be represented as a series of matrices $\mathbf{X}^t=(x_{ij}^t)$ containing dyadic count data. The elements of $\mathbf{X}^t$ can be interpreted as directed edge values  $x_{ij}^t\in \mathbb{N}_0$ among $i,j=1,...,n$ countries  at time points  $t=1,...,T$. We exclude self-loops from our study and,  therefore, elements $x_{ii}^t$ are left undefined for $i=1,...,n$. Accordingly, within-country payments are not regarded. We also assume that the number of nodes $n$ is invariant with respect to time so that at each time point $t$ the number of variables is given by $N=n(n-1)$.
	
	\subsection{Binary Network Structure}
	Although the binary networks structure is readily available if the valued structure is given, both aspects of the network need to be  modelled separately. To account for this aspect, we  also introduce notation for the binary network structure. Let  $\mathbf{Z}^t=(z^t_{ij})$ denote the binary networks, defined via
	\begin{equation*}
	z_{ij}^t=I(x_{ij}^t>0)\text{, for } i  \neq j,
	\end{equation*}
	with elements $z_{ij}^t$ being indicators whether the corresponding entry of the matrix is zero or greater than zero. Let the \textsl{density} (also called the connectivity) of the network be
	\begin{equation*}
	\mathcal{D}^t=\frac{1}{N}\sum_{i\neq j}z_{ij}^t\text{, for } t=1,...,T
	\end{equation*}
	providing the number of non-zero edges in the network relative to the number of possible edges at time point $t$. Additionally, we define the number of outgoing edges to be the \textsl{outdegree} and the number of ingoing edges is measured with the \textsl{indegree}. Formally, the outdegree and the indegree for node $i$ at time point $t$ are given by
	\begin{equation}
	\label{eq:margin_bin}
	\begin{split}
	z_{i \bullet}^t&=\sum_{k \neq i}z_{ik}^t\text{, for }i=1,...,n\\
	z_{\bullet i}^t&=\sum_{k \neq i}z_{ki}^t\text{, for }i=1,...,n.
	\end{split}
	\end{equation}

	\subsection{Valued Network}
	Similarly, we are interested in the row and column sums of the valued network, i.e.\ the \textsl{valued in}- and \textsl{outdegree}. Other names in the network literature describing the same concepts are the in-strength and out-strength or the weighted in- and outdegree.   Let the $i$th valued outdegree and valued indegree be
	\begin{equation}
	\label{eq:sum}
	\begin{split}
	x_{i \bullet}^t&=\sum_{k \neq i}x_{ik}^t\text{, for }i=1,...,n\\
	x_{\bullet i}^t&=\sum_{k \neq i}x_{ki}^t\text{, for }i=1,...,n.
	\end{split}
	\end{equation}
	For a more compact formulation, we stack the row and column sums, resulting in a $2n$-dimensional column vector of marginals
	\begin{equation*}
	\mathbf{y}^t=(x_{1 \bullet}^t,...,x_{n \bullet}^t,x_{ \bullet1}^t,...,x_{ \bullet n}^t)^T \text{, for } t=1,...,T.	
	\end{equation*}
	Furthermore, let
	\begin{equation*}
	\mathbf{x}^t=(x^t_{12},...,x^t_{1n},x^t_{21}...,x^t_{n(n-1)})^T\text{, for } t=1,...,T
	\end{equation*}  be an $N$-dimensional column vector containing the values of the edges (without diagonal elements) and define the known binary $(2n \times N)$ routing matrix $\mathbf{A}^t$ such that the linear relation
	\begin{equation}
	\label{margin}
	\mathbf{y}^t=\mathbf{A}^t \mathbf{x}^t
	\end{equation}
	holds for $t=1,...,T$. Note that relation (\ref{margin}) is just a compact way of writing equations (\ref{eq:sum}) in matrix notation.
	Henceforth, we will refer to relation (\ref{margin}) as \textsl{marginal restrictions}.
	The restriction that all matrix entries are non-negative is referred to as \textsl{non-negativity constraint}.
	If we refer to methods that yield stochastic solutions we adopt the nomenclature from Physics  and 
	label a collection of sampled networks as \textsl{network ensemble} (e.g.\ \citealp{bargigli2014}).
	In the model description we will suppress the time-superscript in most representations for ease of notation.
	
	As a general convention, vectors and matrices are given in bold and (with the exception of the deterministic routing matrix $\mathbf{A}$), random variables are given by upper case and realisations by lower case letters.

	\section{Data description} \label{sec:description}
	\begin{figure}[t!]
		\centering
		\begin{subfigure}{\textwidth}
			\centering			\includegraphics[trim={0cm 0cm 0cm 0cm},clip,width=\textwidth]{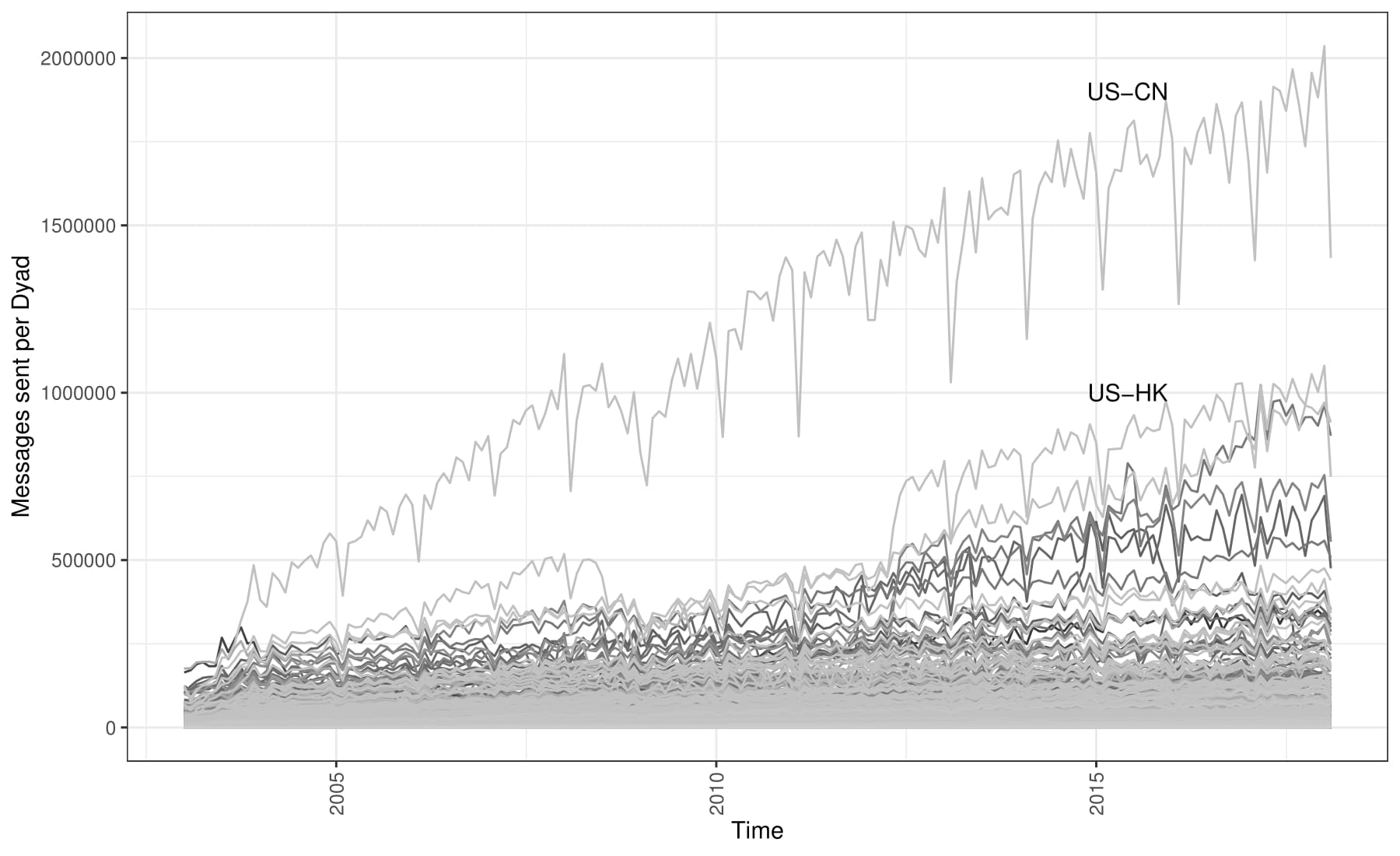}
		\end{subfigure}

		\caption{Time series of valued edges $x^t_{ij}$ in the full MT 103 network on a monthly basis. Messages per edge on the vertical axis, time measured in months on the horizontal axis. \\ Source: SWIFT BI Watch.}
		\label{fig:messages}
	\end{figure}
	The data under study is provided by the Society for Worldwide Interbank Financial Telecommunication (SWIFT, \href{www.swift.com}{www.swift.com}) and provides standardized messages called MT 103, representing payment transfers.	 The data is aggregated to the country level and allow to construct a network $\mathbf{x}^t$ where the countries are the nodes and the directed, valued edges $x_{ij}^t$ between them represent the number of messages sent from country $i$ to country $j$ at time point $t$. The available database covers $T=182$ time points on a monthly basis, ranging from January 2003 to February 2018. 
	
	We restrict our analysis to the $n=203$ countries that are existent during the whole observational period. This includes one entity that is not a country or a territory but represents international market infrastructure, referring to monetary organizations that operate in many countries
	(see the country list in Table \ref{tab:countries} of  Appendix \ref{annex:included}). The network including all 203 countries is set to be the baseline for all models that can deal with ``big'' networks and is labeled \textsl{full network}.
	
	In Figure \ref{fig:messages} we plot all individual edges of the full network against time. Notably, there is a great deal of country-related heterogeneity in the data. The time series with the highest time-averaged amount of messages sent corresponds to the edge United States - China (US-CN) and is on average almost ten times higher than the second-highest valued edge (United States - Hong Kong, US-HK). Furthermore, already the yearly  80\% quantile of the number of messages within each month ranges only between one and five messages per edge. This implies that the major share of all messages is sent and received by a  small subset of countries characterized by a high-intensity exchange. 
	
	To analyze both, the full network and its' ``dense core'' we additionally investigate a reduced dataset, containing the 59 most important countries. This network is labeled to be the  \textsl{reduced network}. Depending on the month, the reduced network accounts for about $85\%$ up to $91\%$ of all messages sent in the whole system. In the following, we show descriptive measures for the full network but in the Annex \ref{annex:reduced} the same descriptives are provided for the reduced network.

	\begin{figure}[t!]
		\centering
		\begin{subfigure}{\textwidth}
			\centering			\includegraphics[trim={0cm 0cm 0cm 0cm},clip,width=\textwidth]{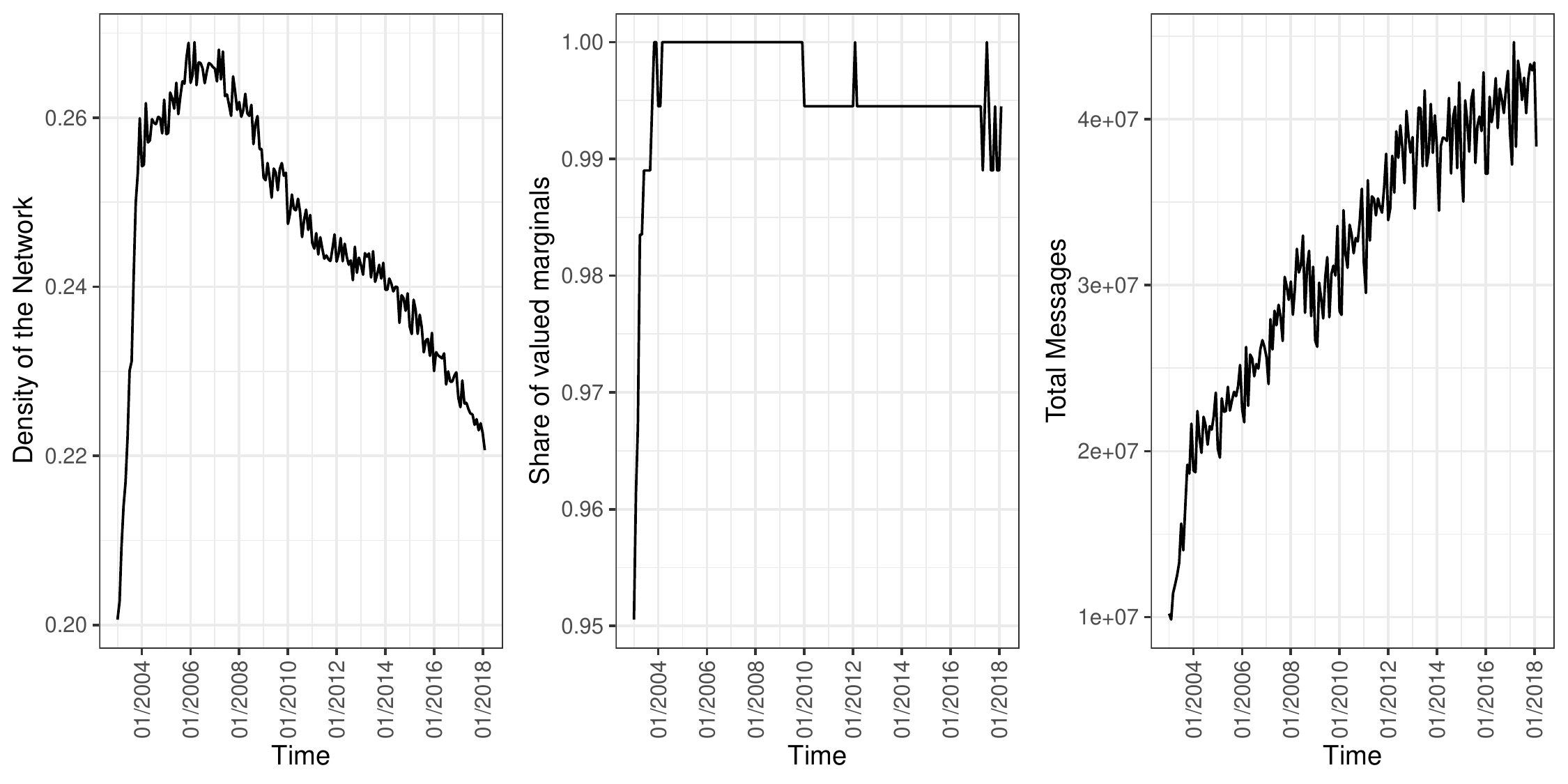}
		\end{subfigure}

		\caption{Summary statistics for the full  MT 103 network as monthly time series. Density of the network (left), share of non-zero marginals $\mathbf{y}^t$ (middle) and cumulative edge values (right). \\ Source: SWIFT BI Watch.}
		\label{fig:nwsummary}
	\end{figure}
	
	The structure of the full binary networks $\mathbf{z}^t$ is summarized in Figure \ref{fig:nwsummary}. On the left-hand side, it can be seen that the density of the network decreases steadily from 2006 on whereas, the development of the total MT 103 messages (right-hand side) follows a clear upward trend. This pattern implies that increasingly more messages are sent per edge. Similarly, but in a more modest form, this can also be concluded for the reduced network (see Figure \ref{fig:nwsummary_small} in the Annex \ref{annex:reduced}). The SWIFT data exclusively contains countries that send or receive MT 103 messages. Therefore, each country can only have either a zero out- or indegree.
	However, if many countries would be restricted to only receive or send messages, the dimensionality of the problem could be greatly reduced. This can be investigated by calculating the share of non-zero marginals $\mathbf{y}^t$ for each year. The resulting plot is given in the middle plot of Figure \ref{fig:nwsummary} and we find that the low density is not mirrored by a low share of valued marginals. This means that almost no information about the density can be inferred from the marginals since the vast majority of them are greater zero. In the reduced network, the density is much higher (about 0.85 averaged over all months) without any zero marginal.
	\begin{figure}[t!]
		\centering
		\begin{subfigure}{\textwidth}
			\centering			\includegraphics[trim={0cm 0cm 0cm 0cm},clip,width=\textwidth]{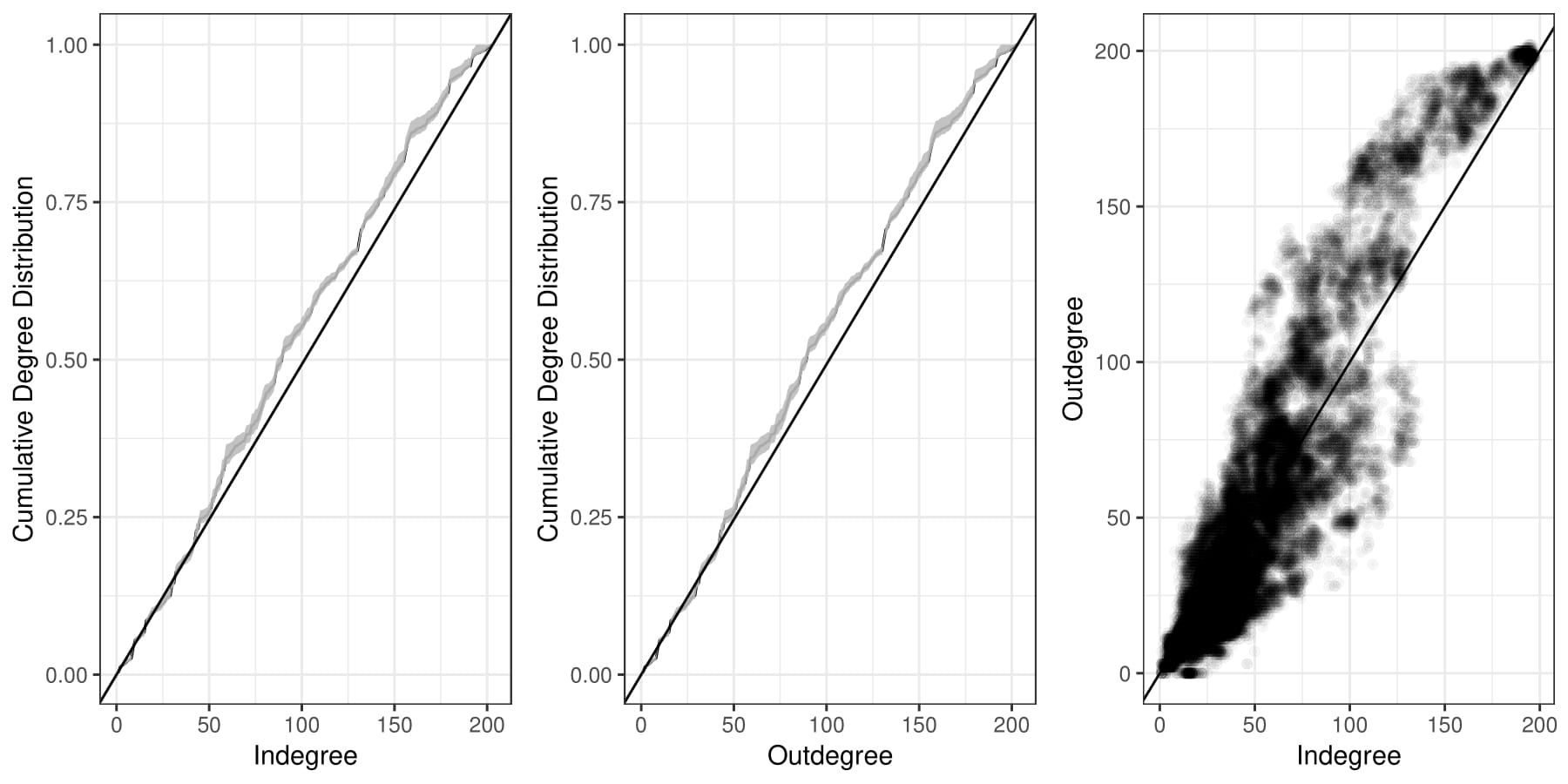}
		\end{subfigure}

		\caption{Binary network topology of the full MT 103 network aggregated for all time points. Cumulative indegree (left) and outdegree distributions (middle) with  maximum and minimum values indicated in grey. Outdegree against indegree (right) for all months in dotted with colour intensity by frequency. 45 degree line in solid black. \\ Source: SWIFT BI Watch.}
		\label{fig:degdist}
	\end{figure}
	
	In Figure \ref{fig:degdist} the degree structure of the full network is visualized. The first two panels show the cumulative degree distribution (indegree on the left and outdegree on the right) aggregated for all months. The realizations between the monthly minimum and maximum values are indicated in grey. It can be seen that both, the indegree as well as the outdegree grow close to linear and are, therefore, almost uniformly distributed. This is a rather uncommon finding and does not match with common  structures like scale-free networks (\citealp{barabasi1999}; \citealp{barabasi2002}) or random graphs (Erd\"os-R\'enyi graphs, \citealp{erdos1959}). This is consistent with the findings of \citet{cook2014global} who noted that the data cannot appropriately be described with standard power-law distributions.
	Again similar and even more pronounced results can be found for the reduced network, shown in  Figure \ref{fig:degdist_small} of Annex \ref{annex:reduced}.
	
	The right panel of Figure  \ref{fig:degdist} plots shows for all nodes and all months the indegree versus the outdegree. This can be thought of a check how ``symmetric'' the network is and it appears that there is a strong positive (non-linear) relationship between the  in- and outdegrees.
	
	In some models, exogenous data can be incorporated. Based on the empirical investigation by \citet{cook2014global}, we assume it to be plausible that financial activity in a given country is related to its' economic size and consider the annual \textsl{Gross Domestic Product} (GDP, in current USD Billions) as a valid covariate. The data is provided by the International Monetary Fund (IMF) and we denote the GDP of country $i$ by $gdp_i$.    
	
	\section{Models for Network Reconstruction} \label{sec:models}
	\subsection{Overview}
	Since almost none of the zeros in the network can be inferred from the marginals, most of the models that provide edge probabilities rely on two crucial assumptions: (i) The true density is known and (ii) the row- and column sums of the valued edges carry information about the binary structure. Both points are highly related and evolve around the basic problem that knowledge of the marginals is not sufficient to provide information about the edge probabilities (\citealp[Proposition 3.1]{gandy2017}). This fundamental identification problem can be overcome only by adding additional constraints (i.e.\ knowledge of the density). For the sake of this article, we assume the true density to be known. In practice, this implies that models that are found to perform very well in our comparison might not do so with an incorrectly specified density. Given knowledge about the true density, the second assumption is less problematic and, depending on the presumed structure of the network under study,  it can be plausible that the marginals provide information that helps to determine the edge probabilities.
	
	Another issue that complicates network reconstruction is the high dimensionality of the full network. With $n=203$ nodes, the number of dyads amounts to $N=41\,006$. This brings many methods to their computational limits. In the reduced network, the problem greatly simplifies as $N$ shrinks to $3\,422$ which allows to apply almost all methods considered in this paper. An exception is the density-corrected directed weighted configuration model (DWCM) by \citet{bargigli2014} that is not considered in this paper because the algorithm failed to converge even in the small network. In the model description, we will mention which methods are computationally tractable in the full network and in case they are not, they are only applied to the reduced network.
	
	All methods used are summarized in Table \ref{tab:model_summary}, including their names, abbreviations and references to the corresponding sections with a detailed description. Additionally, it is shown whether the methods are applied to the full network and whether knowledge of the true density is needed to calibrate the models.

	\begin{table}[t!] \centering 
		\begin{tabular}{@{\extracolsep{5pt}} lcccc} 
			\\[-1.8ex]\hline 
			Method&Abbreviation&  Section & Full netw.& Calibrated   \\
			
			\hline \hline  
			Maximum-Entropy &\textsl{IPFP}&  \ref{sec:ipf}& X&\\
			Maximum-Entropy, GDP &\textsl{IPFP-GDP}&  \ref{sec:ipf}& \\
			Maximum-Entropy, lag. values			& \textsl{IPFP-LAG}&\ref{sec:ipf}& & \\
			Gravity Model &\textsl{GRAVITY}&  \ref{sec:cimi}& X\\
			Dens. cor. Gravity Model	&\textsl{DC-GRAVITY}&  \ref{sec:cimi}& X&X\\
			Dens. cor. Gravity, GDP	&\textsl{DC-GRAVITY-GDP} &\ref{sec:cimi}&  &X \\
			Dens. cor. Gravity,  lag. values	& \textsl{DC-GRAVITY-LAG}&\ref{sec:cimi}&  &X \\
			Tomogravity Model	&\textsl{TOMOGRAVITY}&  \ref{sec:tomo}&\\
			LASSO	&\textsl{LASSO}&  \ref{sec:lasso}& X &X\\
			Hierarchical Erd\"os-R\'enyi Model	&\textsl{H-ER}&  \ref{sec:fit}& X&X\\
			Hierarchical Fitness Model	& \textsl{H-FIT}&\ref{sec:fit}& X&X\\
			
			Minimum Density &\textsl{MINDENS}&  \ref{sec:mindens}& X\\

			\hline 
		\end{tabular}

		\caption{Summary of reconstruction methods used in this article together with abbreviations, their ability to fit the full network (Full netw.) and whether calibration to the true density is needed (Calibrated).} 
		\label{tab:model_summary}
	\end{table}

	\subsection{Iterative Proportional Fitting} \label{sec:ipf}
	A very simplistic, but nevertheless powerful method to reconstruct \textsl{dense} networks is given by the iterative proportional fitting procedure (IPFP, \citealp{deming1940}, \citealp{fienberg1970}). The algorithm has gained much attention in the matrix reconstruction literature under the name maximum-entropy method because it allows for estimating the parameters of the maximum-entropy probability distribution (the methodological backbone of many reconstruction tasks, \citealp{SQUARTINI20181}).
	
	In the Statistics literature, the procedure is originally intended to provide maximum likelihood estimates for parameters of log-linear models in contingency tables (\citealp{bishop1975}, \citealp{haberman1978,haberman1979}). In the given case, this interpretation is convenient because it allows for a specific interpretation of the outcomes as the maximum likelihood estimates for the expectation of a Poisson-distributed random variable 
	\begin{equation}
	\label{eq:pois}
	X_{ij}\sim Poi(\mu_{ij})\text{, for } i \neq j,
	\end{equation}
	with log-linear expectation $\mathbb{E}[X_{ij}]=\mu_{ij}=\exp\{\delta_i + \gamma_j\}$. The two parameters $\delta_i$ and $\gamma_j$ correspond to row- and column-effects. Furthermore, the model provides a model-based possibility to calculate the probability of observing a value $X_{ij}$ greater than zero:
	\begin{equation*}
	p_{ij}:=\mathbb{P}(X_{ij}>0)=1-\mathbb{P}(X_{ij}=0)=1-\exp\{-\mu_{ij}\}\text{, for } i \neq j.
	\end{equation*}
	However, with high values for $X_{ij}$, the probabilities approach zero exponentially fast, meaning that for high marginals most probabilities will be almost or numerically even equal to one. 
	
	The model is a dense reconstruction method that provides edge values for all rows and columns with valued marginals. Besides this drawback, the model has the merit of being computationally efficient (see the \texttt{R} package \texttt{ipfp} by \citealp{blocker2014}) and due to the construction of the algorithm, it is guaranteed that the row- and column sums of the predicted entries match the observed marginals exactly. 
	
	In \citet{lebacher2019regression} the IPFP model from above is extended to incorporate informative dyadic exogenous information, which is labeled here as $C_{ij}$. In particular, the covariates $C_{ij}$ can be included in the log-linear expectation
	\begin{equation*}
	\mathbb{E}[X_{ij}|C_{ij}=c_{ij}]=\mu_{ij}=\exp\{\delta_i + \gamma_j+c_{ij}\beta\}\text{, for } i \neq j.
	\end{equation*}
	If the association between $C_{ij}$ and the unknown $X_{ij}$ is high, the prediction accuracy increases relative to the standard IPFP solution. However, this approach comes at a price since model fitting is based on constrained non-linear optimization making computation significantly more demanding than standard IPFP. Furthermore, only dyadic covariates have the potential to increase the predictive power but in practice often only monadic information is available. We take a pragmatic approach  and use a transformation of the GDP values of countries $i$ and $j$ that is not linearly separable and can be interpreted as dyadic overall GDP, defined through:
	\begin{equation}
	\label{eq:trafo}
	c_{ij}=\log(gdp_i+gdp_j)\text{, for } i\neq j.
	\end{equation}
	This yields the expectation
	\begin{equation*}
	\mathbb{E}[X_{ij}|C_{ij}=c_{ij}]=\mu_{ij}=\exp\{\delta_i + \gamma_j\}+(gdp_i+gdp_j)^\beta\text{, for } i\neq j.
	\end{equation*}
	To test the power of the model in situations where a covariate with a strong association is available, we also include a logarithmic transformation of the lagged edge values:
	\begin{equation}
	\label{eq:lagc}
	c^t_{ij}=\log(1+x^{t-1}_{ij}) \text{, for } i \neq j.
	\end{equation}

	Estimation is pursued as described in \citet{lebacher2019regression}, with a constrained Poisson likelihood. In the following, the IPFP-type models using GDP and the lagged variables are denoted \textsl{IPFP-GDP} and \textsl{IPFP-LAG}, respectively. 
	
	\subsection{Gravity Models} \label{sec:cimi}
    Gravity models are at the heart of many methods related to the analysis of network flow data (\citealp{kolaczyk2009}). Besides their successful application to economic trade data (\citealp{disdier2008}, \citealp{head2014}) they are also among the preferred models for network tomography in Computer Sciences (\citealp{vardi1996}). Network tomography relates to a problem that often appears when analyzing computer networks. Here, the individual edge loads are assumed to be \textsl{known} but the flow is allowed to intersect the nodes in the network. The task is then to provide accurate predictions for flows between arbitrary nodes. Very often the gravity model is found to be among the best algorithms to solve this problem (\citealp{zhang2003fast}). Although the formulation of the problem seems to be very different compared to the network reconstruction task, it leads to the same mathematical structure. 

From a methodological point of view, the gravity model is simply a special case of the IPFP model discussed above and in fact, the gravity model is the immediate maximum-entropy solution in each network reconstruction problem where self-loops are allowed (\citealp{SQUARTINI20181}, \citealp{sheldon1998}). Mathematically, the model  builds on a simple multiplicative structure
\begin{equation}
\label{eq:gravity}
\hat{\mu}_{ij}=\frac{x_{i\bullet}x_{\bullet j}}{x_{\bullet \bullet}} \text{, for }i \neq j,
\end{equation}
with $x_{\bullet \bullet}$ representing the sum over all valued in- or outdegrees.
Though simple in structure and fast to compute, the model has two main drawbacks. First, the model yields biased results if the diagonal elements are restricted to be zero because then the row and column sums of the predictions do not match the marginal restrictions exactly. However, in big networks, the bias is often negligible. Second, as in all maximum-entropy models,  the approach relies on inferring sparseness from the marginals and predicts exclusively non-zero matrix entries if all marginals are greater than zero.

Because economic and financial networks most often exhibit a density smaller than one,  \citet{cimini2015} proposed a model that is designed for reconstructing the binary structure of networks with limited information available. Basically, they extend the gravity model from above towards a two-step procedure. In the first step they propose to model the probabilities of observing an edge with a parameter $\alpha$ such that they match with the pre-defined targeted density
\begin{equation}
\label{eq:cali_alpha}
\mathcal{D}= \frac{1}{N} \sum_{i \neq j}P(X_{ij}>0;\hat{\alpha})=\frac{1}{N}\sum_{i \neq j}\frac{\hat{\alpha} \chi_i \psi_j}{1+\hat{\alpha} \chi_i \psi_j},
\end{equation}
where the parameters $\chi_i$ and $\psi_j$ are node-specific \textsl{fitness variables}. Following the idea that the marginals carry information about the binary network structure, they are typically set equal to the marginals (i.e. $\chi_i=x_{i\bullet}$ and $\psi_j=x_{\bullet j}$) or some transformation of them. Another interpretation is that the economic strength determines the fitness of a country or previous bilateral exchanges influence the fitness of dyadic relations. We include the transformed GDP  values and set $\chi_i\psi_j=c_{ij}$ as defined in equation (\ref{eq:trafo}). For the logarithmic lagged exchange we set $\chi_i\psi_j=\log(1.1+x^{t-1}_{ij})$ as fitness variables. Note that adding $1.1$ instead of $1$ prevents the probabilities from being zero irrespective of $\alpha$ in cases with $x^{t-1}_{ij}=0$.

The parameter $\hat{\alpha}$ can be found by any precise root-search program. In applications with larger dimensionality, the values for $\hat{\alpha}$  might become numerically very small and we use a genetic algorithm (implemented in the \texttt{R} package \texttt{GA} by \citealp{scrucca2013ga}) to overcome this problem.
Given an estimate for $\alpha$ that satisfies (\ref{eq:cali_alpha}), \citet{cimini2015} propose to sample binary networks network ensembles with variables $\hat{z}_{ij}$ and use
a density-corrected version of model (\ref{eq:gravity}) for the edge values
\begin{equation}
\label{eq:fitness}
\hat{\mu}_{ij}=\frac{\hat{\alpha}^{-1}+x_{i \bullet}x_{\bullet j}}{x_{\bullet \bullet}}I(\hat{z}_{ij}>0)\text{, for }i \neq j.
\end{equation}
In the following we refer to the density-corrected gravity model by \textsl{DC-GRAVITY} and the models with GDP and lagged variables are abbreviated by \textsl{DC-GRAVITY-GDP} and \textsl{DC-GRAVITY-LAG}.
	\subsection{Tomogravity model} \label{sec:tomo}
	An important model candidate from the network tomography literature is proposed by \citet{zhang2003}.  In their article, the problem of learning origin-destination flows from link load data in IP networks motivates the estimation of a traffic matrix. The authors regard the problem as an ill-posed regression problem that must be regularized with the Kullback-Leibler divergence from  an independence model.
 The predicted values can be found by minimizing the loss-function
	\begin{equation}
	\label{tomogravity}
	L(\bm{\mu})= (\mathbf{A}\bm{\mu}-\mathbf{y})^T(\mathbf{A}\bm{\mu}-\mathbf{y})+\psi^2\sum_{i \neq j}\frac{\mu_{ij}}{N}\log\bigg{(}\frac{\mu_{ij}}{x_{i\bullet}x_{\bullet j}}\bigg{)} 
	\end{equation}
	with respect to $\bm{\mu}=(\mu_{12},...,\mu_{n (n-1)})^T$ and subject to the non-negativity constraint. The first term is simply the sum of squared deviations from the marginals. In the penalization term, the gravity model serves as a null model together with a regularization parameter $\psi$.
	Note that the model is a dense reconstruction technique and does neither provide probabilities nor do the predictions match with the observed marginals.
	
		Although this appears to be an appealing combination between the successful gravity model and information-theoretic reasoning, the procedure is so far seldom applied to the reconstruction of networks.
	The approach is implemented in the \texttt{R} package \texttt{tomogravity} (see \citealp{blocker2014}). The implementation is computationally expensive and we, therefore, apply this model only for the reduced data set.	\citet{zhang2003} show in a simulation study, that the performance of the algorithm is not very sensitive to varying values of $\psi$ and as a rule of thumb they  recommend to use 
	$\psi=0.01$  if no training data are available and we follow their rule in the application section.  
	
	\subsection{LASSO Model} \label{sec:lasso}
	Regarding the network reconstruction problem again as an ill-posed regression problem, it might not even be necessary to make use of a new penalization term. Instead, the  least absolute shrinkage and selection operator  (\textsl{LASSO}) approach  proposed by \citet{tibshirani1996} can be 
	employed, which uses a $L_1$ penalty to enforce sparsity in the model. Although approaches with some kind of regularization are common in network tomography (\citealp{castro2004network}) the LASSO is applied rather rarely for network reconstruction. An exception is given by \citet{chen2017} who propose a LASSO-type model to predict flows in a bike-sharing network from station traffic (number of ingoing and outgoing bikes at each station).
	
	Technically, the quadratic deviation from the marginals is combined with a regularization term that penalizes the sum of the predicted matrix entries, yielding the following  loss function
	\begin{equation}
	\label{lasso}
	L(\bm{\mu})=(\mathbf{A}\bm{\mu}-\mathbf{y})^T(\mathbf{A}\bm{\mu}-\mathbf{y})+ \tau \sum_{i \neq j}| \mu_{ij}|.
	\end{equation}
	By the non-negativity constraint, the absolute value in the penalization term can be dropped. The \texttt{R} package \texttt{glmnet} by \citet{friedman2009} allows for efficient and scalable estimation.
	
	In principle, the model might appear to be attractive because the regularization shrinks some predictions exactly to zero. However, it is not clear how to derive the penalization parameter $\tau$ because cross-validation aiming at the marginals does not lead to satisfactory results. \citet{chen2017} propose to use a training data set -  information that might not always be available.
	To use the approach nevertheless in the competitive comparison without a training set available, we optimize the penalty parameter $\tau$ on a grid such that the number of non-zero coefficients is consistent with the real density.
	
	Note further, that the predicted marginals are, by construction, always be smaller than the observed ones because of the shrinkage property of the LASSO.
	On the other hand, the model has much potential for exploratory analysis by investigating the path plots of the coefficients, i.e.\ the values of the coefficients against increasing values of $\tau$.
	
	\subsection{Hierarchical Fitness Models} \label{sec:fit}
    A central finding of the study by \citet{anand2018} states that no method works equally well for different reconstruction tasks. Based on this insight, \citet{gandy2017} proposed that a construction method should be adjustable to topological characteristics and especially to the density of a network. To do so, they present a hierarchical model designed for the reconstruction of financial networks. In the hierarchy of the model,  the first step consists of estimating the edge probabilities consistent with the target density $\mathcal{D}$.  As a baseline model, the authors propose an Erd\"os-R\'enyi model with 
\begin{equation*}
p_{ij}=p\text{, for } i \neq j,
\end{equation*} 
treating each edge to be equally likely.
Given the obtained set of probabilities, edge weights are sampled from an exponential distribution with common expectation
\begin{equation}\label{eq:exep}
\mu=\mathbb{E}[X_{ij}|Z_{ij}=1]\text{, for } i \neq j.
\end{equation} The sampling algorithm is constructed such that the sampled networks provide stochastic network ensembles but each realization is consistent with the marginal restrictions.

Additionally, they proposed a model that is inspired by fitness-based approaches similar as in equation (\ref{eq:fitness}). In this model, the edge probability is determined by the logistic function
\begin{equation}
\label{fitness}
p_{ij}(\alpha)=\frac{1}{1+\exp\{-\alpha-\log(x_{\bullet i}+x_{i \bullet})-\log(x_{\bullet j}+x_{j \bullet})\}}\text{, for }i \neq j,
\end{equation}
with $\alpha$ being some constant that is estimated for consistency with the target density. In this model, the marginals serve as log-transformed fitness variables. In principle, any kind of variables could be used for the fitness model but only the marginals are yet implemented in the \texttt{R} package \texttt{systemicrisk}. The software implementation is very efficient and not overstrained by the dimensionality of the full network. Nevertheless, the algorithm is in trouble with the high values of the marginals. In the given application the marginals are scaled down in the estimation procedure and the predictions are then rescaled again.

By construction, the model puts much more emphasis on the binary network structure than on the prediction of the edge values.  This is because the marginals are used directly only in the first step to estimate the edge probabilities. In the second step, all edge values are assumed to share the same expectation (\ref{eq:exep}) and the marginal constraints enter only indirectly as a restriction. 

In the comparison, the hierarchical Erd\"os-R\'enyi model is abbreviated by \textsl{H-ER} and the hierarchical fitness model is called \textsl{H-FIT}.

	\subsection{Minimum Density}	\label{sec:mindens}
    \citet{anand2015} noted that the problem of binary network reconstruction can be viewed as finding a solution between two extreme points in the space of possible networks. Either, a maximally dense solution is searched for (maximum-entropy approaches), or it is the goal to find a solution with a minimal number of non-zero edges that are still consistent with the marginal constraints. Given that financial networks are typically sparse and disassortative,  maximum-entropy solutions almost certainly provide an incorrect binary network structure.

In principle, if the density of the network is driven to the lowest level possible, the allocation of the edge weights might even become a simple task because of the small number of possibilities that are left.
In its original form, the loss function  of the minimum density model is simply given by the number of non-zero edges
\begin{equation*}
L(\bm{\mu})=\sum_{i \neq j}I(\mu_{ij}>0),
\end{equation*}
subject to the marginal constraints and the non-negativity constraint. The loss function is not differentiable and direct minimization is computationally expensive. To circumvent this obstacle, \Citet{anand2015} relax the problem by giving up the assumption that the marginal constraint must hold exactly and shift the focus on the quadratic deviations from the marginals.
Then, the authors propose an algorithm that implements two Markov processes, one adds new edges and weights and the second one deletes edges. Initialized with an arbitrary network the algorithm iterates as long as the loss function does not decrease any more together with a sufficient fit for the marginals.

The proposed algorithm is stochastic with non-unique solutions and generates ensembles of low-density networks. Typically, the realization with the lowest density is taken to be the optimal estimate (called  \textsl{MINDENS} henceforth). By definition, the method does not rely on knowledge of the real density $\mathcal{D}$. Therefore, it is appropriate to regard the model as a lower-bound (in the space of feasible networks that satisfy the marginal constraints) 
instead of viewing it as an accurate reconstruction.  This also has implications for the edge values, because a minimal number of edges in the system leads to maximal concentration of the edge values on a few nodes.

	\section{Evaluation} \label{sec:eval}
	\subsection{Binary Network Reconstruction}
    We evaluate the quality of the binary network reconstruction with different measures. For models that provide edge probabilities, we use the area under the curve (AUC) of the receiver-operating characteristic (ROC) curve and the precision-recall (PR) curve (see \citealp{grau2015}). We regard both measures as complementary for model evaluation. While the ROC curve is, so to speak, ignorant about how good we predict either $Z_{ij}^t=1$ or $Z_{ij}^t=0$, the PR curve describes how well the models do in predicting $Z_{ij}^t=1$. This is relevant because in low-density networks it is simpler to predict a zero than a one.
Further, we look at the Bier score decomposition proposed by \citet{murphy1973new} (see also \citealp{siegert2017simplifying}).
For each time period $t$ and model, we obtain $K^t$ different probabilities $\hat{p}_{ij}^t\in \{\hat{p}_1^t,...,\hat{p}_{K^t}^t\}$ with $n_k^t$ equal probabilities that correspond to edges $z_k^t$ for $k=1,...,K^t$. The Bier score decomposition is given by
\begin{equation*}
\begin{split}
BR_t=& \frac{1}{N}\sum_{i \neq j}(\hat{p}_{ij}^t-z_{ij}^t)^2=\underbrace{\sum_{k=1}^{K^t}\frac{n_k^t}{N}\left( \frac{z_k^t}{n_k^t}-\hat{p}_k^t \right)^2}_{REL_t}+  \underbrace{\sum_{k=1}^{K^t}\frac{n_k^t}{N}\left( \frac{z_k^t}{n_k^t}-\mathcal{D}^t \right)^2}_{RES_t}-\underbrace{ \mathcal{D}^t(1-\mathcal{D}^t)}_{UNC_t}.
\end{split}
\end{equation*}    
The reliability $REL_t$ measures the distance between the estimated probabilities and the average real frequencies, with $0$ being the best value that can be achieved. This means that a \textsl{low reliability} actually is the preferred outcome.
The term labeled $UNC_t$ is called uncertainty and gives the variability of the edges in the sample. 
Resolution ($RES_t$) gives the difference between the different share of empirical probabilities for each of the $K^t$ categories and their overall average. Hence, it is a measure for the ability to discriminate between zero and one. A higher value indicates a better resolution. If the ability to discriminate is at is maximum, all probabilities are either one and zero, in this situation it holds that $RES_t=UNC_t$.
We report these measures aggregated for all years and show the aggregated difference $UNC_t-RES_t$ which can be interpreted as the reduction of the uncertainty due to resolution.

We follow \citet{SQUARTINI20181} and provide graphical representations of the reconstructed networks in the Appendices \ref{annex:adj} and \ref{annex:adj_small}. There, the reconstructed adjacency matrices (based on binarization with a threshold according to the true density) for the most recent full and reduced network are shown.

We are not only interested in the prediction of individual edge occurrences but also in the quality of the reconstructed network topology. Given the strong heterogeneity in the network, the degree distribution can be regarded as a very important measure for the binary structure. We evaluate the fit of the outdegree distribution using the square root of the mean squared error of the real and the reconstructed outdegree distribution
\begin{equation*}
RMSE_{od}^t=\sqrt{\frac{1}{n} \sum_{j=1}^n\left\{  \sum_{i=1}^{n}I(\hat{z}^t_{i\bullet}=j) - \sum_{i=1}^{n}I(z^t_{i\bullet}=j) \right\}^2 } \text{, for }t=1,...,T
\end{equation*}
and correspondingly for the indegree.

\begin{figure}[t!]
	\centering
	\begin{subfigure}{\textwidth}
		\centering            \includegraphics[trim={0cm 0cm 0cm 0cm},clip,width=\textwidth]{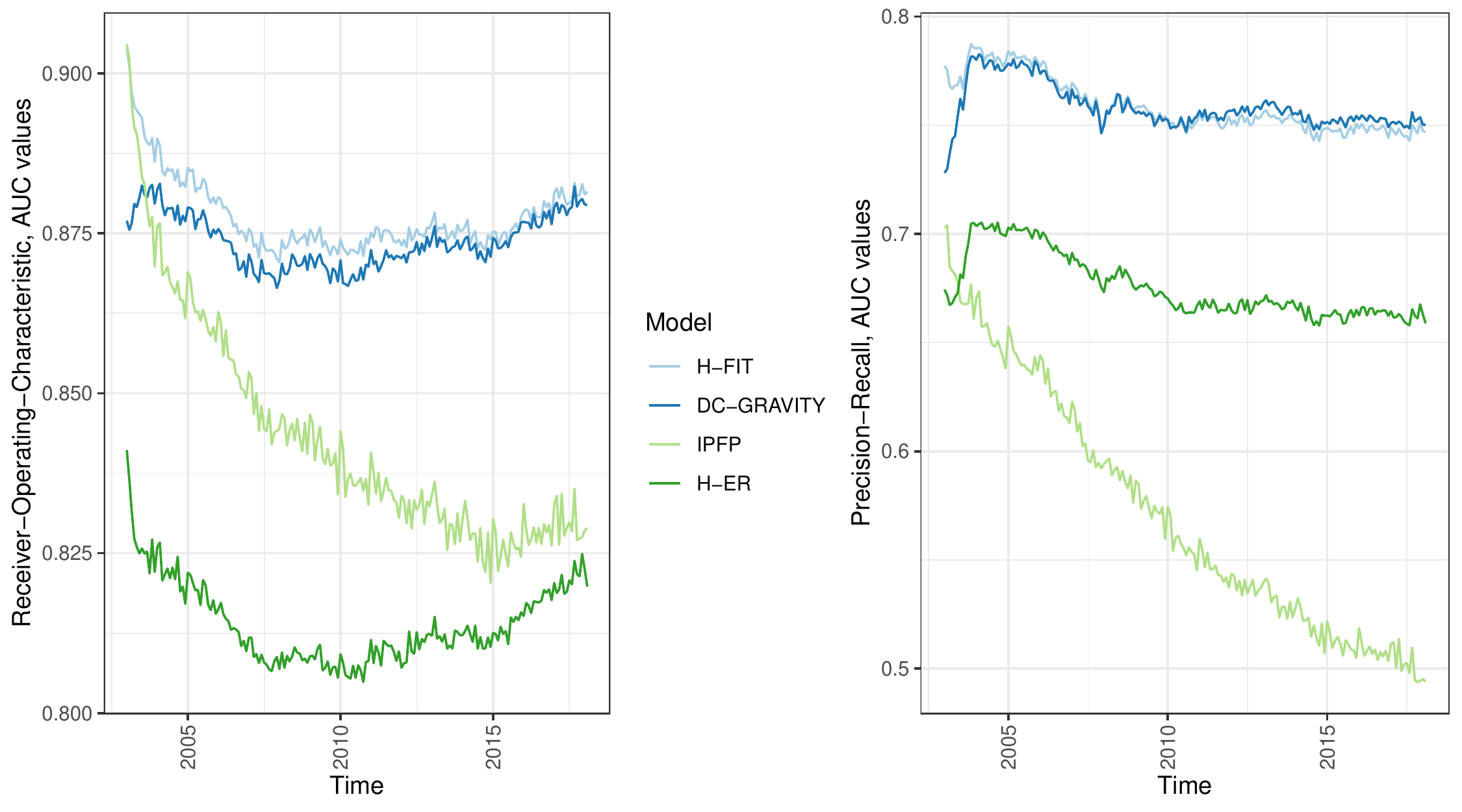}
	\end{subfigure}

	\caption{Evaluation of probabilities in the full network. Time series of the area under the curve (AUC) values for receiver-operating-characteristics (ROC, left panel) curve and precision-recall (PR, right panel) curve for the IPFP model, the degree corrected Gravity model  (DC-GRAVITY), the hierarchical  Erd\"os-R\'enyi model (H-ER) and the hierarchical fitness model. \\ Source: SWIFT BI Watch.}
	\label{fig:pr_roc}
\end{figure}

To make the models comparable, we calibrate all estimates to the same target density. For models that are not scaled to the real density, we use a pragmatic approach and take the highest $\mathcal{D}N$ (the number of edges in the real network) estimates to be one and all other estimates to be zero while in the probability-based models, we use the highest $\mathcal{D}N$ probabilities to predict a one. 

A visual impression of the quality of the degree reconstruction is given in  Appendix \ref{annex:deg}, plotting the predicted outdegree (indegree) against the real outdegree (indegree) for the most recent network observation of the full network  and in \ref{annex:deg_small} for the reduced network.

	\begin{figure}[t!]
		\centering
		\begin{subfigure}{\textwidth}
			\centering			\includegraphics[trim={0cm 0cm 0cm 0cm},clip,width=\textwidth]{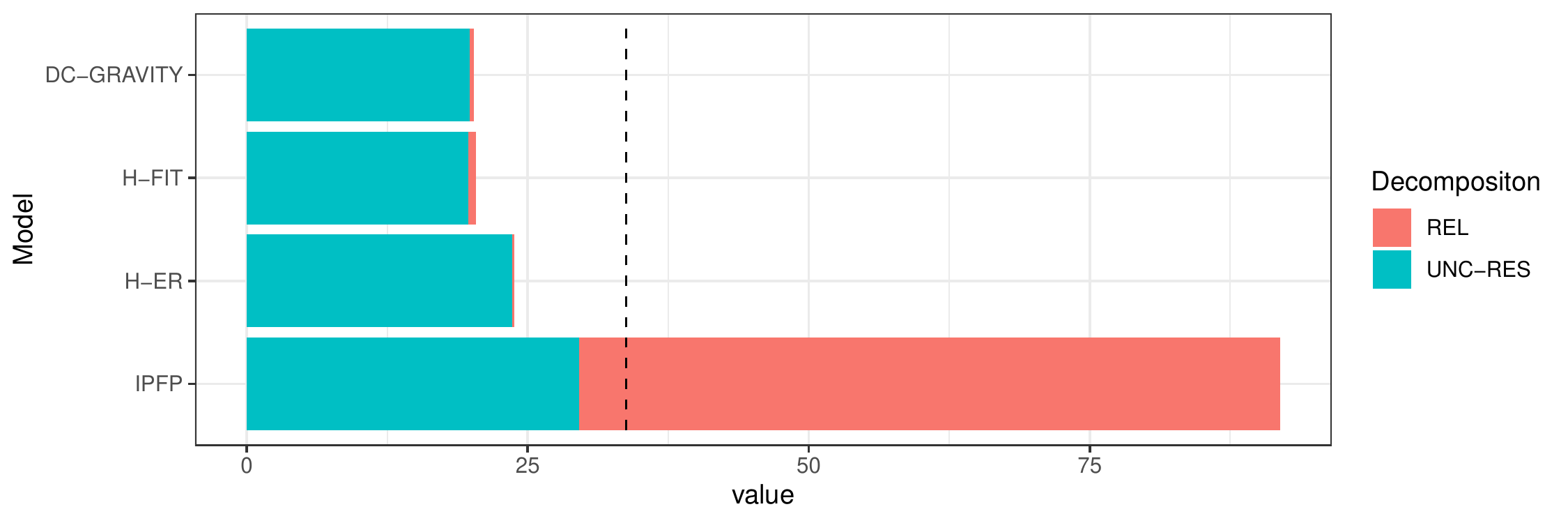}
		\end{subfigure}

		\caption{Decomposition of the Brier score for the full network into reliability (REL) and remaining uncertainty after subtracting resolution (UNC-RES)
			for the IPFP model, the Gravity model (GRAVITY), the degree corrected gravity model model (DC-GRAVITY), the hierarchical fitness models (H-ER, H-FIT). Uncertainty (UNC) as a dashed line.  \\ Source: SWIFT BI Watch.}
		\label{fig:decomp}
	\end{figure}
	
	\subsubsection{Full Network}
    In the full network, four different models can be compared using AUC values and the decomposed Brier score. These four models include the iterative proportional fitting model from Section \ref{sec:ipf} (IPFP),  the density-corrected gravity model by \citet{cimini2015} from Section \ref{sec:cimi} (DC-GRAVITY) and the two hierarchical models from Section \ref{sec:fit}, with edge-probabilities coming either from the Erd\"os-R\'enyi (H-ER) or the fitness model (H-FIT). 
In Figure \ref{fig:pr_roc}, we plot the AUC values for the ROC (left panel) and PR (right panel) curves against time. In Figure \ref{fig:decomp} we show the decomposition of the Brier score,  with uncertainty ($UNC$) as a dashed vertical line.

\begin{figure}[t!]
		\centering
		\begin{subfigure}{\textwidth}
			\centering			\includegraphics[trim={0cm 0cm 0cm 0cm},clip,width=\textwidth]{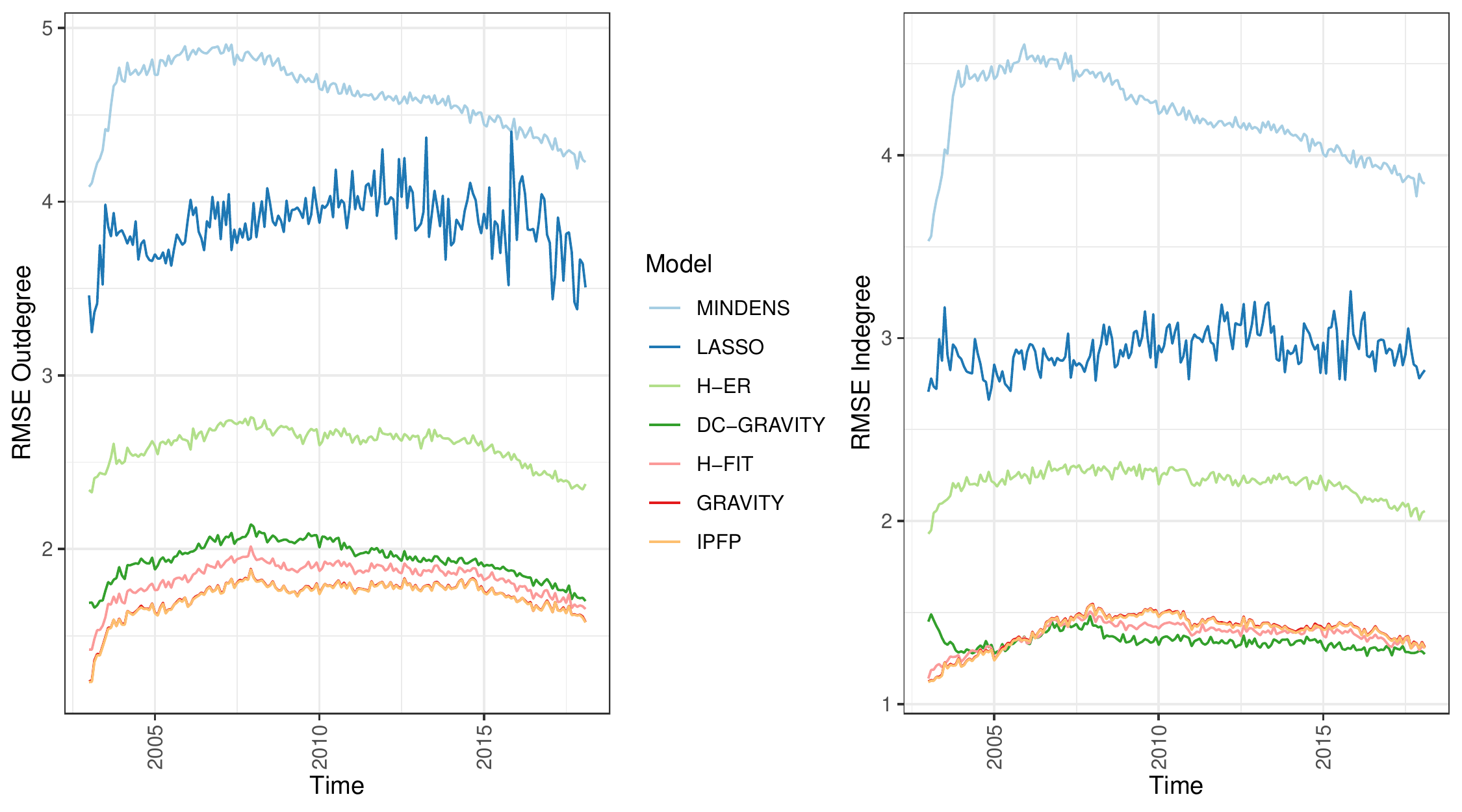}
		\end{subfigure}

		\caption{Time series of root mean squared error (RMSE) for the reconstruction of the outdegree (right) and the indegree (left) of the full network for the IPFP model, the Gravity model (GRAVITY), degree corrected gravity model model (DC-GRAVITY), the hierarchical fitness models (H-ER, H-FIT), the LASSO model and the minimum density model (MINDENS).  \\ Source: SWIFT BI Watch.}
		\label{fig:degree_recon}
	\end{figure}

For the reconstruction of the degrees, additionally the Gravity model (GRAVITY) from Section \ref{sec:cimi}, the LASSO from Section \ref{sec:lasso} and the minimum density (MINDENS) method of Section \ref{sec:mindens} enter the comparison. This is visualized in Figure \ref{fig:degree_recon} with the root mean squared errors for the outdegree in the left panel and for the indegree on the right. In both figures, the abbreviations of models are ordered to approximately match the time-averaged height of the respective measures.

\vspace{0.5cm}
\noindent \textsl{Edge Probabilities}
\\
The hierarchical Erd\"os-R\'enyi (H-ER) model performs worst in the left panel and second-worst in the right panel of Figure \ref{fig:pr_roc}. Seemingly, the assumption of equal probabilities for all dyads is strongly violated in this network. This relates to the discussion in Section \ref{sec:description} where we showed that the patterns of the degree distribution do not match with the Erd\"os-R\'enyi model. 

However, also the IPFP model that allows for differing edge probabilities in its' Poisson interpretation does not perform satisfactorily and we see a declining trend of the prediction accuracy with time. As a consequence, the AUC values of the ROC curve decrease strongly and when evaluated with the PR curves, the model even provides the worst outcomes.
This is a result of the growing values of the marginals, implying that the IPFP probabilities become very close to one or even numerically equal to one, leading to a loss of variation among the probabilities.

The two winners of this comparison, the density-corrected gravity model (DC-GRAVITY) and the hierarchical fitness model (H-FIT), give very similar accuracy measures in both panels of Figure \ref{fig:pr_roc}. The AUC values for the ROC curves provided by the H-FIT model are slightly better than the ones of the DC-GRAVITY model and the other way round when evaluated with the PR curves. The strong similarity of the models' predictive power is, in fact, intuitive and results from the comparable choice of functions for determining the edge probabilities. 

These results can be supported by the decomposition of the aggregated Brier score shown in Figure \ref{fig:decomp}. The DC-GRAVITY model and the H-FIT model both provide a very low reliability measure and a comparatively high resolution. Interestingly,  they are closely followed by the H-ER model that does not appear to be much worse with respect to the Brier score. Different from that, we find that the IPFP model has a low resolution and a high reliability measure, indicating that provided probabilities deviate strongly from the real ones and the ability to separate the predictions into ``0'' and ``1'' is rather low. Again this is because the IPFP model is not calibrated and many predictions are numerically just equal to one.

\vspace{0.5cm}
\noindent \textsl{Degree Structure}
\\
Turning to the reconstruction of the degree structure, the different scaling of the two panels in Figure \ref{fig:degree_recon} shows that it is simpler to reconstruct the indegrees as compared to the outdegrees. The minimum density solution (MINDENS) marks an extreme case, resulting in the worst reconstruction of the out- and indegree structure. However, MINDENS has the comparative disadvantage of not being calibrated to the density and predicts far fewer edges than present in the real networks. Therefore, fewer edges can be allocated to certain nodes. With the exception of the United States, the model predicts no out- or indegrees above 65 at  all (see also Figure \ref{fig:deg_recon_mindens} in the Annex \ref{annex:deg}). 

The LASSO provides the most unstable behavior and exhibits a high variance. Although the model is calibrated to the real density, the edge reconstruction is second-worst and delivers unsatisfactory reconstructions for the out- and the indegree. In Figure  \ref{fig:deg_lasso_mindens} of Annex \ref{annex:adj}  it can be seen that the reconstructed degrees look almost random and Figure \ref{fig:recon_lasso} indicates that the model is not able to make efficient use of the provided information on the row and column sums.

The performance of the hierarchical Erd\"os-R\'enyi model (H-ER) shows that ignorance about the marginals for predicting the binary structure also can lead to unsatisfactory outcomes. The H-ER model ``over-estimates'' the out- and indegree for countries with medium-sized degrees and ``under-estimates'' the out- and indegree for countries with high degrees.  This is clearly a result of the assumption that all edges are equally likely (as long as consistent with the marginals), leading to a random block structure in the network. 

The Gravity model (GRAVITY) and the IPFP model make the best use of the information provided by the marginals to reconstruct the outdegree but not for the indegree. Their predictive quality concerning the degrees is almost identical and it is hard to distinguish both models in the left panel (IPFP overlays GRAVITY in both plots).

The hierarchical fitness model (H-FIT) together with the degree corrected gravity model (DC-GRAVITY) perform slightly worse than the GRAVITY and IPFP models concerning the outdegree but can be said to be the winner in the competition for the indegree reconstruction.

\subsubsection{Reduced Network}
In the reduced network, a greater variety of models can be investigated. Essentially, we can add four additional models in our comparative study, by extending the degree corrected gravity model (DC-GRAVITY) from Section \ref{sec:cimi} with the usage of GDP (DC-GRAVITY-GDP) and the lagged values (DC-GRAVITY-LAG) for determining the edge probabilities as well as the extended IPFP approach from Section \ref{sec:ipf} using the GDP values (IPFP-GDP) and the lagged values (IPFP-LAG) a covariates. In the degree reconstruction part, additionally the TOMOGRAVITY model (Section \ref{sec:tomo}) is considered. The MINDENS model, however, is not considered for the reduced network because it is very dense.

	\begin{figure}[t!]
		\centering
		\begin{subfigure}{\textwidth}
			\centering			\includegraphics[trim={0cm 0cm 0cm 0cm},clip,width=\textwidth]{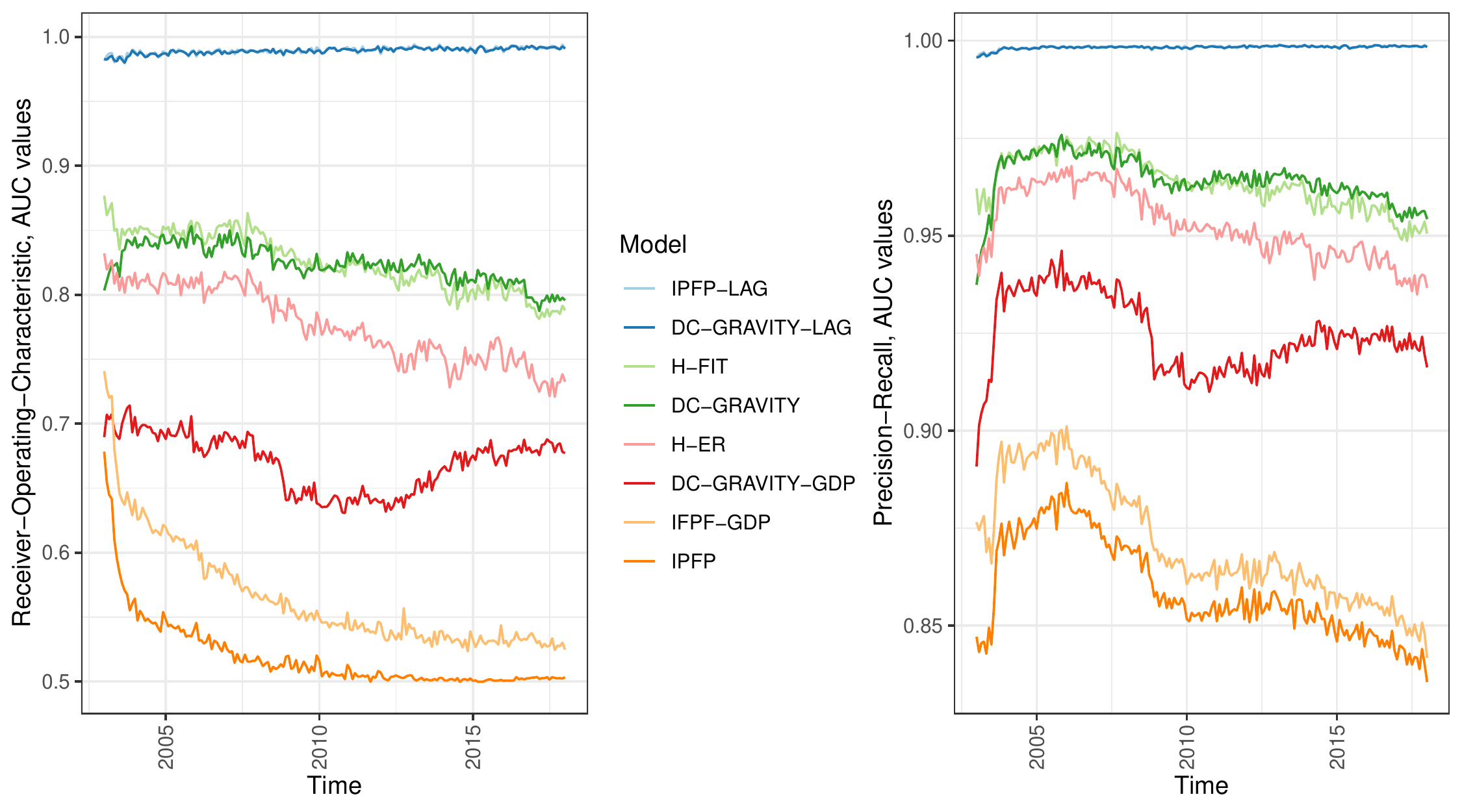}
		\end{subfigure}

		\caption{Evaluation of probabilities in the reduced network. Time series of area under the curve (AUC) values for receiver-operating-characteristics (ROC, left panel) and precision-recall (PR, right panel) for the IPFP model, the degree corrected gravity model (DC-GRAVITY) with covariates (DC-GRAVITY-GDP, DC-GRAVITY-LAG), the hierarchical  Erd\"os-R\'enyi model (H-ER), the hierarchical fitness model (H-FIT) and the IPFP-based models with covariates (IPFP-GDP, IPFP-LAG). \\ Source: SWIFT BI Watch.}
		\label{fig:pr_roc_small}
	\end{figure}
	
	\begin{figure}[t!]
		\centering
		\begin{subfigure}{\textwidth}
			\centering			\includegraphics[trim={0cm 0cm 0cm 0cm},clip,width=\textwidth]{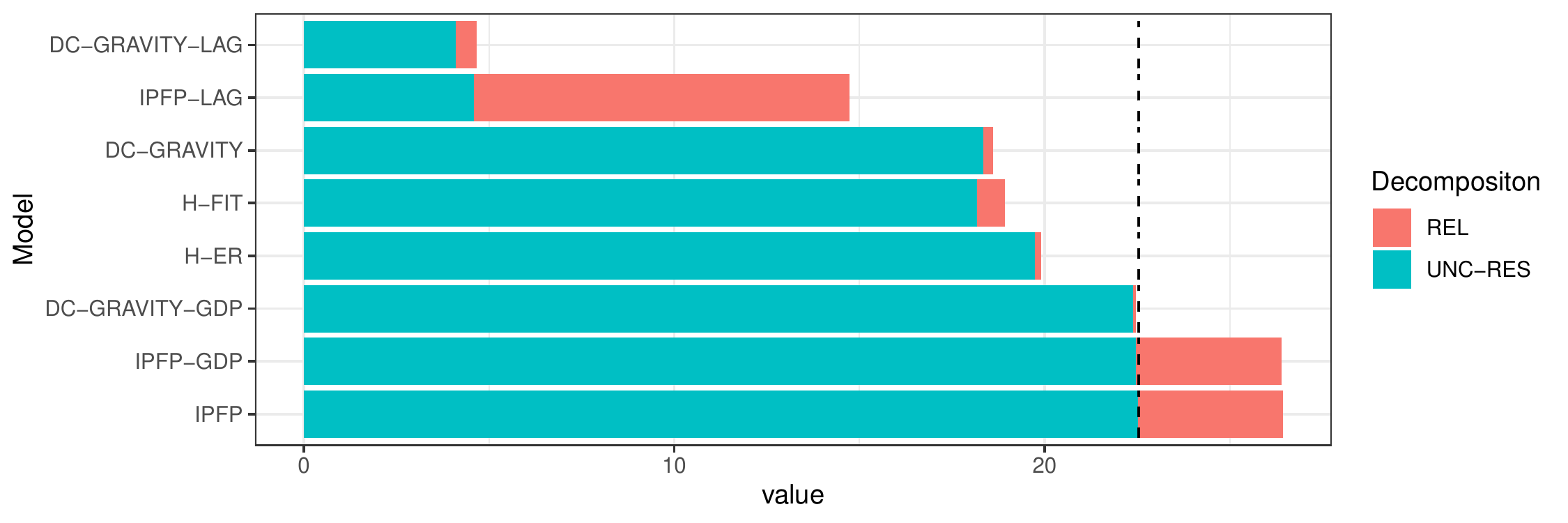}
		\end{subfigure}

		\caption{Decomposition of the Brier score for the reduced network into reliability (REL) and remaining uncertainty after subtracting resolution (UNC-RES)
			for the IPFP model, the degree corrected gravity model (DC-GRAVITY) with covariates (DC-GRAVITY-GDP, DC-GRAVITY-LAG), the hierarchical  Erd\"os-R\'enyi model (H-ER), the hierarchical fitness model (H-FIT) and the IPFP-based models with covariates (IPFP-GDP, IPFP-LAG). Uncertainty (UNC) as a dashed line.  \\ Source: SWIFT BI Watch.}
		\label{fig:decomp_small}
	\end{figure}
	\vspace{0.5cm}
	\noindent\textsl{Edge Probabilities}
	\\
	The IPFP probabilities are among the worst in both panels of Figure \ref{fig:pr_roc_small}.
	Similar to the full network, the AUC values for the ROC curve are strongly decreasing with time. An almost parallel pattern can be found for the IPFP-based reconstruction with GDP values (IPFP-GDP). Although the exogenous information helps to improve the performance relative to IPFP, the outcome is still very bad in comparison to the other models.
	
	While the information on the GDP nevertheless improves the fit in the IPFP-based models, this is not the case for the density-corrected gravity model (DC-GRAVITY). It turns out that the version that includes GDP values (DC-GRAVITY-GDP) performs even worse than without (DC-GRAVITY) with both measures. Again, we find that the DC-GRAVITY and the H-FIT model behave very similar.

	The two models with lagged variables as covariates, the DC-GRAVITY-LAG model and the IPFP model combined with the lagged values (IPFP-LAG), have the unfair advantage of incorporating much more information than all others and reach outstanding AUC values in both panels of Figure \ref{fig:pr_roc_small} (both lines overlay in the plots). In Figures  \ref{fig:recon_cimi_lag_small} and \ref{fig:recon_regression_lag_small} it can be seen that the reconstructed network based on the lagged covariates is almost identical to the original one, showing that having observed an edge in $t-1$ is almost deterministic for predicting an edge in $t$.  
	
	The decomposition of the Brier score in Figure \ref{fig:decomp_small} mirror the results discussed above. However, it is striking that the IPFP-LAG model has a much higher reliability measure in comparison to the DC-GRAVITY-LAG model which results from not being calibrated to the real density. Further note that the three models IPFP, IPFP-GDP and DC-GRAVITY-GDP have a resolution score of almost zero indicating that the knowledge GDP does not contribute much information about the binary edge structure.
	
	\begin{figure}[t!]
		\centering
		\begin{subfigure}{\textwidth}
			\centering			\includegraphics[trim={0cm 0cm 0cm 0cm},clip,width=\textwidth]{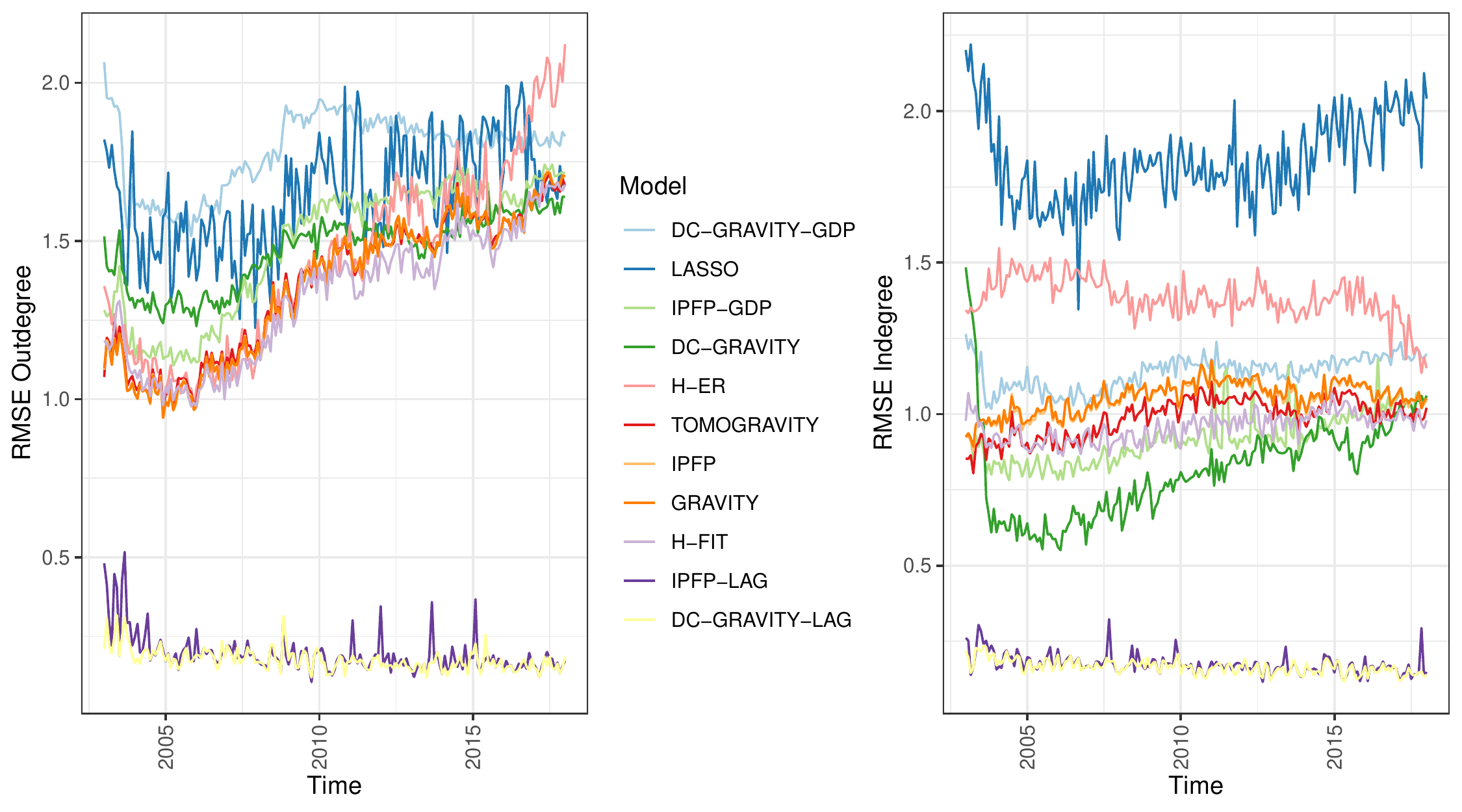}
		\end{subfigure}

		\caption{Time series of root mean squared error (RMSE) for the reconstruction of the outdegree (right) and the indegree (left) of the reduced network for the IPFP model, the Gravity model (GRAVITY), the degree corrected gravity model (DC-GRAVITY) with covariates (DC-GRAVITY-GDP, DC-GRAVITY-LAG), the hierarchical  Erd\"os-R\'enyi model (H-ER), the hierarchical fitness model (H-FIT), the IPFP-based models with covariates (IPFP-GDP, IPFP-LAG), the LASSO model and the TOMOGRAVITY model. \\ Source: SWIFT BI Watch.}
		\label{fig:degree_recon_small}
	\end{figure}
	\vspace{0.5cm}
\noindent \textsl{Degree Structure}
\\
Given that the reduced network is very dense, the degree structures might be more easily reconstructed as compared to the sparse full network. However, the predicted edges still need to be allocated correctly to the corresponding nodes which are not a trivial task. This becomes obvious when regarding the visualization of the degree reconstruction in Supplementary Material. There it can be seen that the reconstruction of the binary degrees is partly very bad. Quantified with the root mean squared errors as shown in Figure \ref{fig:degree_recon_small}, the models can be compared directly.

In both panels of Figure \ref{fig:degree_recon_small} it can be seen very clearly that the models that incorporate the lagged matrix entries  (IPFP-LAG, DC-GRAVITY-LAG)  lead to degree reconstructions that are superior in every respect. Except for some spikes, that might reflect a kind of seasonality pattern, the root mean squared errors are close to zero.

If no information from exogenous covariates is available, the DC-GRAVITY model is found to perform very well for the indegree. Especially regarding the outdegree, almost all methods (amongst others H-FIT, GRAVITY, IPFP) give good and very comparable results.

Again the LASSO proves to be a bad choice for reconstructing the degree structure, exhibiting a high variance over time as well as large deviations from the actual degrees.
\FloatBarrier
\subsection{Valued Network Prediction}
The mechanisms that determine the edge probabilities might differ fundamentally from the ones that lead to certain edge values. Additionally, some models are restricted to the prediction of edge values and the prediction of binary networks constructed with threshold values is not the usage they are originally built for. 
Therefore, we now pay attention to the predictive quality of the valued reconstruction  in terms of the $L_1$ errors
\begin{equation*}
L_1^t=\sum_{i\neq j}|x_{t,ij}-\hat{\mu}_{t,ij}|\text{, for }t=1,...,T
\end{equation*}
and the $L_2$ errors
\begin{equation*}
L_2^t=\sqrt{\sum_{i\neq j}(x_{t,ij}-\hat{\mu}_{t,ij})^2} \text{, for }t=1,...,T.
\end{equation*}
These measures are regarded in terms of overall errors aggregated over all time points as well as their monthly averages and the corresponding standard errors.

\subsubsection{Full Network}
 	\begin{table}[t!] \centering 
 	\begin{tabular}{@{\extracolsep{5pt}} lcccccc} 
 		\\[-1.8ex]\hline 
 		Method& overall $L_1$ & overall $L_2$ & average $L_1$ & SE & average $L_2$ & SE\\
 		\hline \hline  
 		
 		IPFP&$48,443.070$ & $229.987$ & $266.171$ & $68.568$ & $16.225$ & $5.246$ \\
 		GRAVITY &\bf{47,971.710} & \bf{211.125} & \bf{263.581} & \bf{68.321} & \bf{14.886} & \bf{4.843} \\ 
 		DC-GRAVITY&$51,595.870$ & $239.899$ & $283.494$ & $70.855$ & $16.917$ & $5.495$ \\ 
 		LASSO&$452,296.700$ & $824.632$ & $2,485.147$ & $778.114$ & $58.177$ & $18.807$ \\ 
 		H-FIT&$67,171.460$ & $479.524$ & $369.074$ & $100.739$ & $33.669$ & $11.425$ \\ 
 		H-ER&$67,282.440$ & $481.793$ & $369.684$ & $101.241$ & $33.810$ & $11.534$ \\ 
 		MINDENS&$90,610.240$ & $459.455$ & $497.858$ & $114.582$ & $33.021$ & $8.361$ \\ 
 		
 		\hline \\[-1.8ex] 
 	\end{tabular}

 	\caption{Evaluation of the reconstructed valued full MT 103 networks, Method in the first column. Aggregated $L_1$ and $L_2$ errors in columns two and three as well as average errors and their standard errors over time in the last four columns. Minimal values in bold. \\ Source: SWIFT BI Watch.} 
 	\label{tab:L_errors}
 \end{table}

	In Table \ref{tab:L_errors}, it can be seen that the two dense reconstruction models IPFP and GRAVITY give the best reconstruction evaluated with the $L_1$ and $L_2$ errors with the GRAVITY model being slightly ahead. The third-best prediction quality is delivered by the DC-GRAVITY model. It can be inferred that the risk of guessing the wrong edges to be zero or one (and placing a high weight or no weight to the false edges) strongly counterweights the seeming disadvantage of the dense reconstruction methods. This effect is pronounced in the MINDENS model and even more so in the LASSO model that comes with extremely high errors. However, also the hierarchical fitness model (H-FIT), one of the best models for binary network reconstruction is found to provide edge value predictions that are by far worse compared to the GRAVITY solution.

\FloatBarrier
\subsubsection{Reduced Network}

	\begin{table}[t!] \centering \small
		\begin{tabular}{@{\extracolsep{5pt}} lcccccc} 
			\\[-1.8ex]\hline 
			Method& overall $L_1$ & overall $L_2$ & average $L_1$ & SE & average $L_2$ & SE\\
			\hline \hline  
			IPFP&$39,087.480$ & $216.339$ & $214.766$ & 54.409 & $15.270$ & $4.912$ \\ 
			GRAVITY &\bf{38,646.430} & \bf{203.317} & \bf{212.343} & \bf{54.268} & \bf{14.321} & \bf{4.709} \\ 
			DC-GRAVITY&$41,470.410$ & $232.719$ & $227.859$ & $55.489$ & $16.396$ & $5.378$ \\ 
			TOMOGRAVITY&$39,691.210$ & $215.081$ & $218.084$ & $55.635$ & $15.182$ & $4.878$ \\ 
			LASSO&$290,061.700$ & $1,042.029$ & $1,593.746$ & $486.963$ & $73.577$ & $23.569$ \\ 
			H-FIT&$55,979.590$ & $471.112$ & $307.580$ & $83.433$ & $33.426$ & $10.136$ \\ 
			H-ER&$56,071.170$ & $471.710$ & $308.083$ & $83.580$ & $33.460$ & $10.178$ \\  
			\hline
			DC-GRAVITY-GDP&$39,759.120$ & $232.127$ & $218.457$ & $53.322$ & $16.337$ & $5.415$ \\ 
			DC-GRAVITY-LAG&$40,110.850$ & $232.127$ & $220.389$ & $53.532$ & $16.351$ & $5.373$ \\ 
			IPFP-GDP&38,231.110& 214.741 & 210.061 & $54.639$ & 15.018 & $5.290$ \\ 
			IPFP-LAG&\bf{4,137.484} & \bf{34.178} & \bf{22.859} & \bf{11.523} & \bf{2.128} & \bf{1.391} \\  
			%
			\hline \\[-1.8ex] 
		\end{tabular}

		\caption{Evaluation of the reconstructed valued reduced MT 103 networks, Method in the first column. Aggregated $L_1$ and $L_2$ errors in columns two and three as well as average errors and their standard errors over time in the last four columns. The last four rows give models with exogenous information included.   		Minimal values in bold. \\ Source: SWIFT BI Watch.}
		\label{tab:L_errors2}
	\end{table} 
	
	\noindent In the reduced network, we conclude that IPFP, the GRAVITY model and the TOMOGRAVITY model results in very similar aggregated $L_1$ and $L_2$ errors. These models are closely followed by the DC-GRAVITY-GDP model. The hierarchical models (H-FIT, G-ER) perform comparable and by far better than the LASSO.
	
    Models that include exogenous information are separated and given in the last four rows of Table \ref{tab:L_errors2}. Among these models, the second-best result is given by the IPFP-GDP model, showing that the GDP values provide useful information that improves the quality of the edge value reconstruction, for example, relative to IPFP or the GRAVITY model.  The IPFP model that incorporates lagged edge values (IPFP-LAG) as covariates performs outstandingly well. But again it might unrealistic to assume the availability of lagged data points. Interestingly, both density-corrected gravity models with exogenous covariates (DC-GRAVITY-GDP, DC-GRAVITY-LAG) are only slightly better than the DC-GRAVITY model. This can be explained by the fact, that the exogenous information is only used to determine the edge probabilities.
    
	\section{Discussion} \label{sec:conclusion}
In this paper, we have compared different models for network construction using the SWIFT MT 103 networks. The models are compared along different dimensions, including the accuracy of edge prediction, degree reconstruction, and edge value estimation.  Overall, four conclusions that can be drawn from this competitive comparison.

(i) The task of reconstructing edge values differs fundamentally from the task of estimating edge probabilities. Technically, this is very intuitive because the marginals give exclusively information about the edge values and all approaches that output edge probabilities are necessarily dependent on further restrictions (the real density). 

Even if the true density is assumed to be known, no model emerged that can be said to be great in achieving outstanding predictions of the edge probabilities and their values. This conclusion is also in line with the findings of the extensive comparison by \citet{anand2018}. We, therefore, recommend that the model choice should be governed by the specific use case and depending on the importance attached to either reconstruction. If the binary structure is of interest and the model is presumed to be sparse, the hierarchical fitness model (H-FIT) and the density-corrected gravity model (DC-GRAVITY) are good choices. 
While \citet{anand2018} highlight the ability of the minimum density method (MINDENS) to detect absent edges we must supplement this by noting that the method nevertheless performs not that good if interest lies in detecting present edges.

Regarding the quality of the edge value prediction, either in sparse or dense networks the maximum entropy models (IPFP and GRAVITY) work very well. The same was found by \citet{anand2018}, pointing on the good quality of maximum entropy solutions. However, in contrast to their findings, we highlight here more clearly the potential shortfalls for sparse reconstruction methods for the prediction of edge values.

(ii) Other than \citet{anand2018}, we do not find that the preferred models change when either a dense or a sparse network is to be reconstructed. However, this statement must be taken with care since in our analysis the dense network is, in fact, a subset of the sparse one.

(iii) Including exogenous information can help to improve both, the binary and the valued network reconstruction and partly leads to dramatic increases in the predicted performance. However, this increase in predictive accuracy is not guaranteed. If variables with a low association to the unknown edge values are chosen, the quality of the reconstruction might even decline (see also \citealp{lebacher2019regression}). Especially  regarding the binary network reconstruction, the inclusion of GDP led to mixed results.

(iv) As an ``off the shelf'' model in situations without exogenous information available, the density-corrected gravity model (DC-GRAVITY) can be recommended because it is found to work well on the big sparse network as well as on the small dense network with respect to the edge probabilities and the edge values. 
A similar conclusion can be found in \citet[p. 116]{anand2018}, stating that among the probabilistic methods the model is the  \textsl{``clear winner across all measures of interest''}. Similarly, \citet{gandy2018} report that this model is performing very well in binary and valued reconstruction. Further, the model can be extended towards the inclusion of exogenous information in a simple way.

For further research, it seems to be necessary to compare the performance of edge probabilities when using calibration densities that differ from the real one. Another important research question relates to the ability of reconstruction models to provide uncertainty quantification.  Many approaches introduced above results in network ensembles or come with an associated stochastic structure that can be used to construct prediction intervals. 

\section*{Acknowledgement}	
We would like to thank Peter Ware and Nancy Murphy for providing the SWIFT data.

\section*{Declaration of Interest}
The project was supported by the European Cooperation in Science and Technology [COST Action CA15109 (COSTNET)]. We also gratefully acknowledge funding provided by the German Research Foundation (DFG) for the project  KA 1188/10-1: \textit{International Trade of Arms: A Network Approach.}
\bibliographystyle{Chicago}

\bibliography{literature}
\newpage
\appendix
\pagenumbering{Roman}
\section{Countries included} \label{annex:included}
	
	\begin{table}[!htbp] \centering \tiny
		\resizebox{\textwidth}{!}{
			
		\begin{tabular}{@{\extracolsep{5pt}} ccccccccc} 
			\\[-1.8ex]\hline 
			\hline \\[-1.8ex] 
			ISO & name & reduced &ISO& name & reduced & ISO & name & reduced  \\ 
			\hline \\[-1.8ex] 
			AD & Andorra & 0 & GQ & Equatorial Guinea & 0 & PE & Peru & 0 \\ 
			AE & United Arab Emirates & 1 & GR & Greece & 1 & PF & French Polynesia & 0 \\ 
			AG & Antigua \& Barbuda & 0 & GT & Guatemala & 0 & PG & Papua New Guinea & 0 \\ 
			AI & Anguilla & 0 & GY & Guyana & 0 & PH & Philippines & 1 \\ 
			AL & Albania & 0 & HK & Hong Kong SAR China & 1 & PK & Pakistan & 0 \\ 
			AM & Armenia & 0 & HN & Honduras & 0 & PL & Poland & 1 \\ 
			AO & Angola & 0 & HR & Croatia & 1 & PR & Puerto Rico & 0 \\ 
			AR & Argentina & 0 & HT & Haiti & 0 & PS & Palestinian Territories & 0 \\ 
			AT & Austria & 1 & HU & Hungary & 1 & PT & Portugal & 1 \\ 
			AU & Australia & 1 & ID & Indonesia & 1 & PY & Paraguay & 0 \\ 
			AW & Aruba & 0 & IE & Ireland & 1 & QA & Qatar & 0 \\ 
			AZ & Azerbaijan & 0 & IL & Israel & 1 & RE & Réunion & 0 \\ 
			BA & Bosnia \& Herzegovina & 0 & IM & Isle of Man & 0 & RO & Romania & 1 \\ 
			BB & Barbados & 0 & IMI & Internat. Market Infrastrct. & 0 & RU & Russia & 1 \\ 
			BD & Bangladesh & 0 & IN & India & 1 & RW & Rwanda & 0 \\ 
			BE & Belgium & 1 & IR & Iran & 0 & SA & Saudi Arabia & 1 \\ 
			BF & Burkina Faso & 0 & IS & Iceland & 0 & SB & Solomon Islands & 0 \\ 
			BG & Bulgaria & 1 & IT & Italy & 1 & SC & Seychelles & 0 \\ 
			BH & Bahrain & 0 & JE & Jersey & 0 & SD & Sudan & 0 \\ 
			BI & Burundi & 0 & JM & Jamaica & 0 & SE & Sweden & 1 \\ 
			BJ & Benin & 0 & JO & Jordan & 0 & SG & Singapore & 1 \\ 
			BM & Bermuda & 0 & JP & Japan & 1 & SI & Slovenia & 1 \\ 
			BN & Brunei & 0 & KE & Kenya & 0 & SK & Slovakia & 1 \\ 
			BO & Bolivia & 0 & KG & Kyrgyzstan & 0 & SL & Sierra Leone & 0 \\ 
			BR & Brazil & 1 & KH & Cambodia & 0 & SM & San Marino & 0 \\ 
			BS & Bahamas & 0 & KN & St. Kitts \& Nevis & 0 & SN & Senegal & 0 \\ 
			BW & Botswana & 0 & KR & South Korea & 1 & SR & Suriname & 0 \\ 
			BY & Belarus & 1 & KW & Kuwait & 1 & SV & El Salvador & 0 \\ 
			BZ & Belize & 0 & KY & Cayman Islands & 0 & SY & Syria & 0 \\ 
			CA & Canada & 1 & KZ & Kazakhstan & 1 & TC & Turks \& Caicos Islands & 0 \\ 
			CF & Central African Republic & 0 & LA & Laos & 0 & TG & Togo & 0 \\ 
			CH & Switzerland & 1 & LB & Lebanon & 0 & TH & Thailand & 1 \\ 
			CI & Côte d’Ivoire & 0 & LC & St. Lucia & 0 & TJ & Tajikistan & 0 \\ 
			CL & Chile & 0 & LI & Liechtenstein & 0 & TL & Timor-Leste & 0 \\ 
			CM & Cameroon & 0 & LK & Sri Lanka & 0 & TM & Turkmenistan & 0 \\ 
			CN & China & 1 & LS & Lesotho & 0 & TN & Tunisia & 0 \\ 
			CO & Colombia & 0 & LT & Lithuania & 1 & TO & Tonga & 0 \\ 
			CR & Costa Rica & 0 & LU & Luxembourg & 1 & TR & Turkey & 1 \\ 
			CU & Cuba & 0 & LV & Latvia & 1 & TT & Trinidad \& Tobago & 0 \\ 
			CV & Cape Verde & 0 & LY & Libya & 0 & TW & Taiwan & 1 \\ 
			CY & Cyprus & 1 & MA & Morocco & 0 & TZ & Tanzania & 0 \\ 
			CZ & Czechia & 1 & MC & Monaco & 0 & UA & Ukraine & 1 \\ 
			DE & Germany & 1 & MD & Moldova & 0 & UG & Uganda & 0 \\ 
			DJ & Djibouti & 0 & MG & Madagascar & 0 & US & United States & 1 \\ 
			DK & Denmark & 1 & MK & Macedonia & 0 & UY & Uruguay & 0 \\ 
			DM & Dominica & 0 & ML & Mali & 0 & UZ & Uzbekistan & 0 \\ 
			DO & Dominican Republic & 0 & MN & Mongolia & 0 & VC & St. Vincent \& Grenadines & 0 \\ 
			DZ & Algeria & 0 & MO & Macau SAR China & 0 & VE & Venezuela & 0 \\ 
			EC & Ecuador & 0 & MR & Mauritania & 0 & VG & British Virgin Islands & 0 \\ 
			EE & Estonia & 1 & MS & Montserrat & 0 & VI & U.S. Virgin Islands & 0 \\ 
			EG & Egypt & 0 & MT & Malta & 0 & VN & Vietnam & 1 \\ 
			ES & Spain & 1 & MU & Mauritius & 0 & VU & Vanuatu & 0 \\ 
			ET & Ethiopia & 0 & MV & Maldives & 0 & WS & Samoa & 0 \\ 
			FI & Finland & 1 & MW & Malawi & 0 & YE & Yemen & 0 \\ 
			FJ & Fiji & 0 & MX & Mexico & 1 & YT & Mayotte & 0 \\ 
			FO & Faroe Islands & 0 & MY & Malaysia & 1 & ZA & South Africa & 1 \\ 
			FR & France & 1 & MZ & Mozambique & 0 & ZM & Zambia & 0 \\ 
			GA & Gabon & 0 & NA & Namibia & 0 & ZW & Zimbabwe & 0 \\ 
			GB & United Kingdom & 1 & NC & New Caledonia & 0 & GF & French Guiana & 0 \\ 
			GD & Grenada & 0 & NE & Niger & 0 & KI & Kiribati & 0 \\ 
			GE & Georgia & 0 & NG & Nigeria & 1 & CD & Congo - Kinshasa & 0 \\ 
			GG & Guernsey & 0 & NI & Nicaragua & 0 & CG & Congo - Brazzaville & 0 \\ 
			GH & Ghana & 0 & NL & Netherlands & 1 & MQ & Martinique & 0 \\ 
			GI & Gibraltar & 0 & NO & Norway & 1 & SZ & Swaziland & 0 \\ 
			GL & Greenland & 0 & NP & Nepal & 0 & CK & Cook Islands & 0 \\ 
			GM & Gambia & 0 & NZ & New Zealand & 1 & VA & Holy See & 0 \\ 
			GN & Guinea & 0 & OM & Oman & 0 & TD & Chad & 0 \\ 
			GP & Guadeloupe & 0 & PA & Panama & 1 &  &  &  \\ 
			\hline \\[-1.8ex] 
		\end{tabular} 
	}
		\caption{Countries included in the analysis with ISO 2 country code (ISO), name of the country (name) and occurrence in the small MT 103 network set (reduced=1).  \\ Source: SWIFT BI Watch.}
		\label{tab:countries} 
	\end{table} 
	\FloatBarrier
	\newpage
	\section{Descriptives for the reduced data set}\label{annex:reduced}
	\FloatBarrier
	\begin{figure}[!htbp]
		\centering
		\begin{subfigure}{\textwidth}
			\centering			\includegraphics[trim={0cm 0cm 0cm 0cm},clip,width=0.97\textwidth]{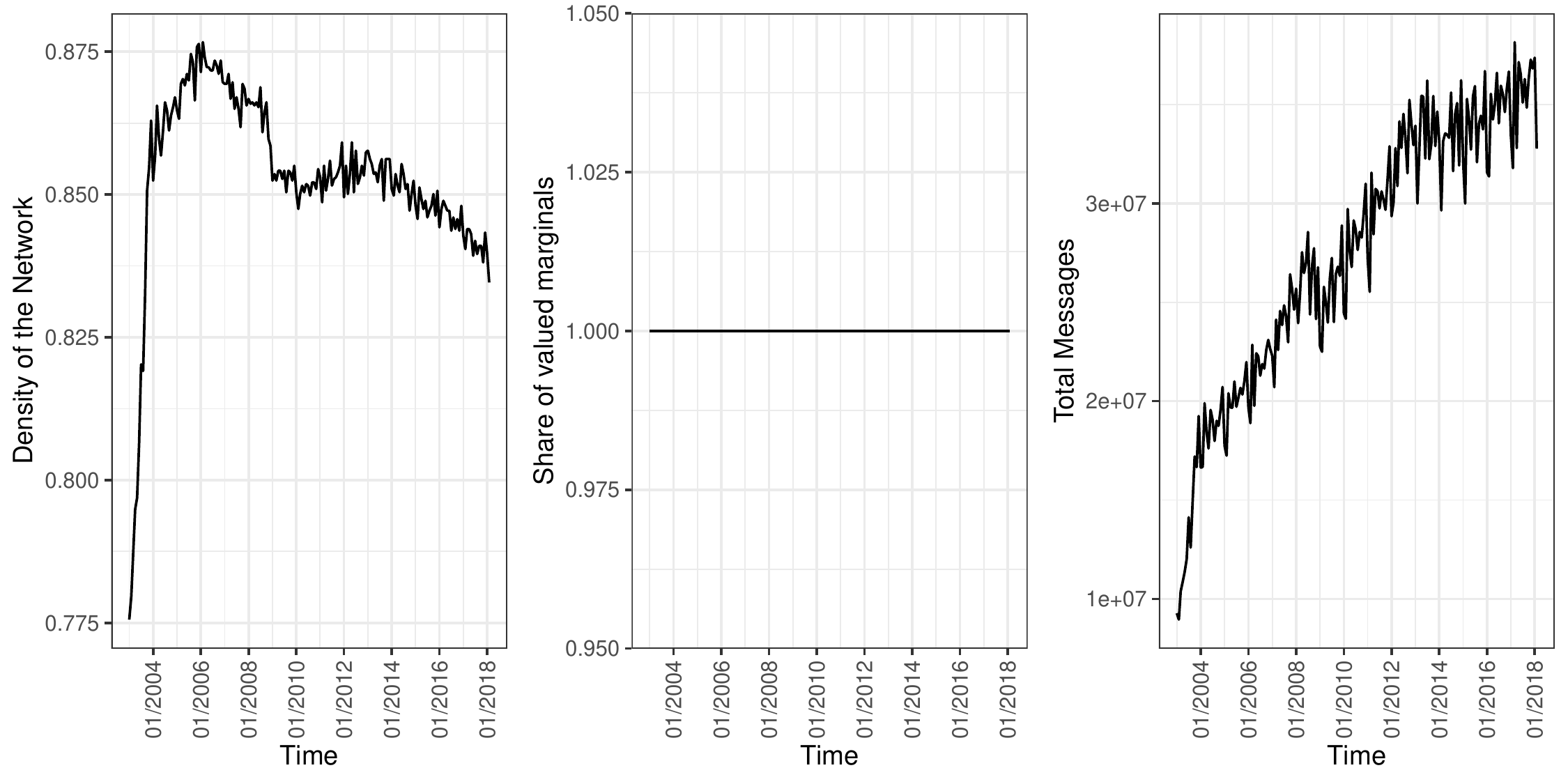}
		\end{subfigure}

		\caption{Summary statistics for the reduced MT 103 network as monthly time series. Density of the network (left), share of non-zero marginals (middle) and cumulative edge values (right). \\ Source: SWIFT BI Watch.}
		\label{fig:nwsummary_small}
	\end{figure}	
	\begin{figure}[!htbp]
		\centering
		\begin{subfigure}{\textwidth}
			\centering			\includegraphics[trim={0cm 0cm 0cm 0cm},clip,width=0.98\textwidth]{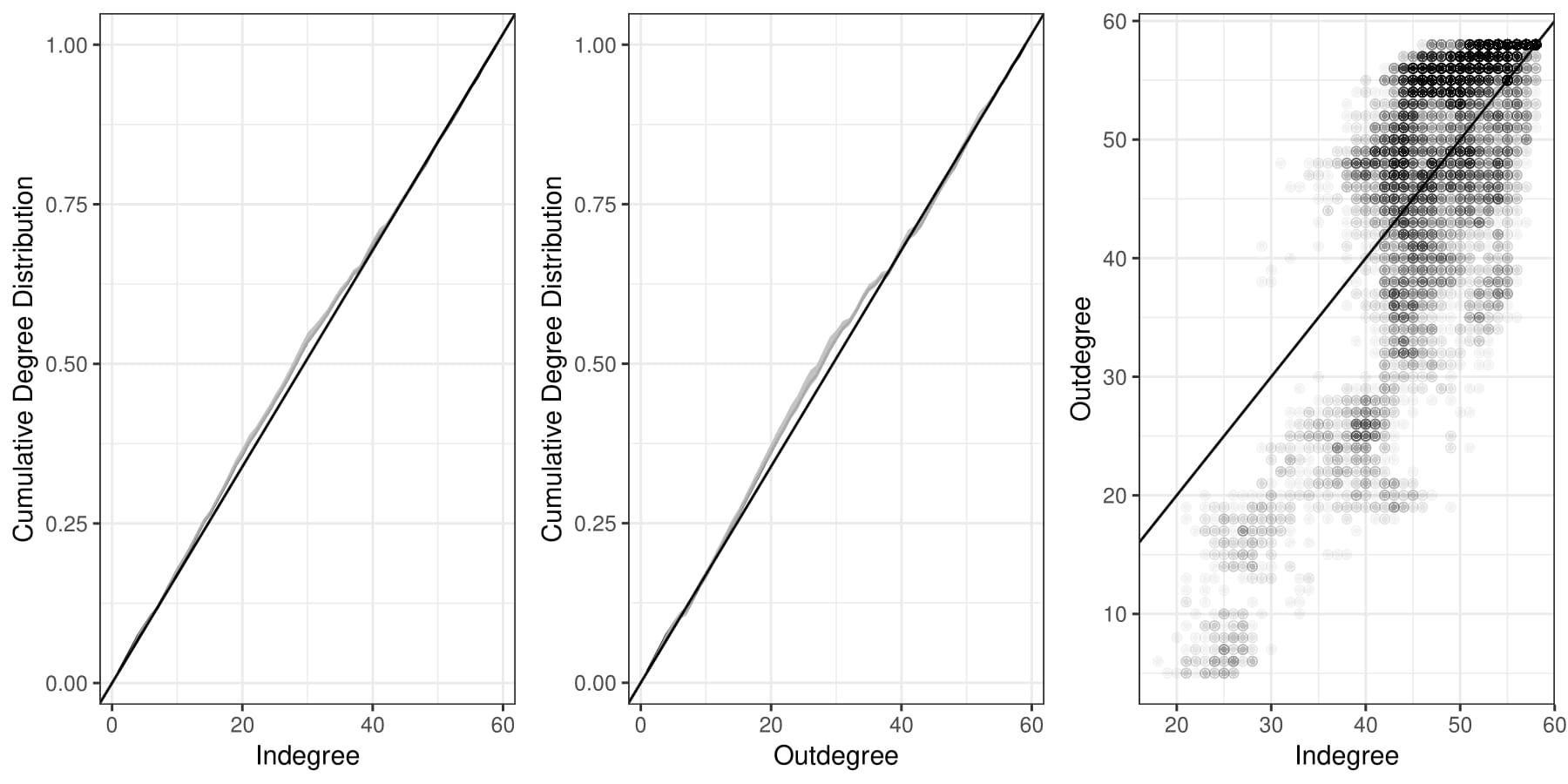}
		\end{subfigure}

		\caption{Binary network topology of the reduced MT 103 network aggregated for all time points. Cumulative indegree (left) and outdegree distributions (middle) with  maximum and minimum values indicated in grey. Outdegree against indegree (right) for all months in dotted with colour intensity by frequency. 45 degree line in solid black. \\ Source: SWIFT BI Watch.}
		\label{fig:degdist_small}
	\end{figure}
	\FloatBarrier
	\newpage
	\section{Binary Reconstruction: Full Network}
	\subsection{Predicted Adjacency Matrices: Full Network} \label{annex:adj}
	\begin{figure}[!htbp]
		\centering
		\begin{subfigure}{\textwidth}
			\centering			\includegraphics[trim={2cm 2.5cm 1.5cm 2cm},clip,width=\textwidth]{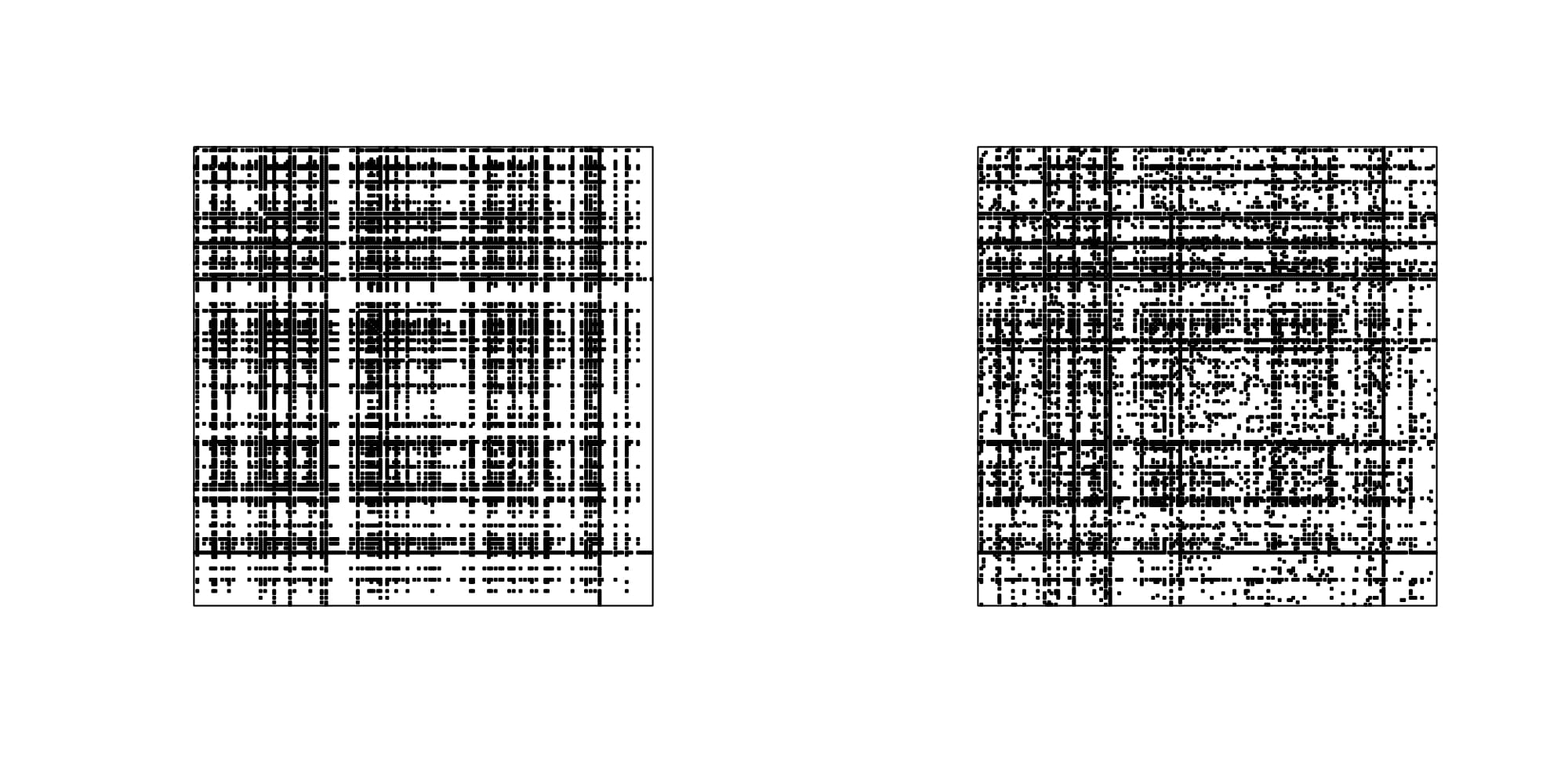}
		\end{subfigure}

		\caption{Adjacency matrices, representing the full MT 103 network in February 2018. IFPF reconstruction (left), real network (right). \\ Source: SWIFT BI Watch.}
		\label{fig:recon_ipfp}
	\end{figure}
	
	\begin{figure}[!htbp]
		\centering
		\begin{subfigure}{\textwidth}
			\centering			\includegraphics[trim={2cm 2.5cm 1.5cm 2cm},clip,width=\textwidth]{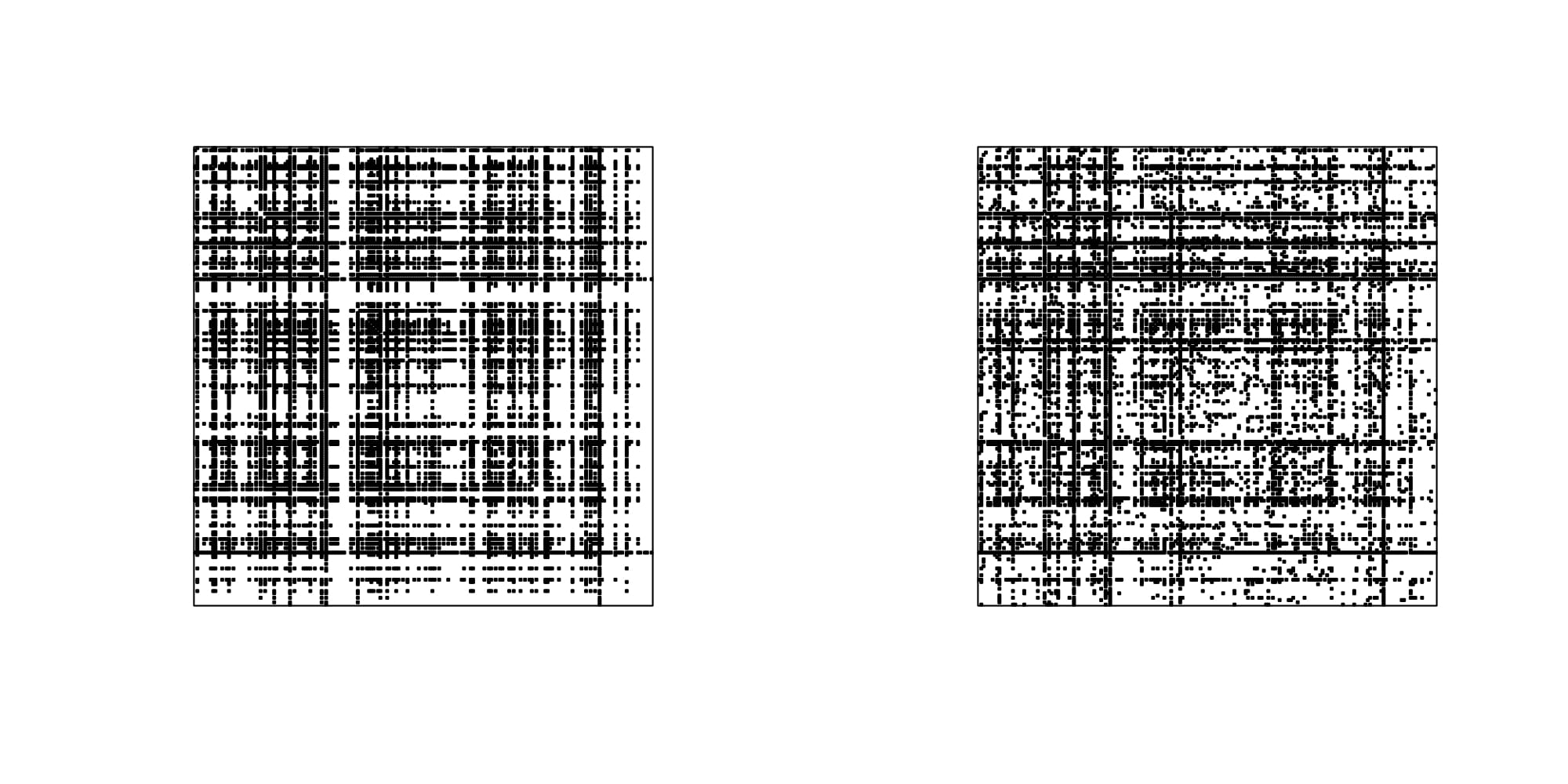}
		\end{subfigure}

		\caption{Adjacency matrices, representing the full MT 103 network in February 2018. GRAVITY reconstruction (left), real network (right). \\ Source: SWIFT BI Watch.}
		\label{fig:recon_gravity}
	\end{figure}
	
	\begin{figure}[!htbp]
		\centering
		\begin{subfigure}{\textwidth}
			\centering			\includegraphics[trim={2cm 2.5cm 1.5cm 2cm},clip,width=\textwidth]{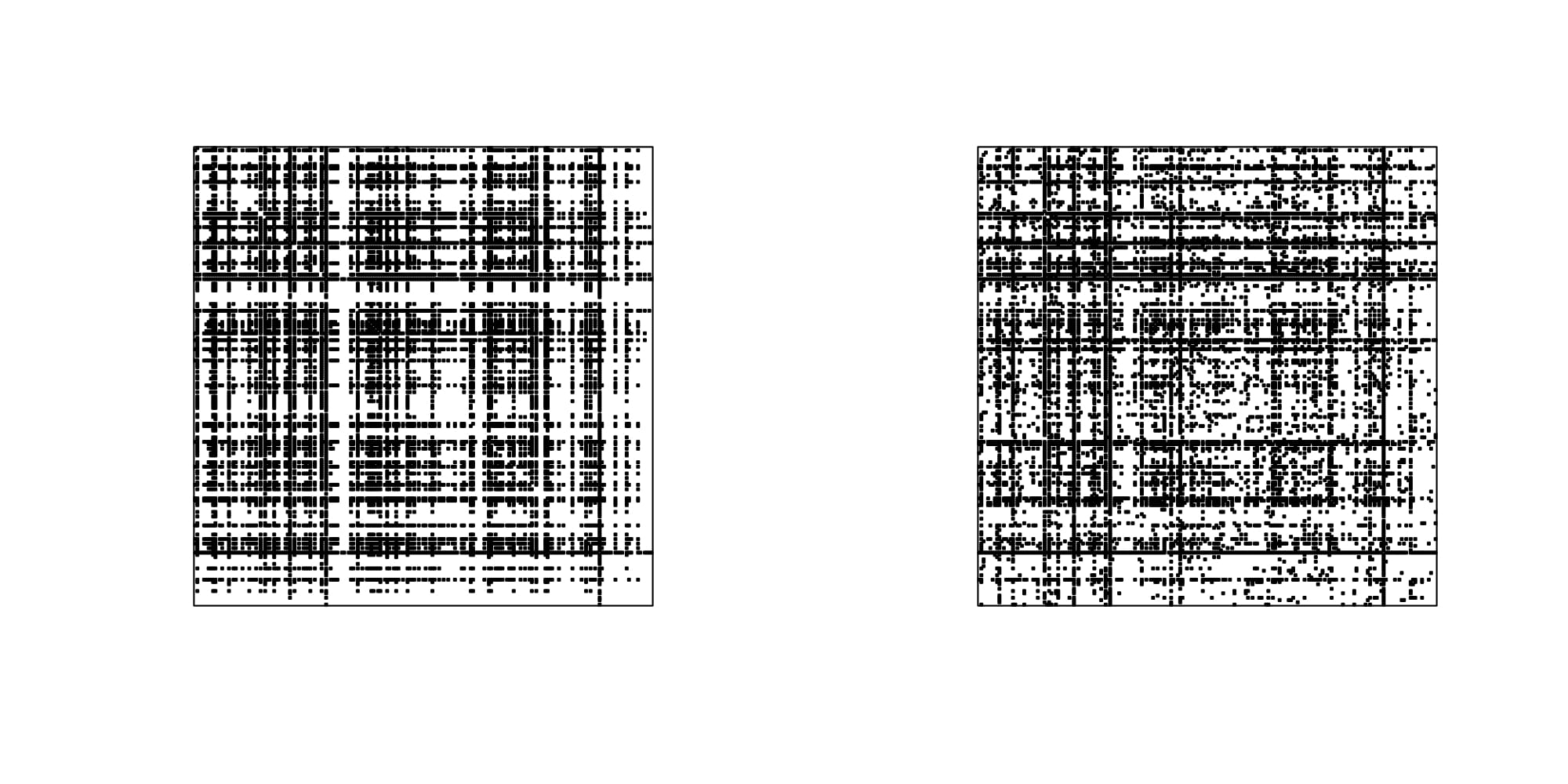}
		\end{subfigure}

		\caption{Adjacency matrices, representing the full MT 103 network in February 2018. DC-GRAVITY reconstruction (left), real network (right). \\ Source: SWIFT BI Watch.}
		\label{fig:recon_cimi}
	\end{figure}
	
	\begin{figure}[!htbp]
		\centering
		\begin{subfigure}{\textwidth}
			\centering			\includegraphics[trim={2cm 2.5cm 1.5cm 2cm},clip,width=\textwidth]{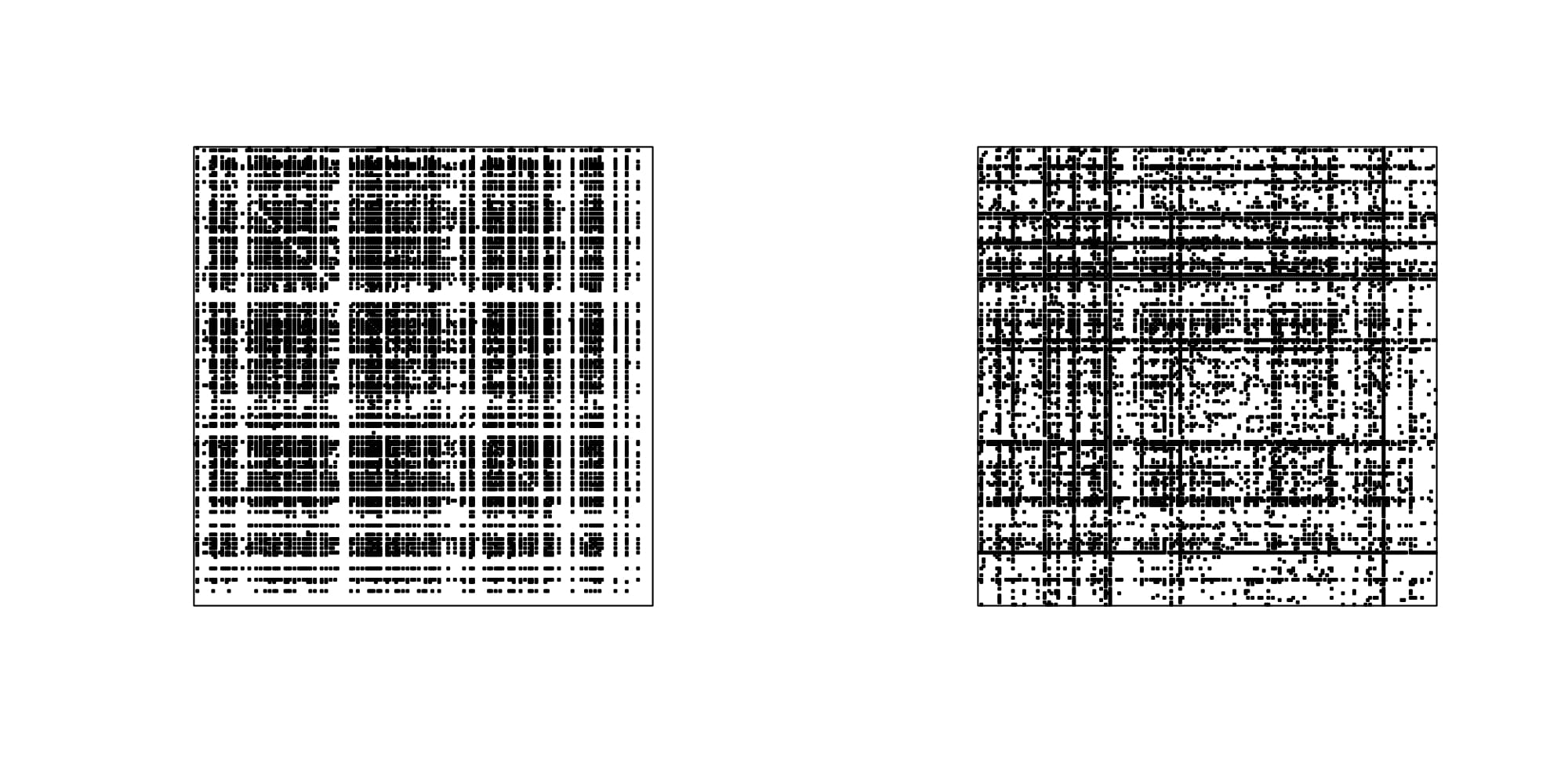}
		\end{subfigure}

		\caption{Adjacency matrices, representing the full MT 103 network in February 2018. H-ER reconstruction (left), real network (right). \\ Source: SWIFT BI Watch.}
		\label{fig:recon_er}
	\end{figure}
	
	\begin{figure}[!htbp]
		\centering
		\begin{subfigure}{\textwidth}
			\centering			\includegraphics[trim={2cm 2.5cm 1.5cm 2cm},clip,width=\textwidth]{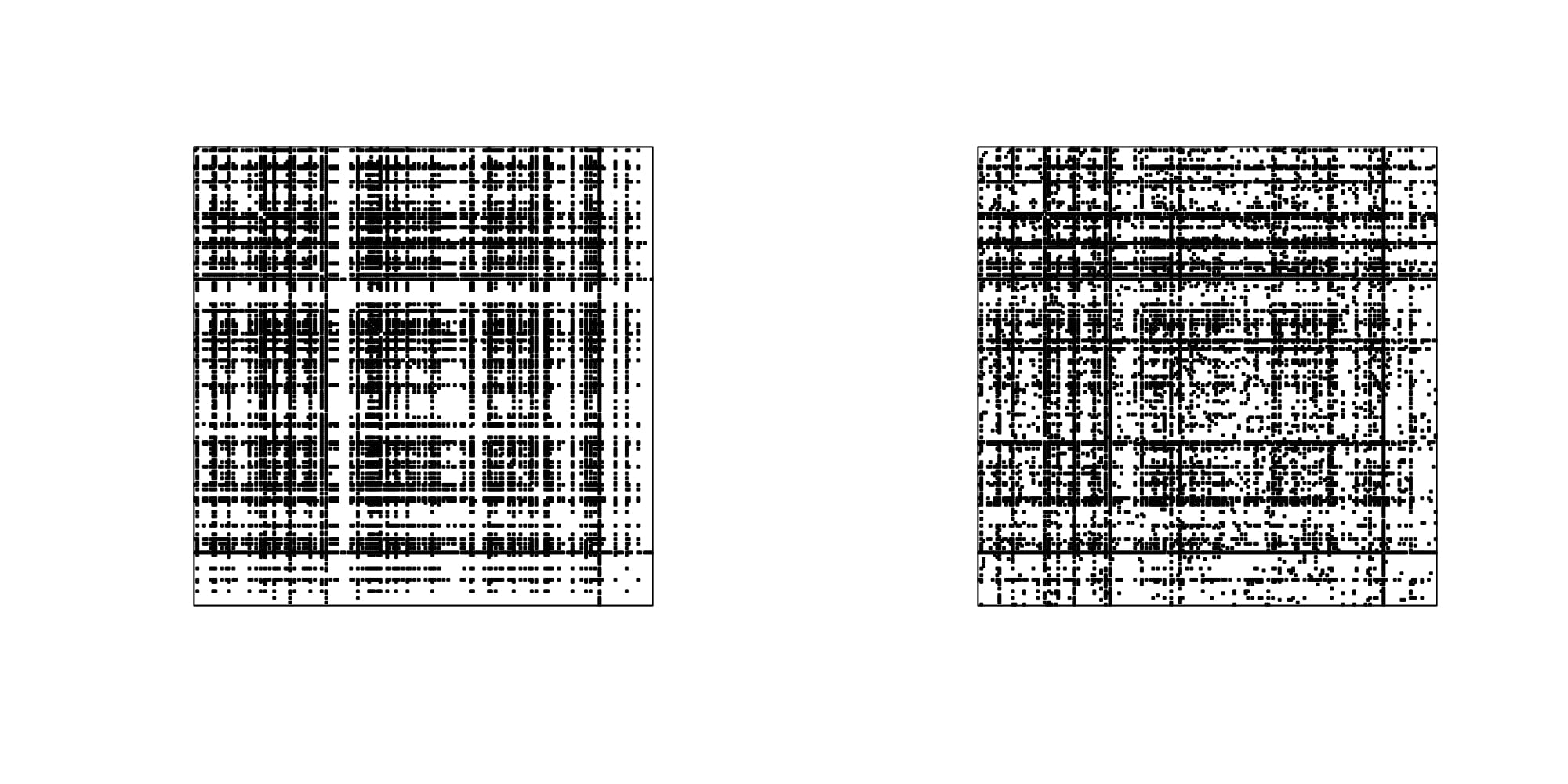}
		\end{subfigure}

		\caption{Adjacency matrices, representing the full MT 103 network in February 2018. H-FIT reconstruction (left), real network (right). \\ Source: SWIFT BI Watch.}
		\label{fig:recon_fit}
	\end{figure}
	
	\begin{figure}[!htbp]
		\centering
		\begin{subfigure}{\textwidth}
			\centering			\includegraphics[trim={2cm 2.5cm 1.5cm 2cm},clip,width=\textwidth]{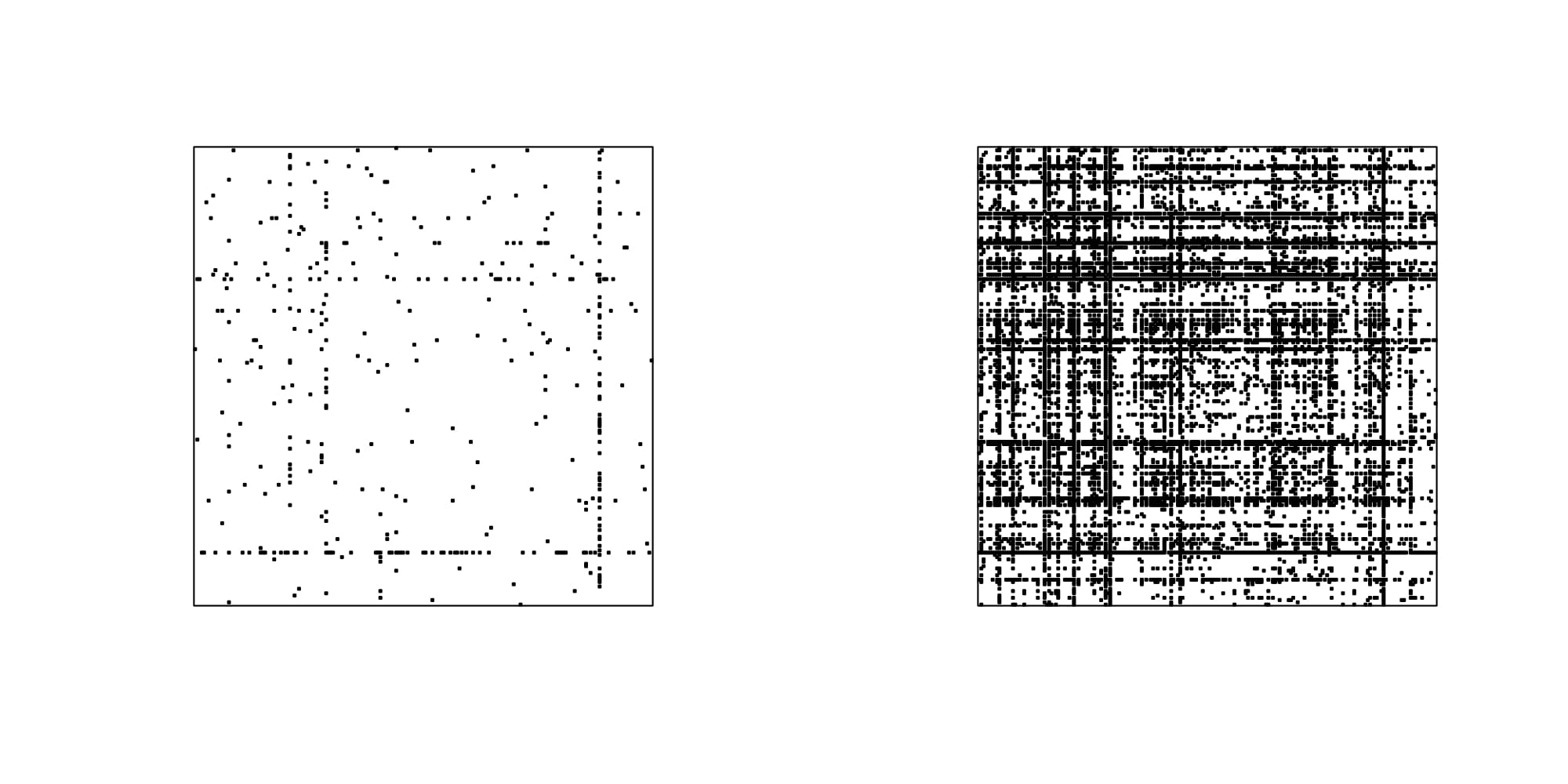}
		\end{subfigure}

		\caption{Adjacency matrices, representing the full MT 103 network in February 2018. MINDENS reconstruction (left), real network (right). \\ Source: SWIFT BI Watch.}
		\label{fig:recon_mindens}
	\end{figure}
	
	\begin{figure}[!htbp]
		\centering
		\begin{subfigure}{\textwidth}
			\centering			\includegraphics[trim={2cm 2.5cm 1.5cm 2cm},clip,width=\textwidth]{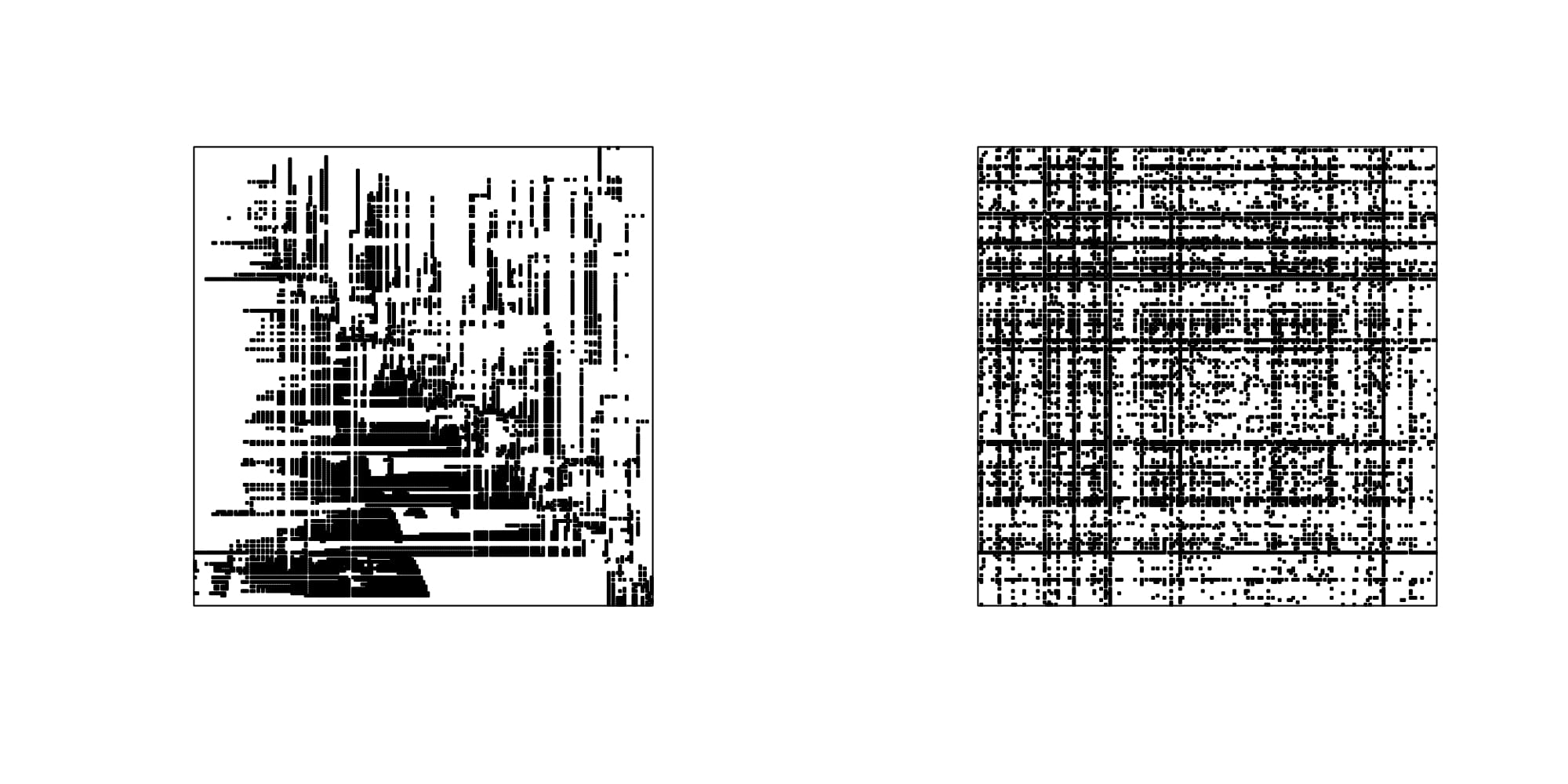}
		\end{subfigure}

		\caption{Adjacency matrices, representing the full MT 103 network in February 2018. LASSO reconstruction (left), real network (right). \\ Source: SWIFT BI Watch.}
		\label{fig:recon_lasso}
	\end{figure}

	\FloatBarrier
	\subsection{Degree Reconstruction: Full Network}\label{annex:deg}
	\begin{figure}[!htbp]
		\centering
		\begin{subfigure}{\textwidth}
			\centering			\includegraphics[trim={0cm 0cm 0cm 0cm},clip,width=\textwidth]{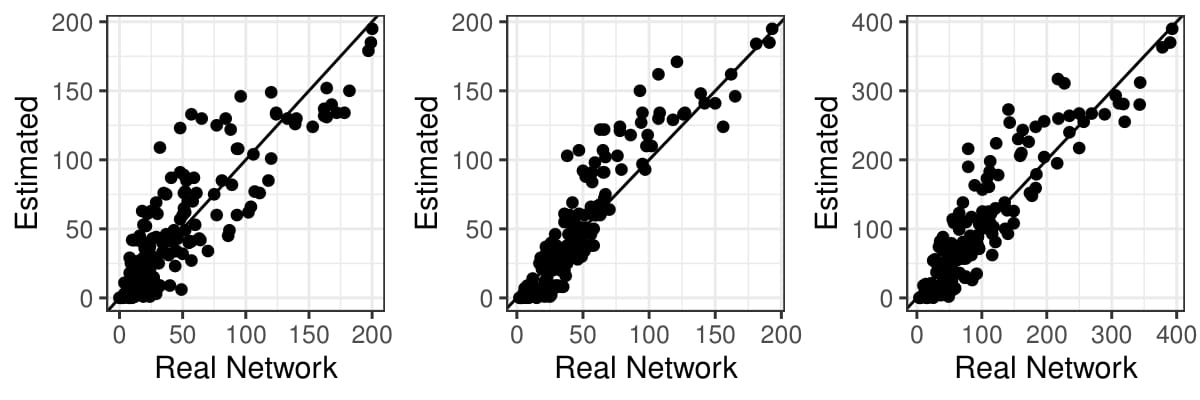}
		\end{subfigure}

		\caption{Degree Reconstruction in the full MT 103 network in February 2018. IPFP reconstruction of the outdegree (left), outdegree (middle) and in- and outdegree (right). \\ Source: SWIFT BI Watch.}
		\label{fig:deg_recon_ipfp}
	\end{figure}
	
	\begin{figure}[!htbp]
		\centering
		\begin{subfigure}{\textwidth}
			\centering			\includegraphics[trim={0cm 0cm 0cm 0cm},clip,width=\textwidth]{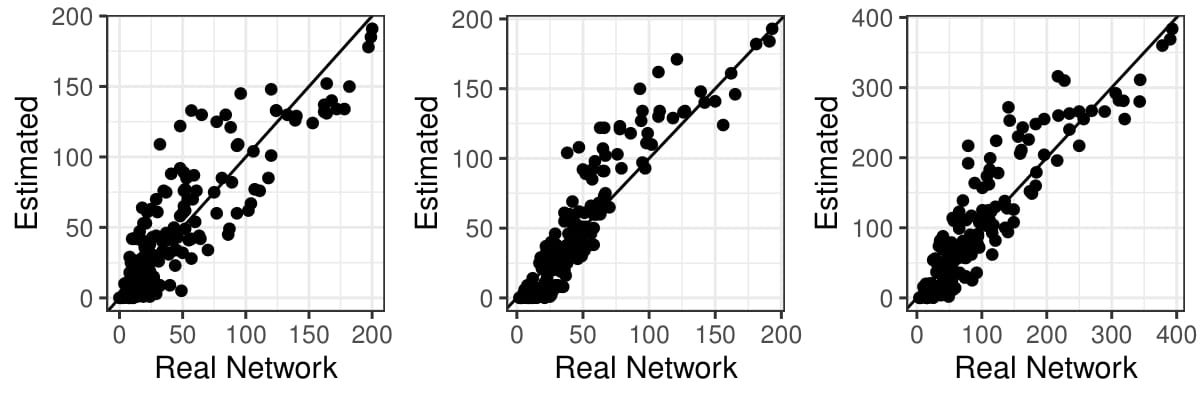}
		\end{subfigure}

		\caption{Degree Reconstruction in the full MT 103 network in February 2018. GRAVITY reconstruction of the outdegree (left), outdegree (middle) and in- and outdegree (right).
		 \\ Source: SWIFT BI Watch.}
		\label{fig:deg_recon_gravity}
	\end{figure}
	
	\begin{figure}[!htbp]
		\centering
		\begin{subfigure}{\textwidth}
			\centering			\includegraphics[trim={0cm 0cm 0cm 0cm},clip,width=\textwidth]{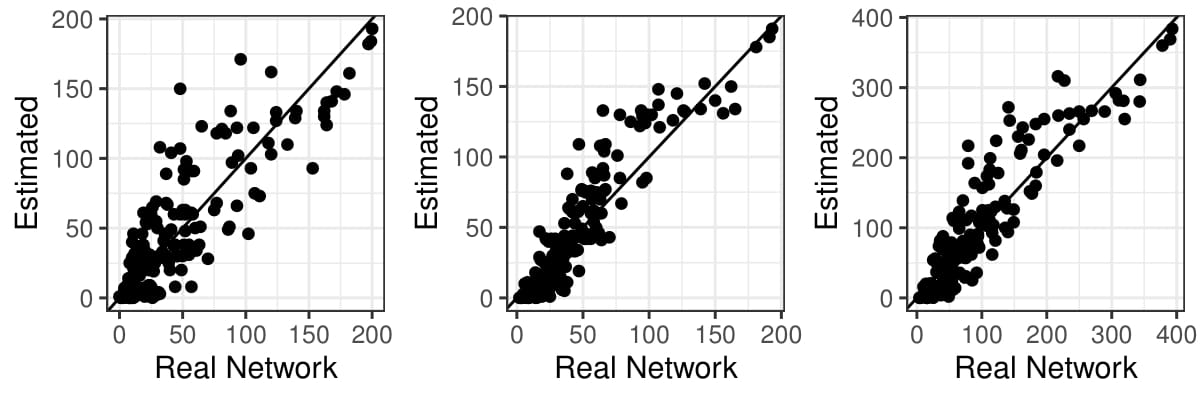}
		\end{subfigure}

		\caption{Degree Reconstruction in the full MT 103 network in February 2018. DC-GRAVITY reconstruction of the outdegree (left), outdegree (middle) and in- and outdegree (right). \\ Source: SWIFT BI Watch.}
		\label{fig:deg_recon_cimi}
	\end{figure}
	
	\begin{figure}[!htbp]
		\centering
		\begin{subfigure}{\textwidth}
			\centering			\includegraphics[trim={0cm 0cm 0cm 0cm},clip,width=\textwidth]{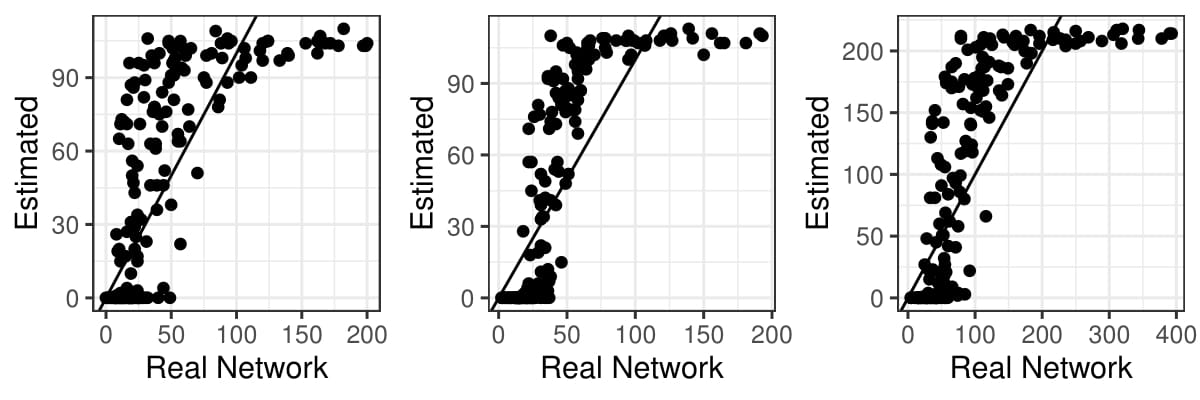}
		\end{subfigure}

		\caption{Degree Reconstruction in the full MT 103 network in February 2018. H-ER reconstruction of the outdegree (left), outdegree (middle) and in- and outdegree (right). \\ Source: SWIFT BI Watch.}
		\label{fig:deg_recon_er}
	\end{figure}
	
	\begin{figure}[!htbp]
		\centering
		\begin{subfigure}{\textwidth}
			\centering			\includegraphics[trim={0cm 0cm 0cm 0cm},clip,width=\textwidth]{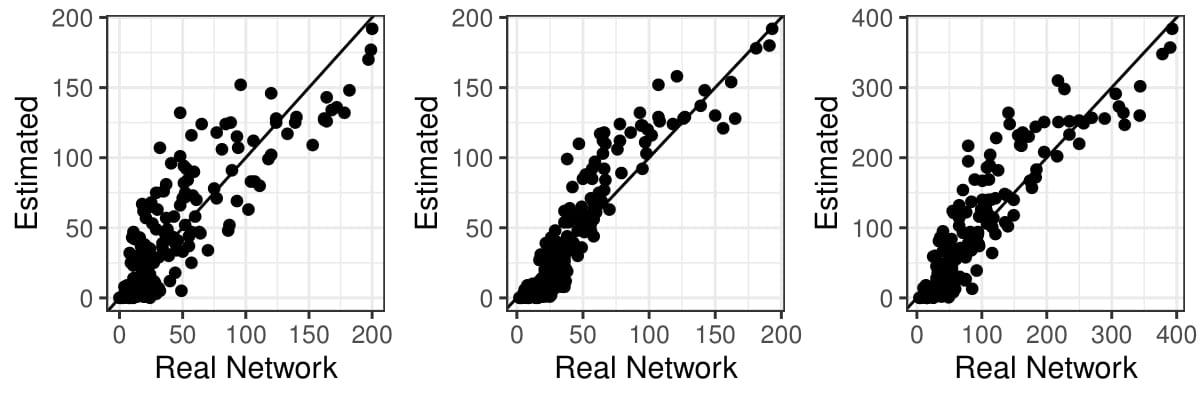}
		\end{subfigure}

		\caption{Degree Reconstruction in the full MT 103 network in February 2018. H-FIT reconstruction of the outdegree (left), outdegree (middle) and in- and outdegree (right). \\ Source: SWIFT BI Watch.}
		\label{fig:deg_recon_fit}
	\end{figure}
	
	\begin{figure}[!htbp]
		\centering
		\begin{subfigure}{\textwidth}
			\centering			\includegraphics[trim={0cm 0cm 0cm 0cm},clip,width=\textwidth]{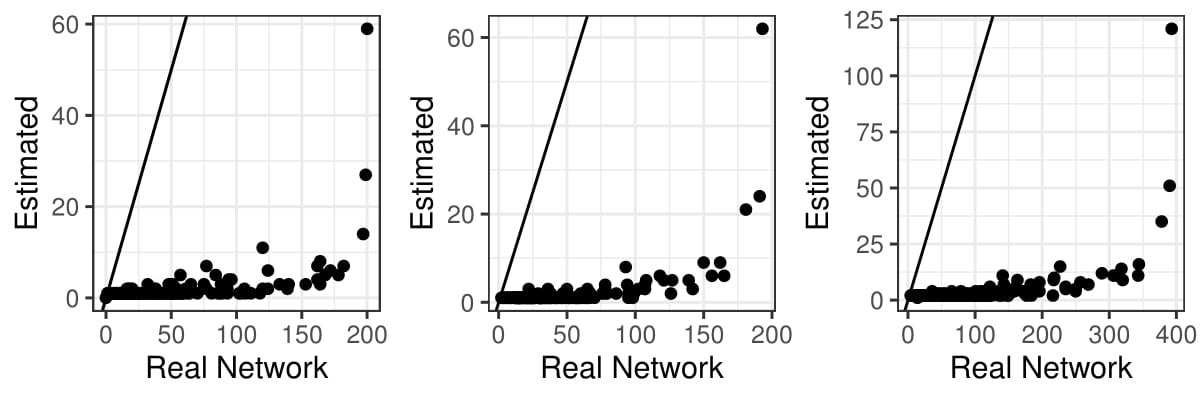}
		\end{subfigure}

		\caption{Degree Reconstruction in the full MT 103 network in February 2018. MINDENS reconstruction of the outdegree (left), outdegree (middle) and in- and outdegree (right). \\ Source: SWIFT BI Watch.}
		\label{fig:deg_recon_mindens}
	\end{figure}
	
	\begin{figure}[!htbp]
		\centering
		\begin{subfigure}{\textwidth}
			\centering			\includegraphics[trim={0cm 0cm 0cm 0cm},clip,width=\textwidth]{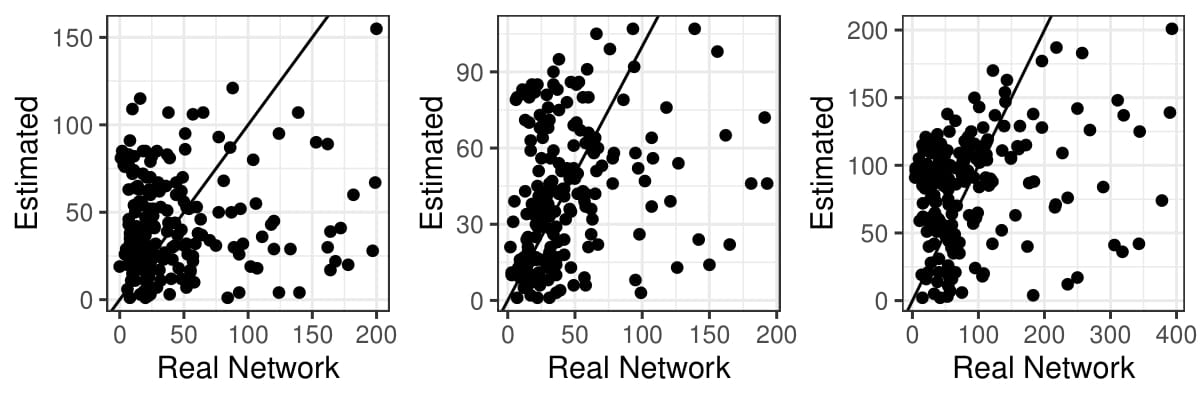}
		\end{subfigure}

		\caption{Degree Reconstruction in the full MT 103 network in February 2018. LASSO reconstruction of the outdegree (left), outdegree (middle) and in- and outdegree (right). \\ Source: SWIFT BI Watch.}
		\label{fig:deg_lasso_mindens}
	\end{figure}

	\newpage
	\section{Binary Reconstruction: Reduced Network}
	\subsection{Predicted Adjacency Matrices: Reduced Network} \label{annex:adj_small}
	\begin{figure}[!htbp]
		\centering
		\begin{subfigure}{\textwidth}
			\centering			\includegraphics[trim={2cm 2.5cm 1.5cm 2cm},clip,width=\textwidth]{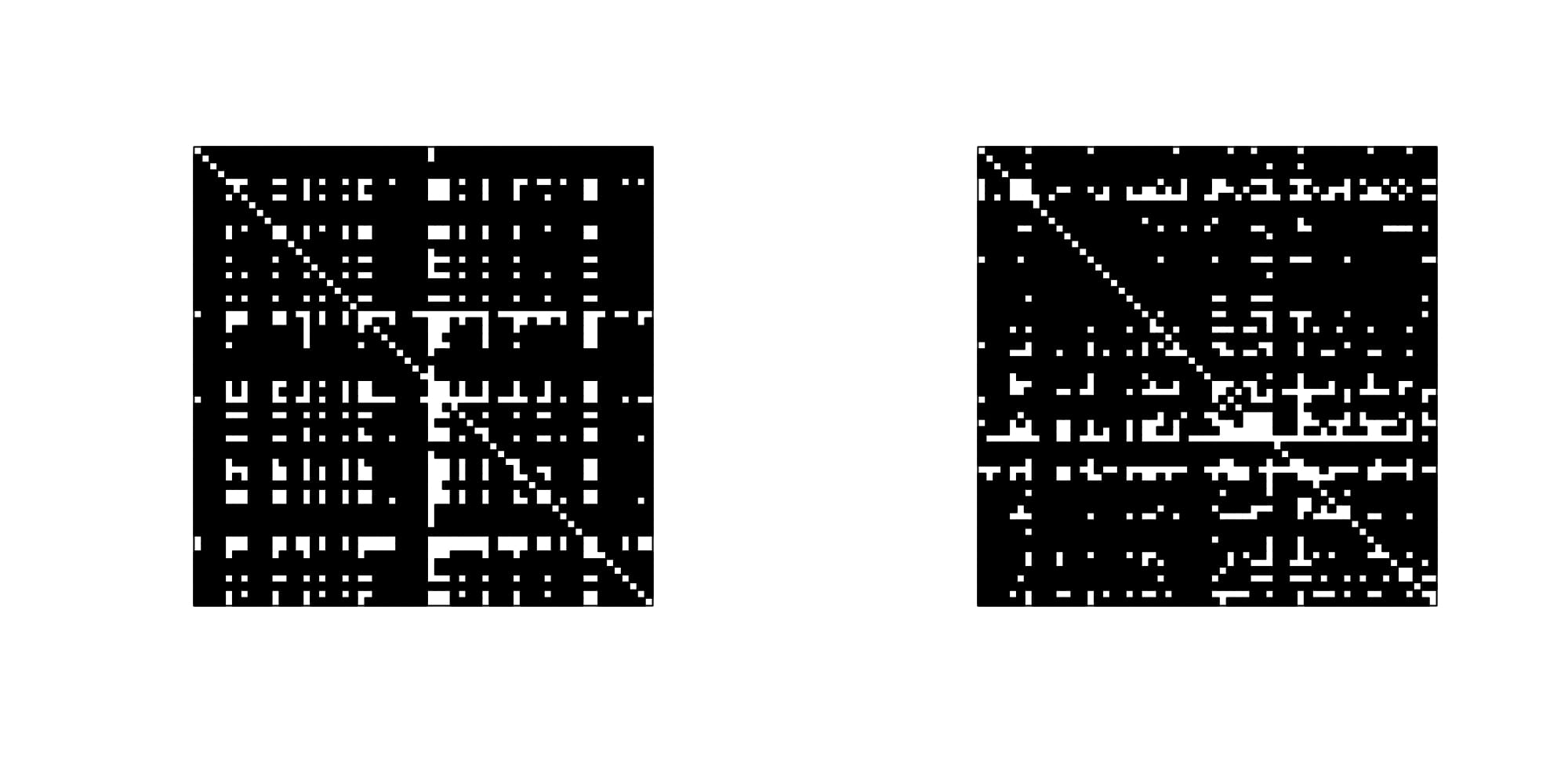}
		\end{subfigure}

		\caption{Adjacency matrices, representing the reduced MT 103 network in February 2018. IPFP reconstruction (left), real network (right). \\ Source: SWIFT BI Watch.}
		\label{fig:recon_ipfp_small}
	\end{figure}
	
	\begin{figure}[!htbp]
		\centering
		\begin{subfigure}{\textwidth}
			\centering			\includegraphics[trim={2cm 2.5cm 1.5cm 2cm},clip,width=\textwidth]{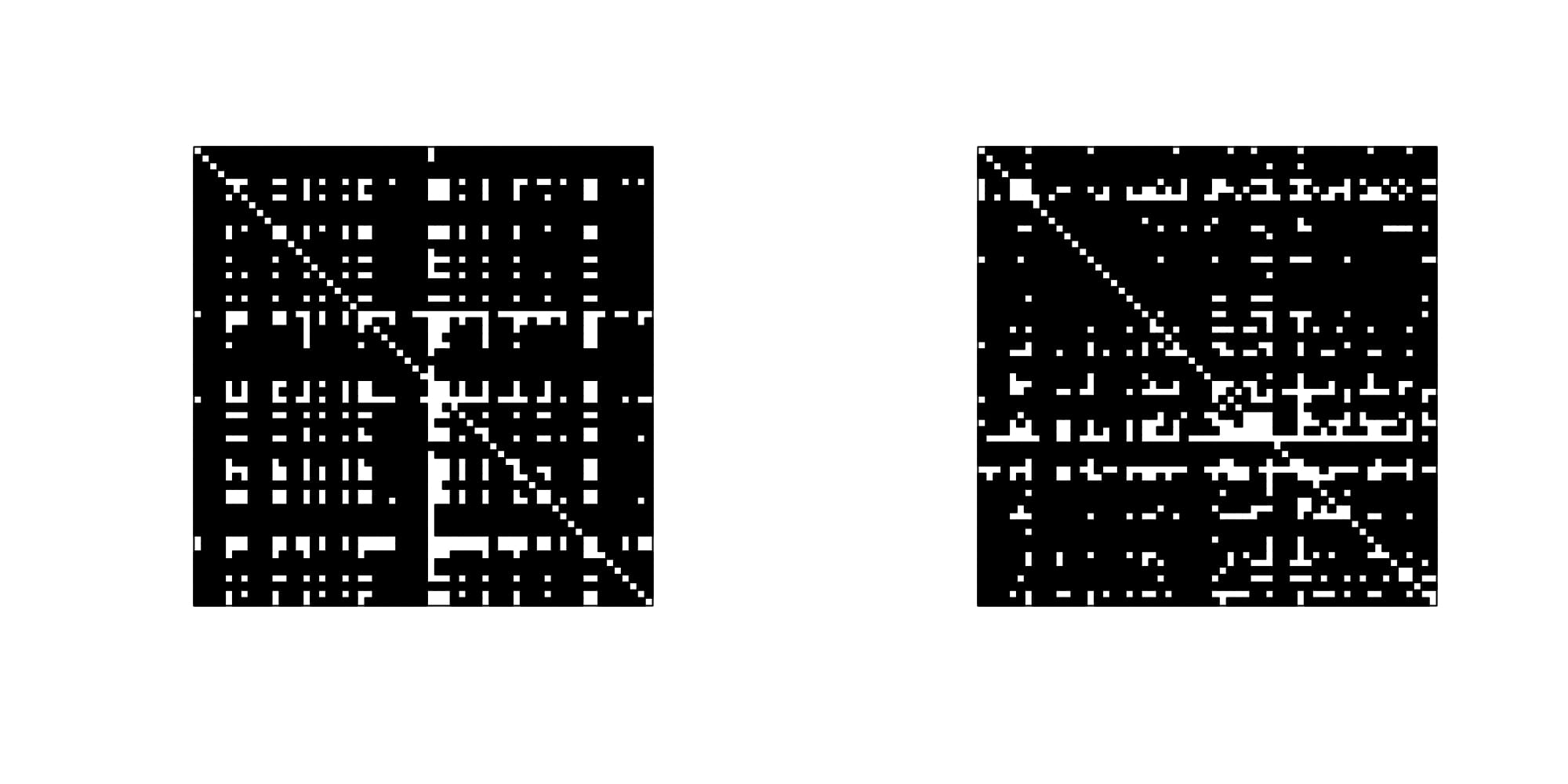}
		\end{subfigure}

		\caption{Adjacency matrices, representing the reduced MT 103 network in February 2018. GRAVITY reconstruction (left), real network (right). \\ Source: SWIFT BI Watch.}
		\label{fig:recon_gravity_small}
	\end{figure}
	
	\begin{figure}[!htbp]
		\centering
		\begin{subfigure}{\textwidth}
			\centering			\includegraphics[trim={2cm 2.5cm 1.5cm 2cm},clip,width=\textwidth]{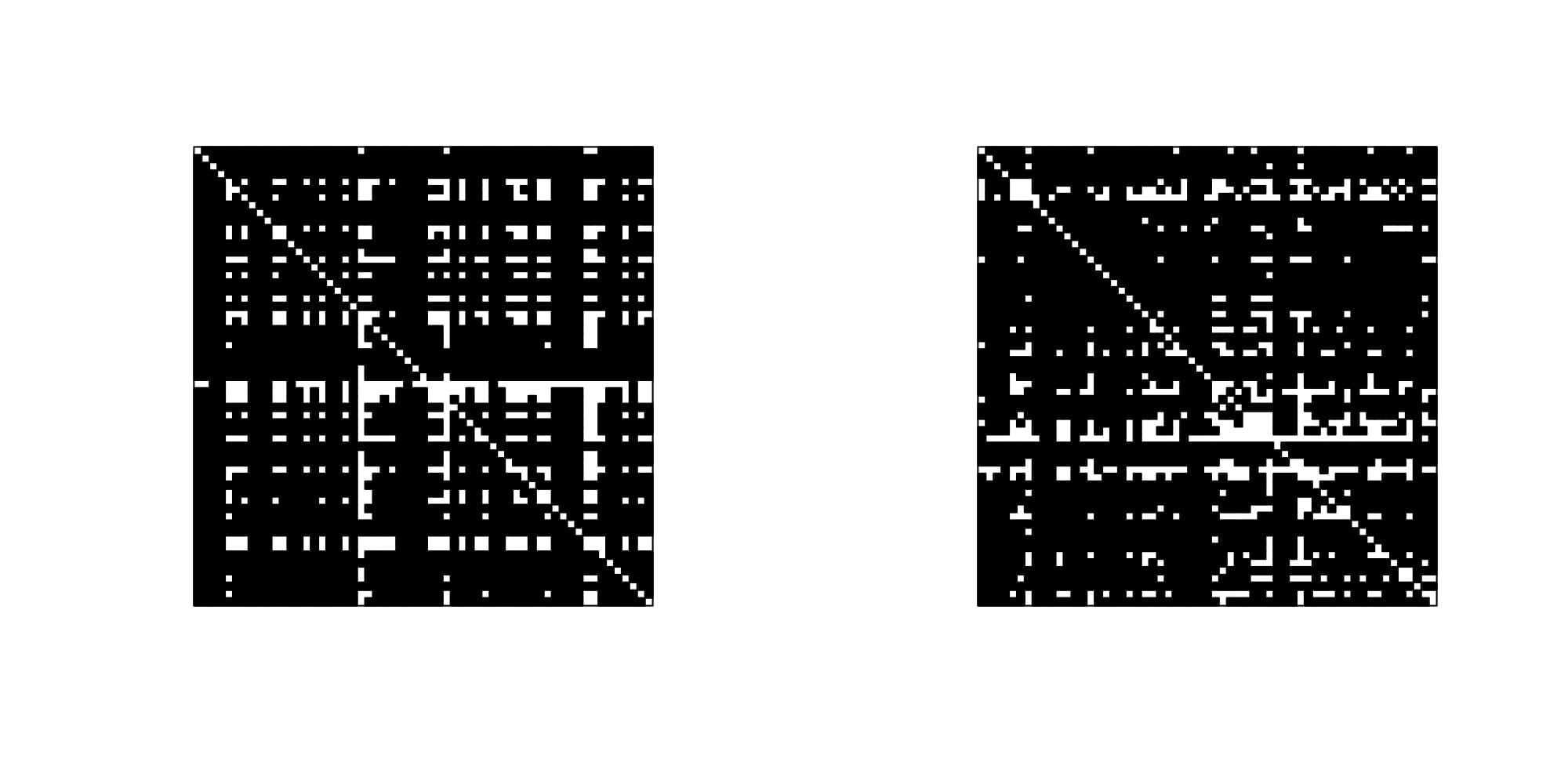}
		\end{subfigure}

		\caption{Adjacency matrices, representing the reduced MT 103 network in February 2018. DC-GRAVITY reconstruction (left), real network (right). \\ Source: SWIFT BI Watch.}
		\label{fig:recon_cimi_small}
	\end{figure}

	\begin{figure}[!htbp]
		\centering
		\begin{subfigure}{\textwidth}
			\centering			\includegraphics[trim={2cm 2.5cm 1.5cm 2cm},clip,width=\textwidth]{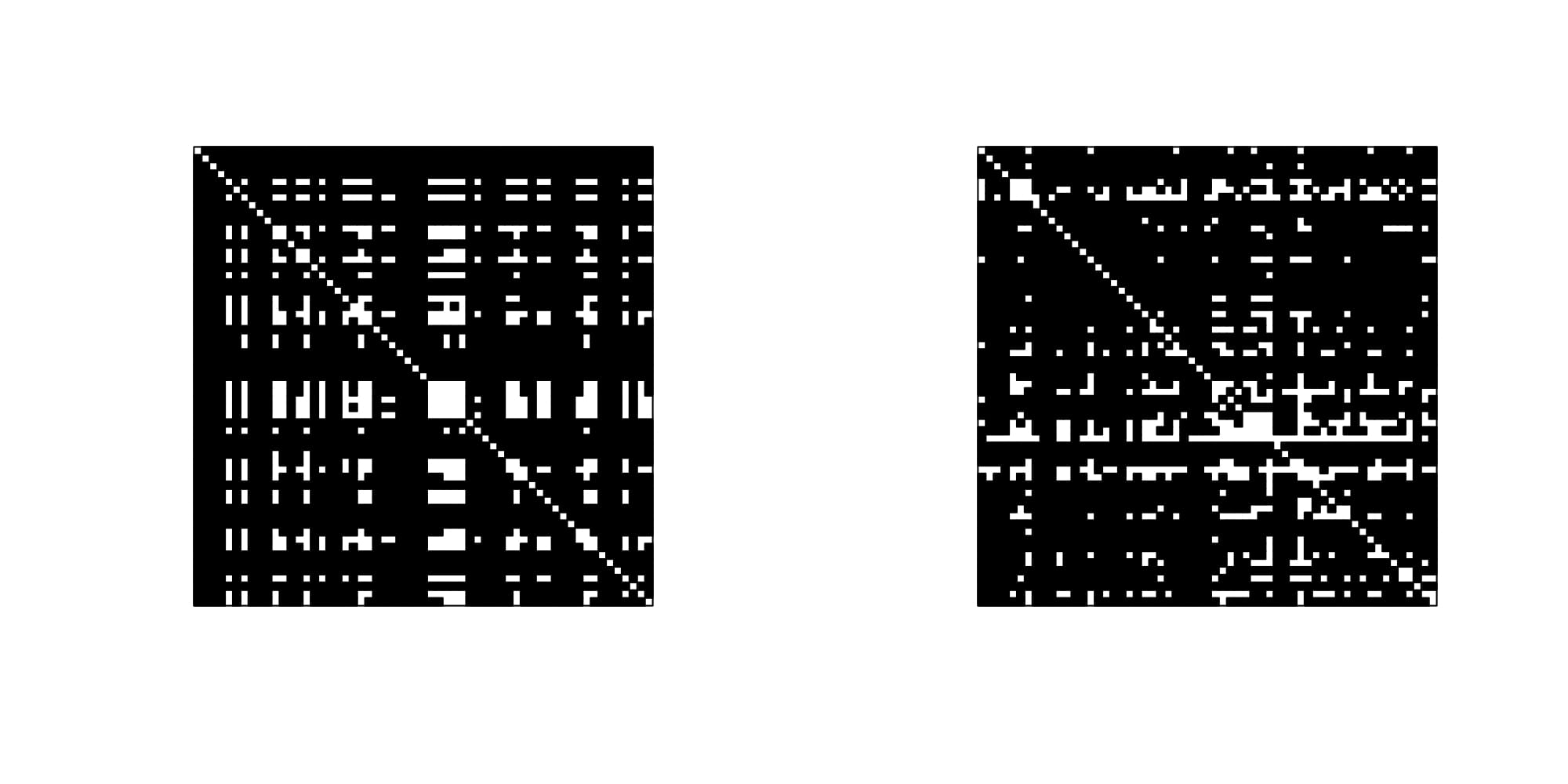}
		\end{subfigure}

		\caption{Adjacency matrices, representing the reduced MT 103 network in February 2018. DC-GRAVITY-GDP reconstruction (left), real network (right). \\ Source: SWIFT BI Watch.}
		\label{fig:recon_cimi_gdp_small}
	\end{figure}
	
	\begin{figure}[!htbp]
		\centering
		\begin{subfigure}{\textwidth}
			\centering			\includegraphics[trim={2cm 2.5cm 1.5cm 2cm},clip,width=\textwidth]{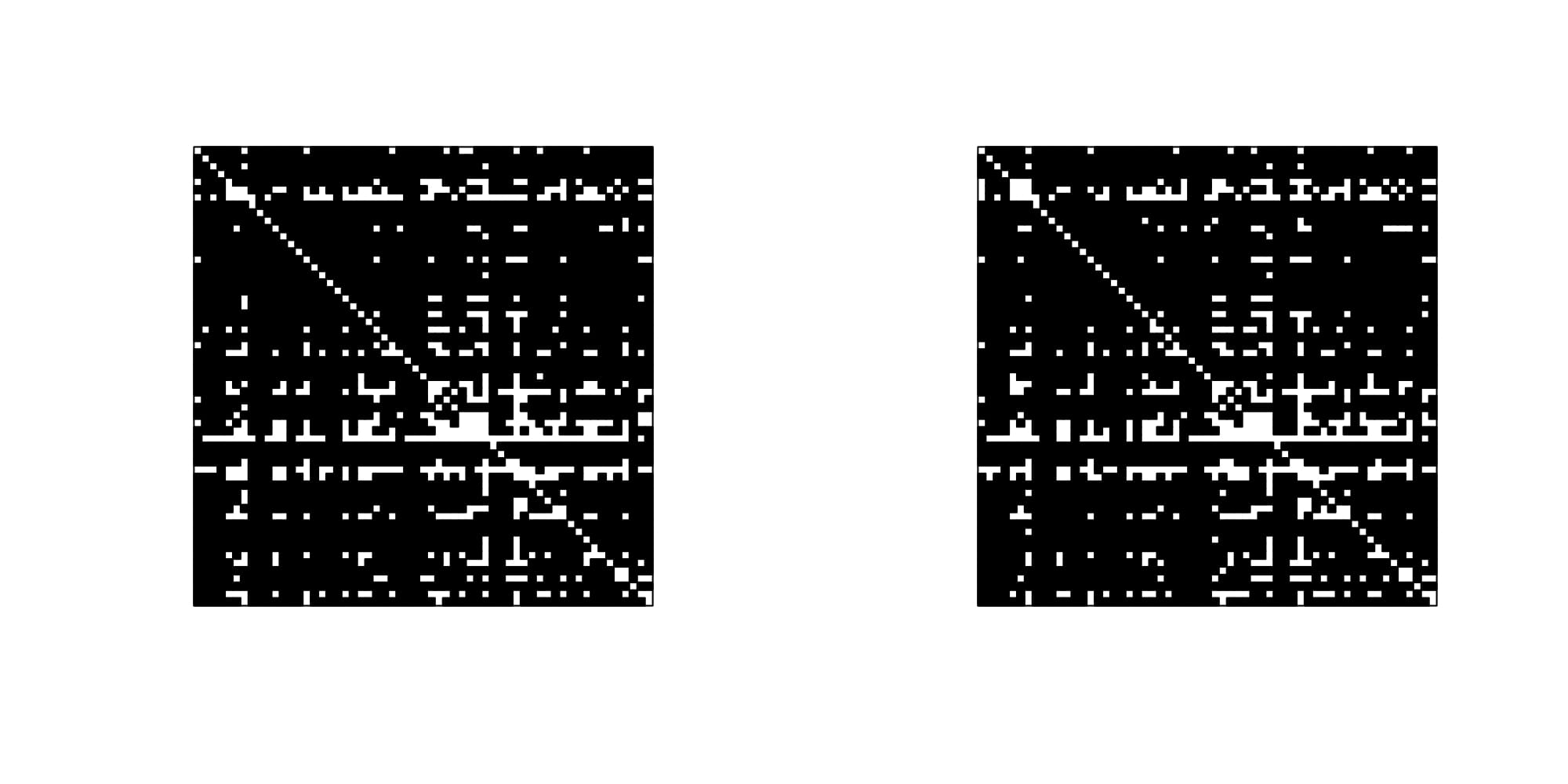}
		\end{subfigure}

		\caption{Adjacency matrices, representing the reduced MT 103 network in February 2018. DC-GRAVITY-LAG reconstruction (left), real network (right). \\ Source: SWIFT BI Watch.}
		\label{fig:recon_cimi_lag_small}
	\end{figure}
	
	\begin{figure}[!htbp]
		\centering
		\begin{subfigure}{\textwidth}
			\centering			\includegraphics[trim={2cm 2.5cm 1.5cm 2cm},clip,width=\textwidth]{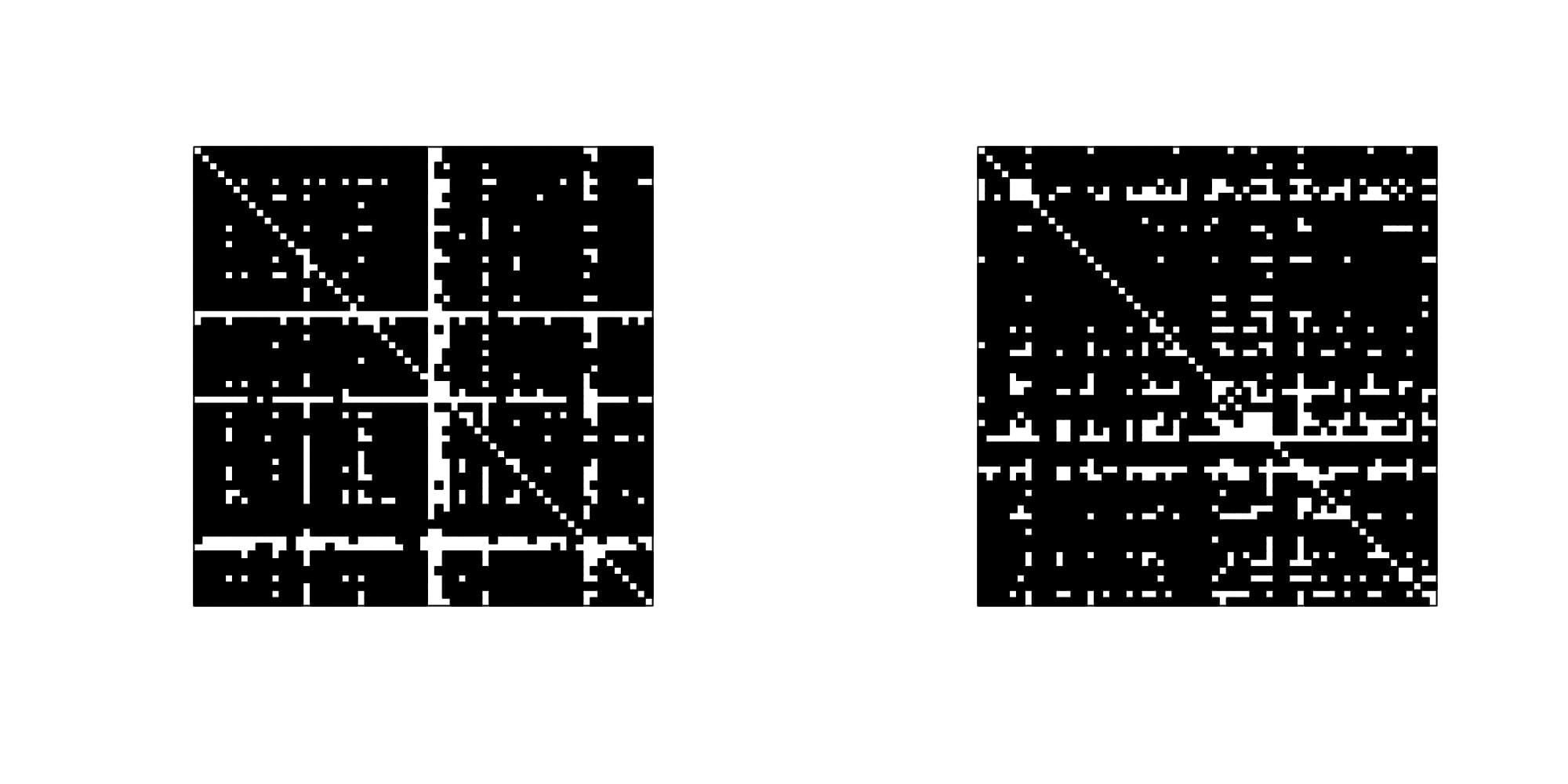}
		\end{subfigure}

		\caption{Adjacency matrices, representing the reduced MT 103 network in February 2018. H-ER reconstruction (left), real network (right). \\ Source: SWIFT BI Watch.}
		\label{fig:recon_er_small}
	\end{figure}
	
	\begin{figure}[!htbp]
		\centering
		\begin{subfigure}{\textwidth}
			\centering			\includegraphics[trim={2cm 2.5cm 1.5cm 2cm},clip,width=\textwidth]{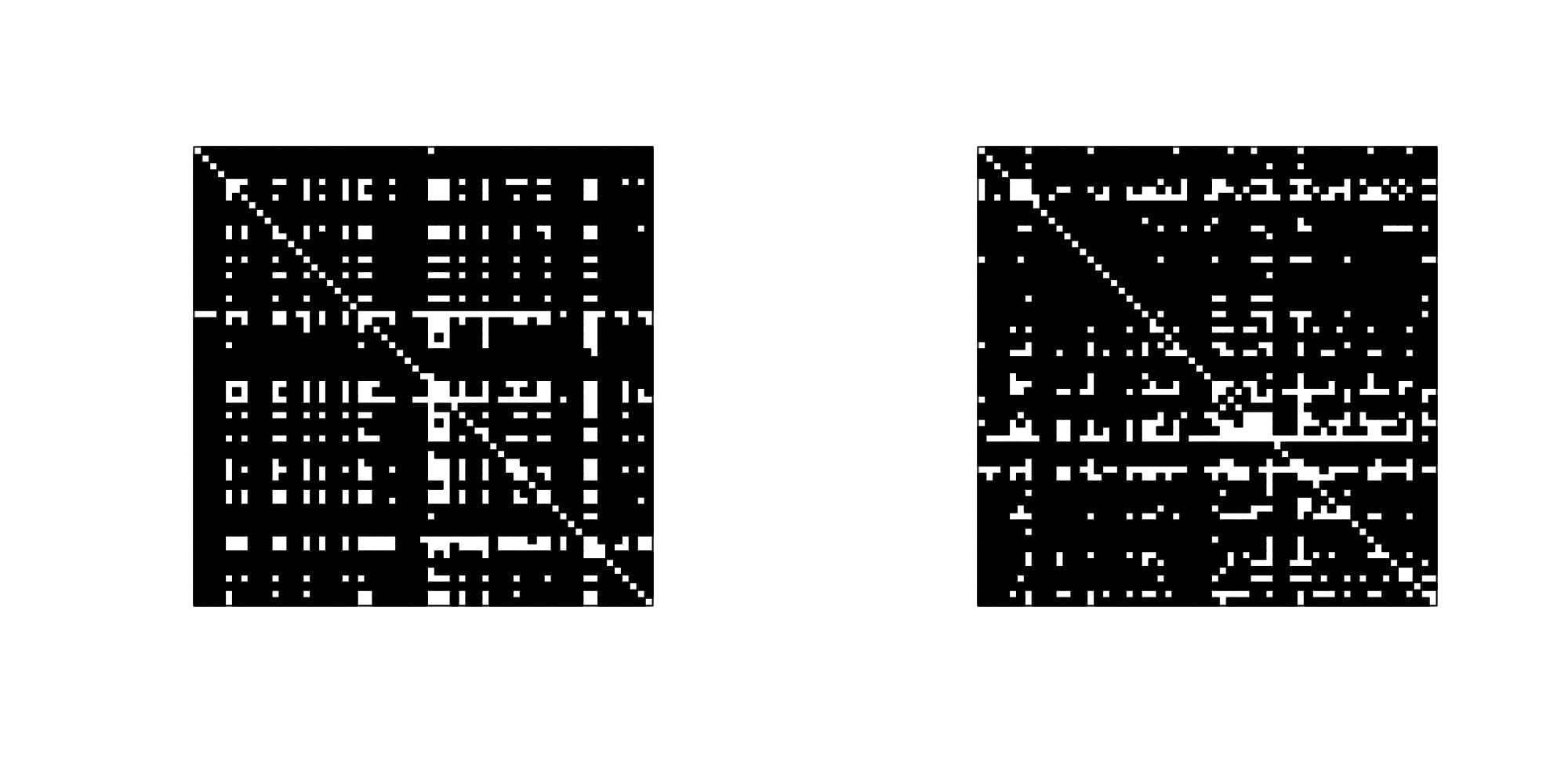}
		\end{subfigure}

		\caption{Adjacency matrices, representing the reduced MT 103 network in February 2018. H-FIT reconstruction (left), real network (right). \\ Source: SWIFT BI Watch.}
		\label{fig:recon_fit_small}
	\end{figure}

	\begin{figure}[!htbp]
		\centering
		\begin{subfigure}{\textwidth}
			\centering			\includegraphics[trim={2cm 2.5cm 1.5cm 2cm},clip,width=\textwidth]{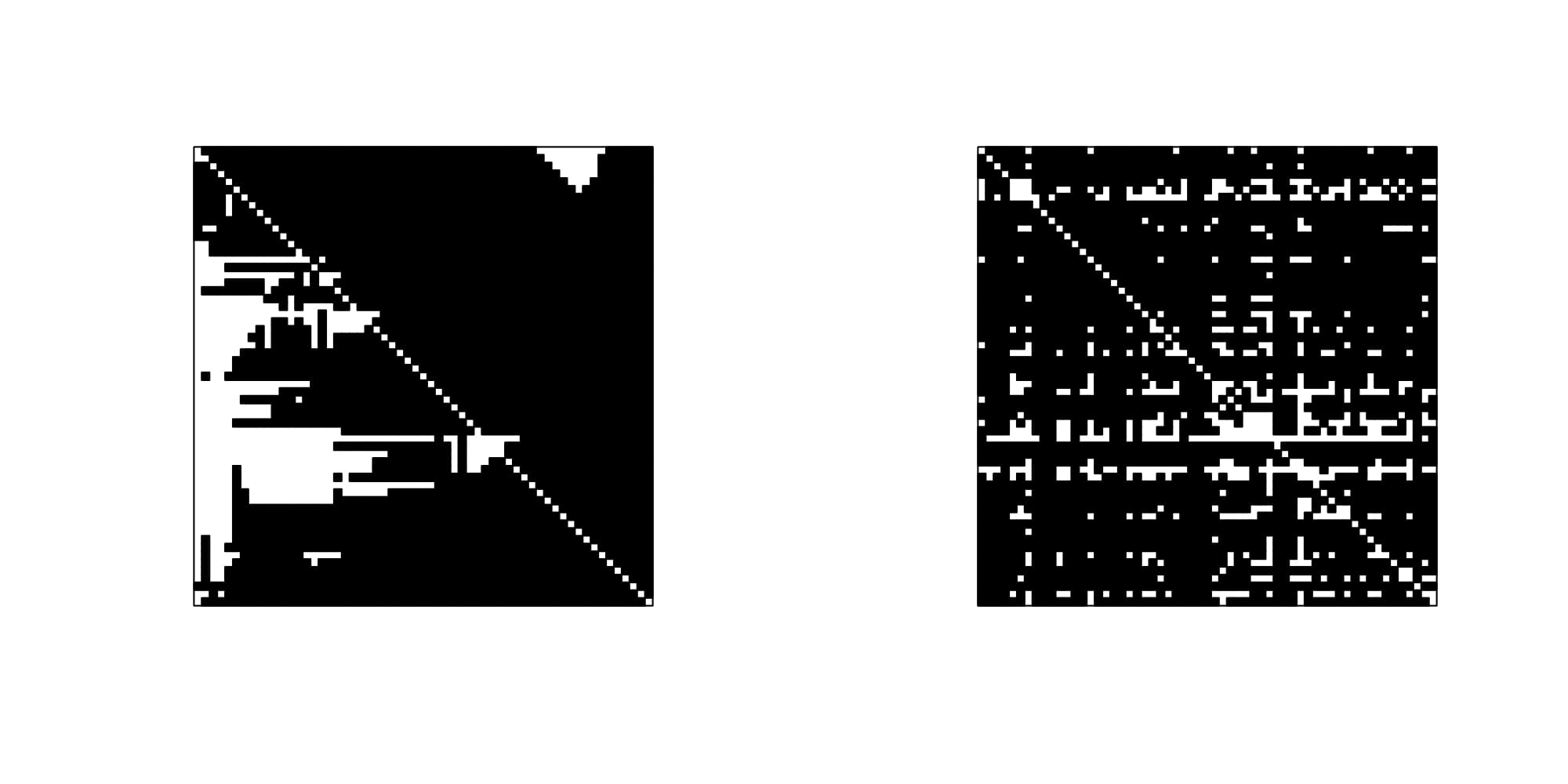}
		\end{subfigure}

		\caption{Adjacency matrices, representing the reduced MT 103 network in February 2018. LASSO reconstruction (left), real network (right). \\ Source: SWIFT BI Watch.}
		\label{fig:recon_lasso_small}
	\end{figure}
	
	\begin{figure}[!htbp]
		\centering
		\begin{subfigure}{\textwidth}
			\centering			\includegraphics[trim={2cm 2.5cm 1.5cm 2cm},clip,width=\textwidth]{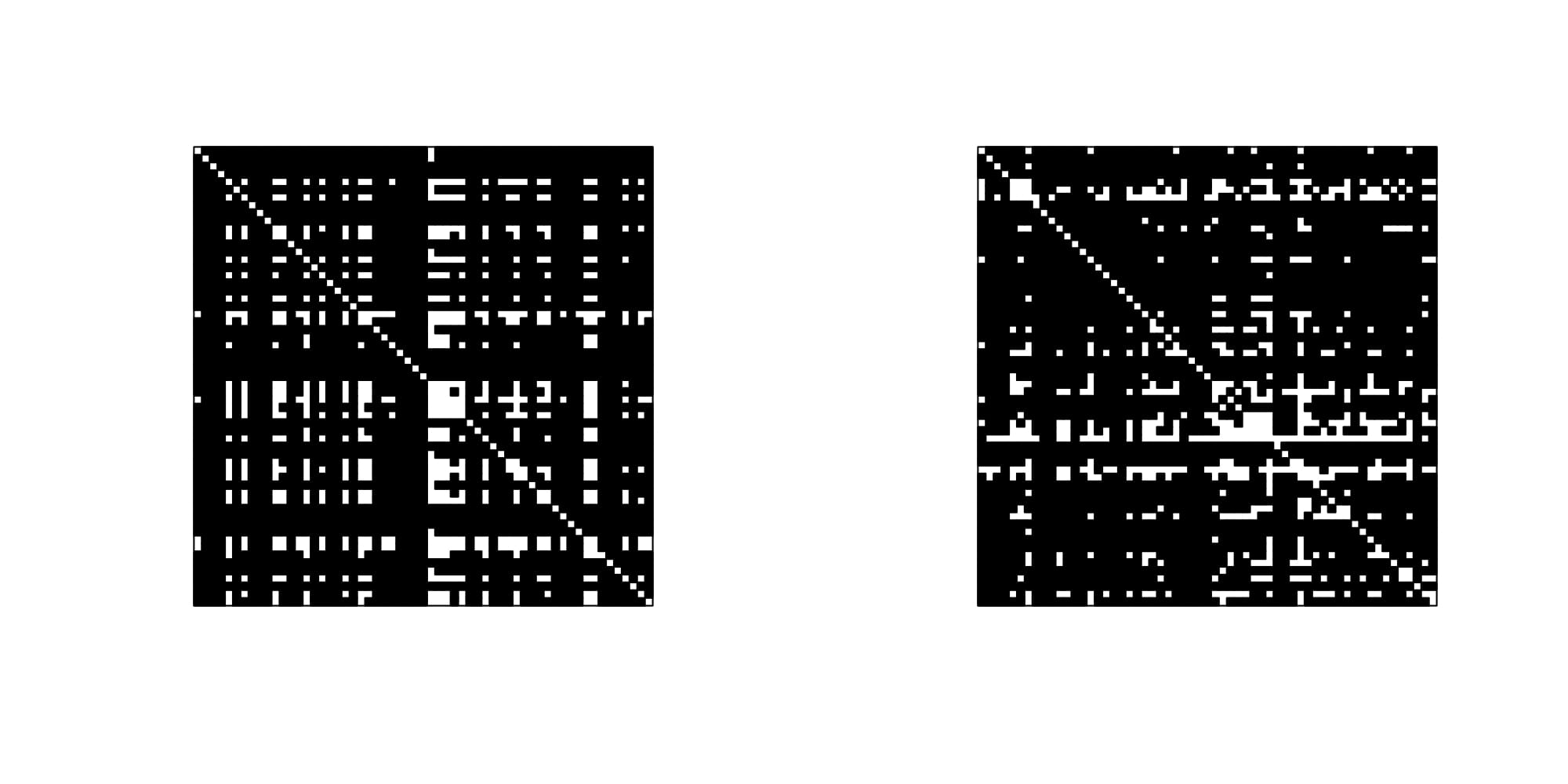}
		\end{subfigure}

		\caption{Adjacency matrices, representing the reduced MT 103 network in February 2018. IPFP-GDP reconstruction (left), real network (right). \\ Source: SWIFT BI Watch.}
		\label{fig:recon_regression_gdp_small}
	\end{figure}
	\begin{figure}[!htbp]
		\centering
		\begin{subfigure}{\textwidth}
			\centering			\includegraphics[trim={2cm 2.5cm 1.5cm 2cm},clip,width=\textwidth]{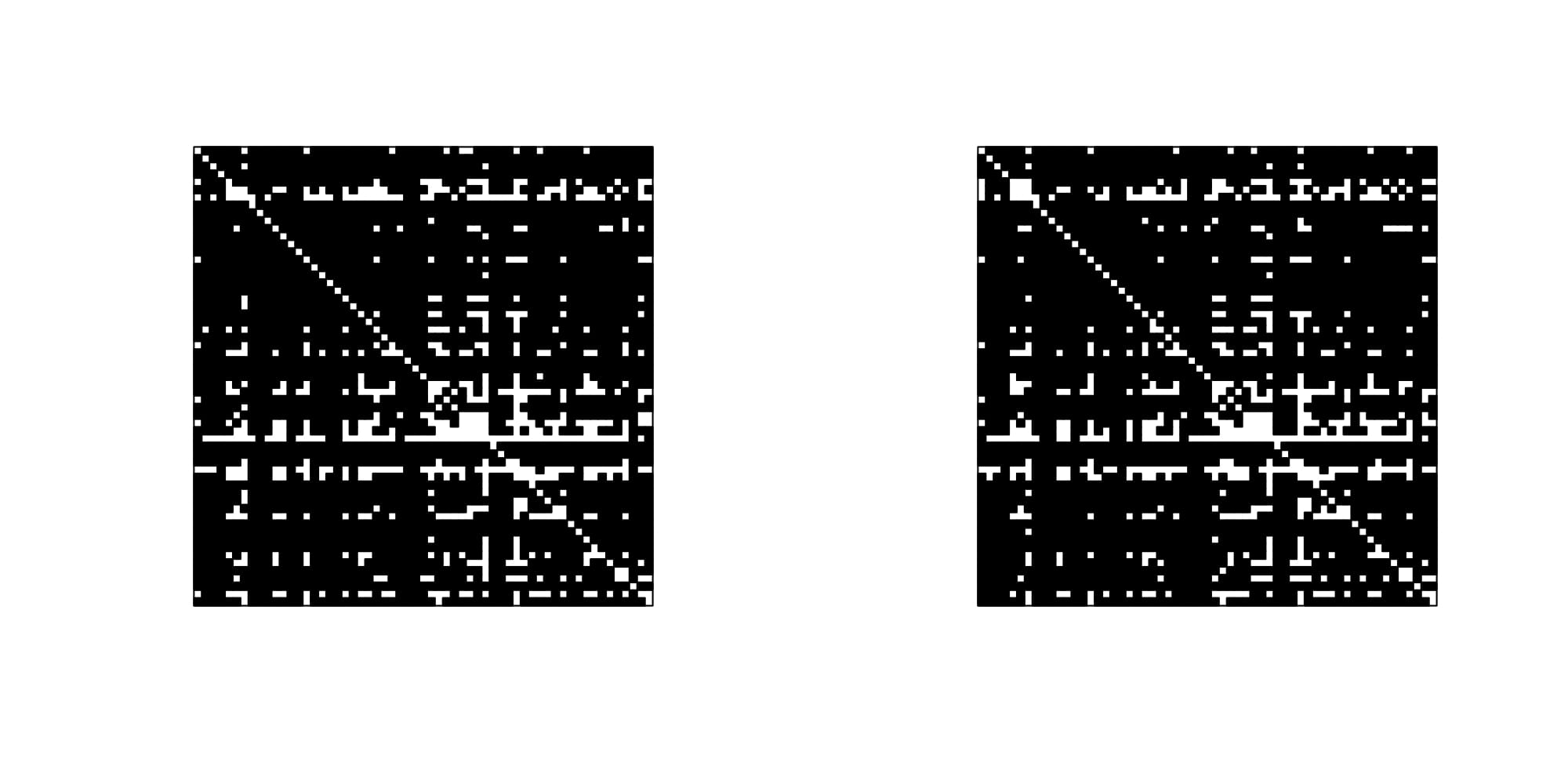}
		\end{subfigure}

		\caption{Adjacency matrices, representing the reduced MT 103 network in February 2018. IPFP-LAG reconstruction (left), real network (right). \\ Source: SWIFT BI Watch.}
		\label{fig:recon_regression_lag_small}
	\end{figure}
	
	\begin{figure}[!htbp]
		\centering
		\begin{subfigure}{\textwidth}
			\centering			\includegraphics[trim={2cm 2.5cm 1.5cm 2cm},clip,width=\textwidth]{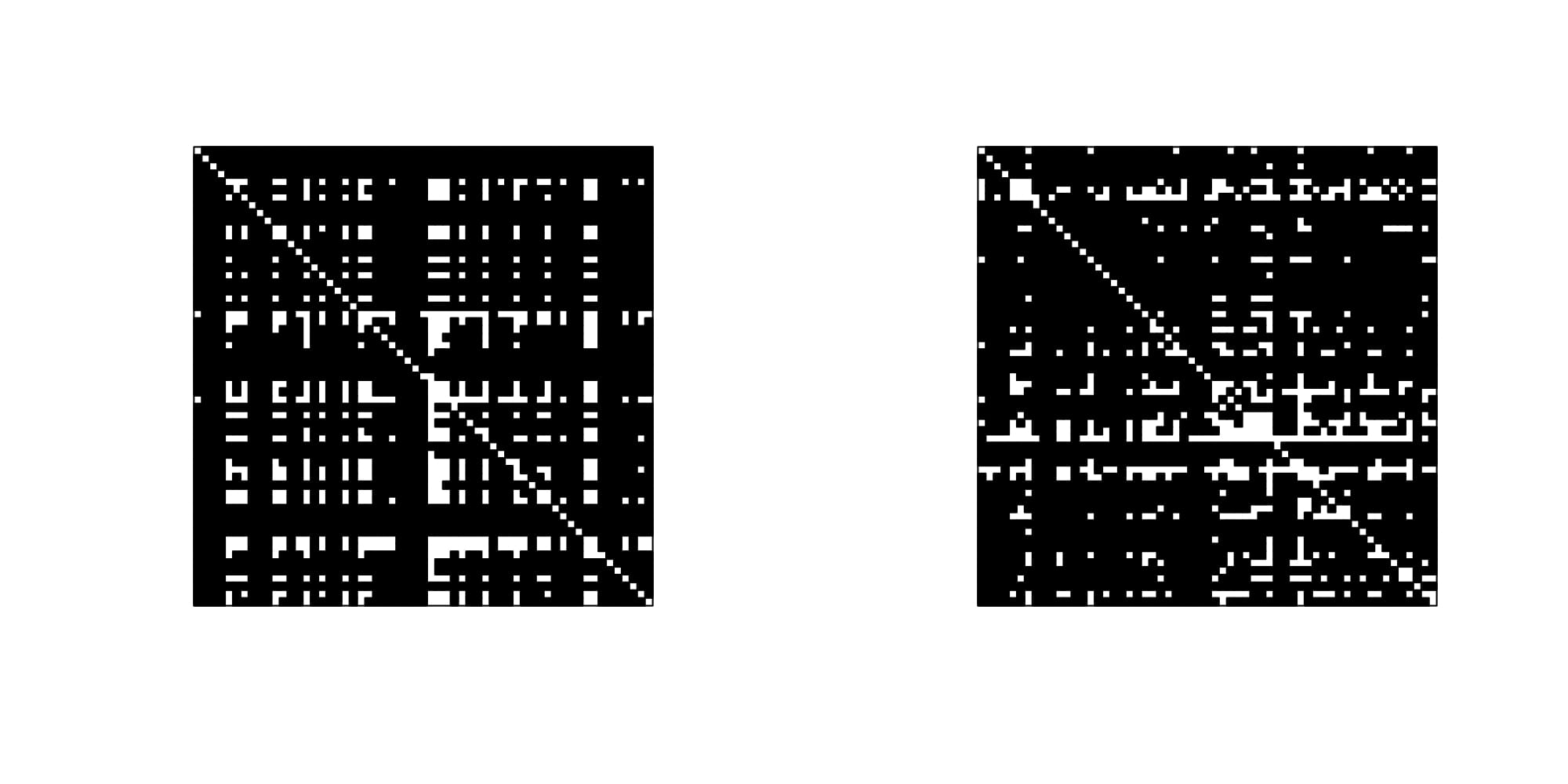}
		\end{subfigure}

		\caption{Adjacency matrices, representing the reduced MT 103 network in February 2018. TOMOGRAVITY reconstruction (left), real network (right). \\ Source: SWIFT BI Watch.}
		\label{fig:recon_tomo_small}
	\end{figure}

	\FloatBarrier
\newpage	
	\subsection{Degree Reconstruction: Reduced Network}\label{annex:deg_small}
	\begin{figure}[!htbp]
		\centering
		\begin{subfigure}{\textwidth}
			\centering			\includegraphics[trim={0cm 0cm 0cm 0cm},clip,width=\textwidth]{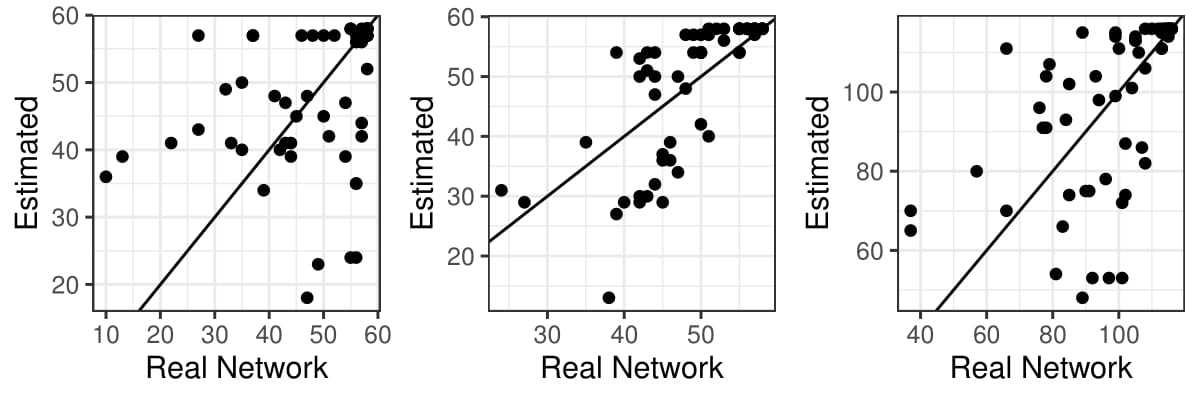}
		\end{subfigure}

		\caption{Degree Reconstruction in the reduced MT 103 network in February 2018. IPFP reconstruction of the outdegree (left), outdegree (middle) and in- and outdegree (right). \\ Source: SWIFT BI Watch.}
		\label{fig:deg_recon_ipfp_small}
	\end{figure}
	
	\begin{figure}[!htbp]
		\centering
		\begin{subfigure}{\textwidth}
			\centering			\includegraphics[trim={0cm 0cm 0cm 0cm},clip,width=\textwidth]{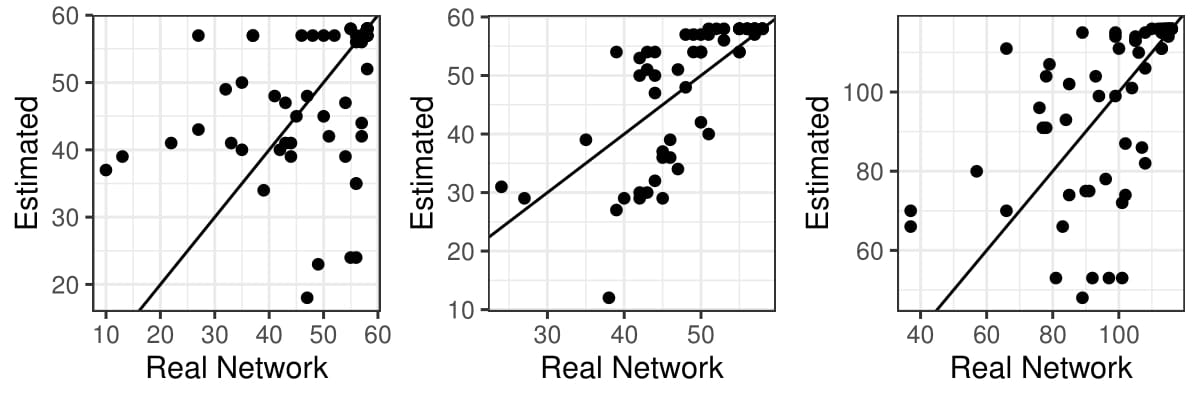}
		\end{subfigure}

		\caption{Degree Reconstruction in the reduced MT 103 network in February 2018. GRAVITY reconstruction of the outdegree (left), outdegree (middle) and in- and outdegree (right). \\ Source: SWIFT BI Watch.}
		\label{fig:deg_recon_gravity_small}
	\end{figure}
	
	\begin{figure}[!htbp]
		\centering
		\begin{subfigure}{\textwidth}
			\centering			\includegraphics[trim={0cm 0cm 0cm 0cm},clip,width=\textwidth]{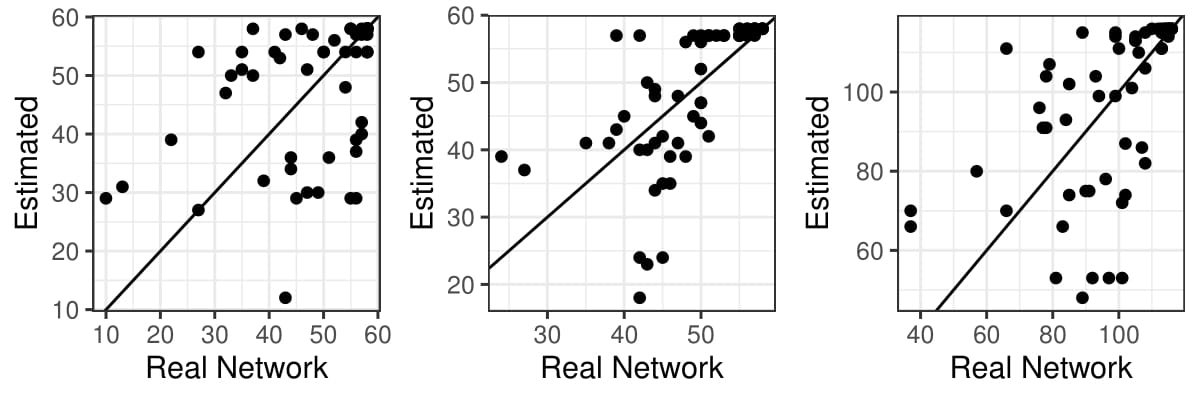}
		\end{subfigure}

		\caption{Degree Reconstruction in the reduced MT 103 network in February 2018. DC-GRAVITY reconstruction of the outdegree (left), outdegree (middle) and in- and outdegree (right). \\ Source: SWIFT BI Watch.}
		\label{fig:deg_recon_cimi_small}
	\end{figure}
	
	\begin{figure}[!htbp]
		\centering
		\begin{subfigure}{\textwidth}
			\centering			\includegraphics[trim={0cm 0cm 0cm 0cm},clip,width=\textwidth]{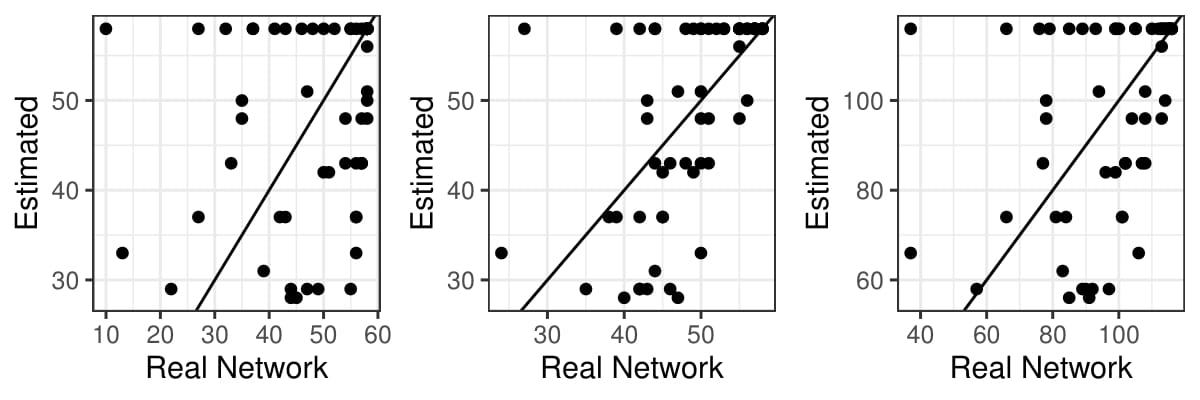}
		\end{subfigure}

		\caption{Degree Reconstruction in the reduced MT 103 network in February 2018. DC-GRAVITY-GDP reconstruction of the outdegree (left), outdegree (middle) and in- and outdegree (right). \\ Source: SWIFT BI Watch.}
		\label{fig:deg_recon_cimi_gdp_small}
	\end{figure}
	
	\begin{figure}[!htbp]
		\centering
		\begin{subfigure}{\textwidth}
			\centering			\includegraphics[trim={0cm 0cm 0cm 0cm},clip,width=\textwidth]{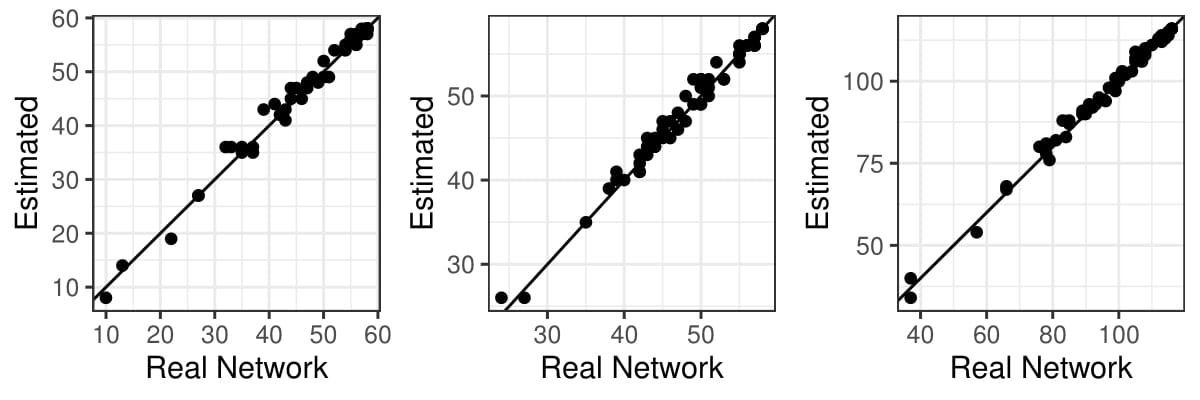}
		\end{subfigure}

		\caption{Degree Reconstruction in the reduced MT 103 network in February 2018. DC-GRAVITY-LAG reconstruction of the outdegree (left), outdegree (middle) and in- and outdegree (right). \\ Source: SWIFT BI Watch.}
		\label{fig:deg_recon_cimi_lag_small}
	\end{figure}

	\begin{figure}[!htbp]
		\centering
		\begin{subfigure}{\textwidth}
			\centering			\includegraphics[trim={0cm 0cm 0cm 0cm},clip,width=\textwidth]{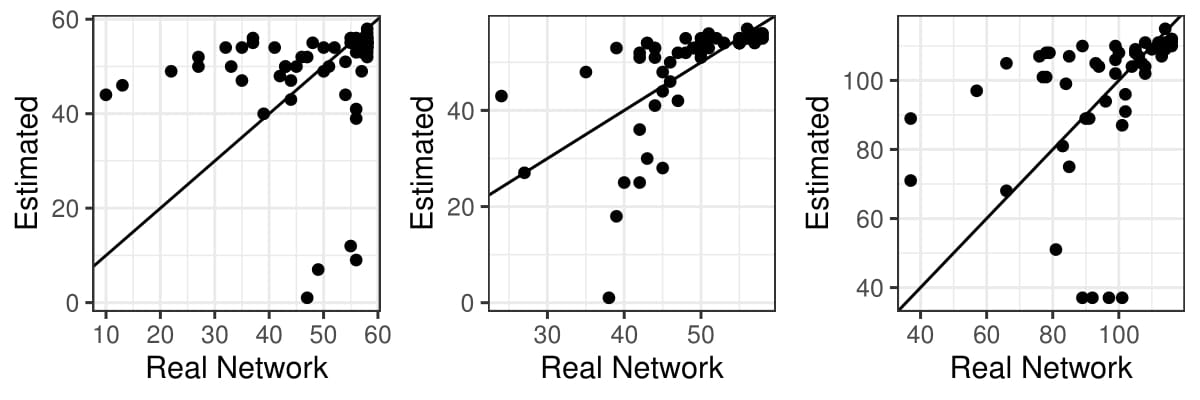}
		\end{subfigure}

		\caption{Degree Reconstruction in the reduced MT 103 network in February 2018. H-ER reconstruction of the outdegree (left), outdegree (middle) and in- and outdegree (right). \\ Source: SWIFT BI Watch.}
		\label{fig:deg_recon_er_small}
	\end{figure}
	
	\begin{figure}[!htbp]
		\centering
		\begin{subfigure}{\textwidth}
			\centering			\includegraphics[trim={0cm 0cm 0cm 0cm},clip,width=\textwidth]{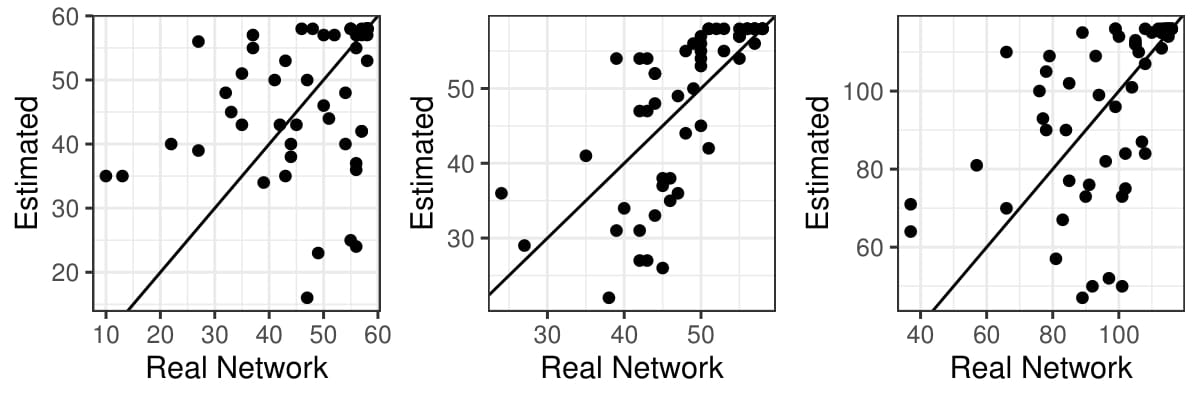}
		\end{subfigure}

		\caption{Degree Reconstruction in the reduced MT 103 network in February 2018. H-FIT reconstruction of the outdegree (left), outdegree (middle) and in- and outdegree (right). \\ Source: SWIFT BI Watch.}
		\label{fig:deg_recon_fit_small}
	\end{figure}

	\begin{figure}[!htbp]
		\centering
		\begin{subfigure}{\textwidth}
			\centering			\includegraphics[trim={0cm 0cm 0cm 0cm},clip,width=\textwidth]{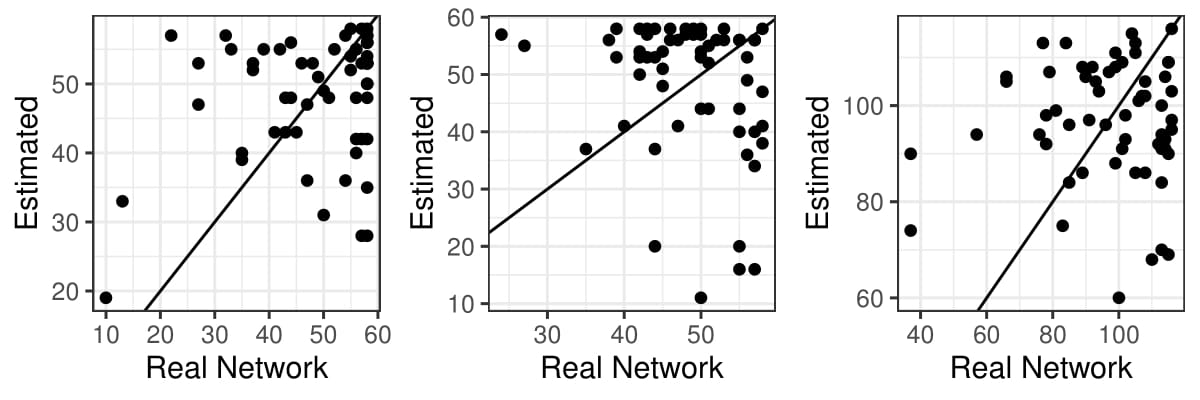}
		\end{subfigure}

		\caption{Degree Reconstruction in the reduced MT 103 network in February 2018. LASSO reconstruction of the outdegree (left), outdegree (middle) and in- and outdegree (right). \\ Source: SWIFT BI Watch.}
		\label{fig:deg_recon_lasso_small}
	\end{figure}
	
	\begin{figure}[!htbp]
		\centering
		\begin{subfigure}{\textwidth}
			\centering			\includegraphics[trim={0cm 0cm 0cm 0cm},clip,width=\textwidth]{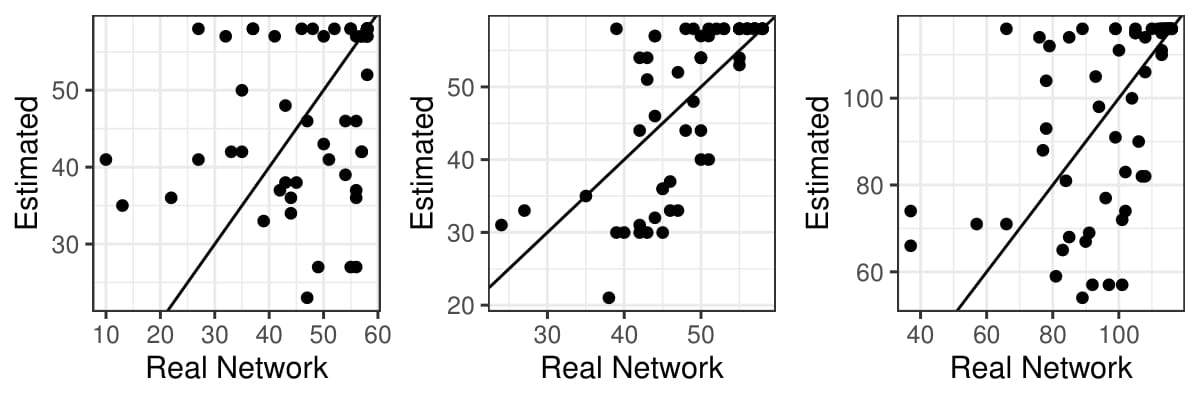}
		\end{subfigure}

		\caption{Degree Reconstruction in the reduced MT 103 network in February 2018. IPFP-GDP reconstruction of the outdegree (left), outdegree (middle) and in- and outdegree (right). \\ Source: SWIFT BI Watch.}
		\label{fig:deg_recon_regression_gdp_small}
	\end{figure}
	
	\begin{figure}[!htbp]
		\centering
		\begin{subfigure}{\textwidth}
			\centering			\includegraphics[trim={0cm 0cm 0cm 0cm},clip,width=\textwidth]{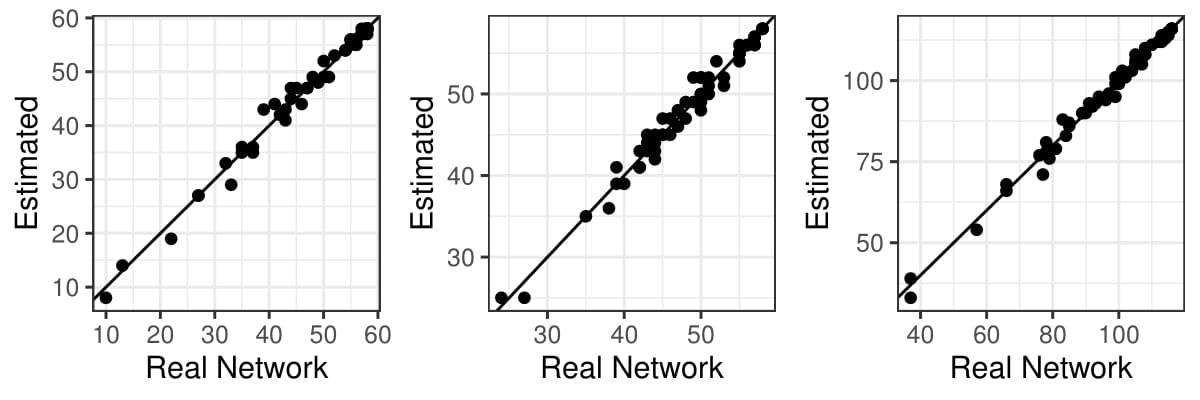}
		\end{subfigure}

		\caption{Degree Reconstruction in the reduced MT 103 network in February 2018. IPFP-LAG reconstruction of the outdegree (left), outdegree (middle) and in- and outdegree (right). \\ Source: SWIFT BI Watch.}
		\label{fig:deg_recon_regression_lag_small}
	\end{figure}
	
	\begin{figure}[!htbp]
		\centering
		\begin{subfigure}{\textwidth}
			\centering			\includegraphics[trim={0cm 0cm 0cm 0cm},clip,width=\textwidth]{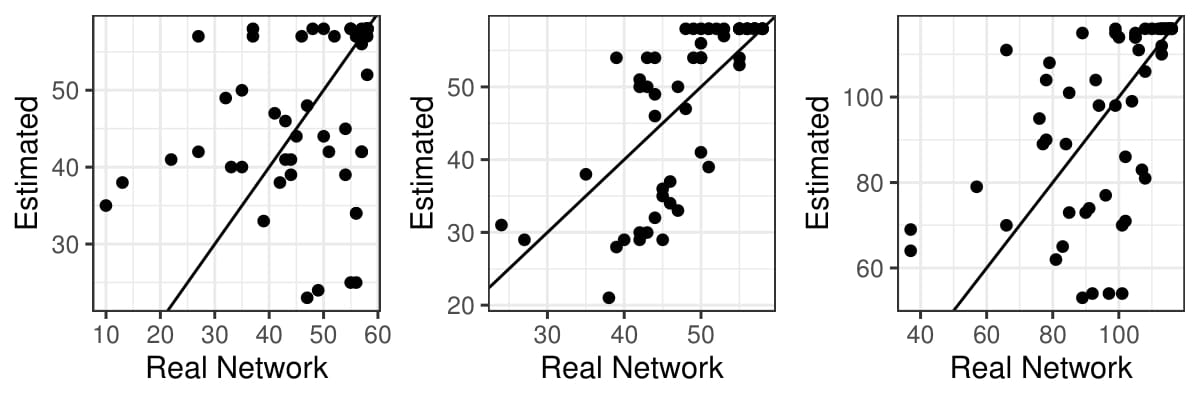}
		\end{subfigure}

		\caption{Degree Reconstruction in the reduced MT 103 network in February 2018. TOMOGRAVTIY reconstruction of the outdegree (left), outdegree (middle) and in- and outdegree (right). \\ Source: SWIFT BI Watch.}
		\label{fig:deg_recon_tomo_small}
	\end{figure}

\end{document}